\documentclass[manuscript,screen]{acmart}

\AtBeginDocument{%
  }

\usepackage{booktabs}
\usepackage{array}
\usepackage{ragged2e}
\usepackage{graphicx}
\usepackage{enumitem}
\usepackage{makecell}
\usepackage{adjustbox}
\usepackage{silence}
\usepackage{tikz}
\usetikzlibrary{shapes.geometric, arrows.meta, positioning, calc, fit, backgrounds}
\usepackage{forest}
\usepackage[table]{xcolor} % Required for \rowcolor

\definecolor{upstreamGray}{RGB}{240, 240, 240}
\definecolor{downstreamBlue}{RGB}{230, 240, 255}
\definecolor{feedbackRed}{RGB}{180, 0, 0}

\usepackage{hyperref}

\usepackage{longtable}
\usepackage{placeins}

\newcolumntype{L}[1]{>{\RaggedRight\arraybackslash}p{#1}}

\newcolumntype{Y}[1]{>{\RaggedRight\arraybackslash}p{#1}}

\setcopyright{none}
\acmJournal{CSUR}
\graphicspath{{./}}

\title[AI-Driven Alert Screening in SOCs]{AI-Driven Security Alert Screening and Alert Fatigue Mitigation in Security Operations Centers: A Survey}
\author{Samuel Ndichu}
\authornote{Corresponding author.}
\orcid{0000-0001-9632-2407}
\affiliation{%
  \institution{National Institute of Information and Communications Technology}
  \city{Tokyo}
  \country{Japan}
}
\email{ndichu@nict.go.jp}

\author{Tao Ban}
\orcid{0000-0002-9616-3212}
\affiliation{%
  \institution{National Institute of Information and Communications Technology}
  \city{Tokyo}
  \country{Japan}
}

\author{Seiichi Ozawa}
\orcid{0000-0002-0965-0064}
\affiliation{%
  \institution{Kobe University}
  \city{Kobe}
  \country{Japan}
}

\author{Takeshi Takahashi}
\orcid{0000-0002-6477-7770}
\affiliation{%
  \institution{National Institute of Information and Communications Technology}
  \city{Tokyo}
  \country{Japan}
}

\author{Daisuke Inoue}
\orcid{0000-0002-4373-0834}
\affiliation{%
  \institution{National Institute of Information and Communications Technology}
  \city{Tokyo}
  \country{Japan}
}

\begin{document}
\sloppy
\begin{abstract}
Security alert screening is the downstream task of filtering, prioritizing, correlating, and contextualizing alerts for analyst attention in Security Operations Centers. This survey reviews artificial-intelligence-driven alert screening and alert-fatigue mitigation from 2015--2026. We synthesize 119 records, including 87 core studies, into a four-stage workflow taxonomy covering filtering, triage, correlation, and generative augmentation. We find persistent gaps in operational validation, adversarial robustness, cross-environment generalization, and evaluation practice. The survey concludes with a research agenda toward trustworthy Cognitive Security Operations Centers.
\end{abstract}



\maketitle

\noindent\textit{This is the author's preprint of a manuscript submitted to ACM Computing Surveys (May 2026). The version of record will appear at {\upshape\texttt{https://doi.org/10.1145/...}} once published.}

\section{Introduction}

The Security Operations Center (SOC) is the operational hub of modern cybersecurity defense. Intrusion detection systems (IDS), endpoint detection and response (EDR) platforms, and security information and event management (SIEM) engines generate alerts upstream. This survey addresses the downstream task, security alert screening: filtering, prioritizing, correlating, and contextualizing already-generated alerts to determine which warrant analyst attention. The upstream/downstream distinction is formalized in Section~\ref{sec:formal-task}; Artificial Intelligence (AI) techniques applied to upstream detection improvement are a separate and well-surveyed field \cite{S1,S2,S3,S4,ref24}. The downstream problem remains comparatively under-surveyed despite its direct link to alert fatigue, analyst burnout \cite{ref5,ref99}, and extended breach dwell times \cite{ref1}.

Academic and practitioner evidence align on the operational burden. Industry reports describe high daily alert volumes, widespread false-positive (FP) burden, and growing investigation backlogs in enterprise SOCs \cite{ref3,ref4,ref5,ref6,ref15}. The IBM 2024 Cost of a Data Breach Report places average breach cost at \$4.88M \cite{ref1}. Under sustained low-signal alert streams, analysts adopt coping behaviors such as blanket suppression rules, severity-based skimming, and over-reliance on tool severity scores, creating blind spots exploitable by sophisticated adversaries \cite{ref9}.

AI offers a structural response. Machine learning (ML) models can identify FP patterns in historical alert data; deep learning (DL) can correlate multi-source telemetry into multi-stage attack narratives; reinforcement learning (RL) can adapt prioritization policies to analyst feedback; and large language models (LLMs) can summarize incidents, draft playbooks, and augment investigation through conversational interfaces.

\subsection{Scope and Contributions}

This survey reviews AI techniques applied to SOC alert fatigue and downstream screening over a 2015--2026 systematic search window. Relative to prior surveys on human--AI teaming (HAT), prioritization, or LLM use cases, our focus is cross-workflow synthesis: we map benchmark datasets to taxonomic categories, trace the Evidence-Quality Gap between Tier~1 field evidence and Tier~2/3 benchmark or conceptual evidence across all four workflow stages, and examine how downstream screening outcomes can inform upstream tuning. Section~\ref{sec:competing-surveys} positions this contribution against directly competing surveys. Our primary contributions are:

\begin{itemize}[leftmargin=*]
\item A workflow-aligned four-dimensional taxonomy of AI approaches to the security alert screening task, enabling practitioners to identify solutions by operational need. The taxonomy covers the complete screening pipeline: noise reduction before the analyst queue, prioritized triage of queued alerts, multi-alert correlation into incident hypotheses, and Generative AI (GenAI) augmentation of analyst workflows.
\item Paper-level comparison tables for each taxonomic category, encoding method, dataset/setting, key metric, and a notable limitation for representative studies, together with a complete mapping of all 87 core studies to categories and source types in the supplementary material.
\item A systematic review of 22 benchmark and alert-level datasets, classified by primary research orientation (LLM, ML/DL, or upstream-only) and annotated with publicly reported class-imbalance ranges and representational gaps relative to real SOC environments.
\item An integrated analysis of HAT architectures, explainable AI (XAI) adoption barriers, reinforcement learning from human feedback (RLHF), and privacy-preserving federated learning (FL), cross-cutting themes not jointly emphasized at this depth in any single existing survey \cite{T1,T2,T3,T4,T5,T6,T7}.
\item A structured nine-direction research agenda derived from identified open problems, with operationally grounded prioritization criteria.
\end{itemize}

\subsection{Survey Methodology}
\label{sec:methodology}
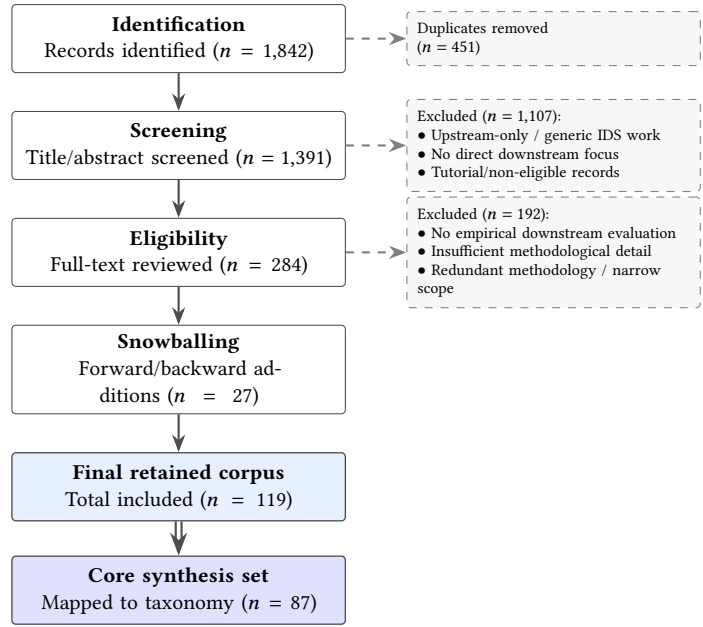
\begin{figure}[!tbp]
\centering
\begin{tikzpicture}[
    node distance=0.5cm,
    block/.style={rectangle, draw=black!70, fill=white, text width=4.2cm, text centered, minimum height=0.9cm, font=\small, line width=0.4pt, rounded corners=1.5pt},
    excl/.style={rectangle, draw=black!50, dashed, fill=gray!6, text width=3.6cm, font=\scriptsize, align=left, inner sep=4pt, rounded corners=1.5pt},
    final/.style={rectangle, draw=black!70, fill=downstreamBlue, text width=4.2cm, text centered, minimum height=0.9cm, font=\small, line width=0.5pt, rounded corners=1.5pt},
    core/.style={rectangle, draw=black!70, fill=blue!12, text width=4.2cm, text centered, minimum height=0.9cm, font=\small, line width=0.5pt, rounded corners=1.5pt},
    arrow/.style={-Stealth, thick, draw=black!70},
    excarrow/.style={-Stealth, thick, draw=black!50, dashed}
]
    % Nodes
    \node (id) [block] {\textbf{Identification}\\Records identified ($n = 1{,}842$)};
    \node (dup) [excl, right=0.8cm of id] {Duplicates removed\\($n = 451$)};

    \node (screen) [block, below=of id] {\textbf{Screening}\\Title/abstract screened ($n = 1{,}391$)};
    \node (ex_screen) [excl, right=0.8cm of screen] {Excluded ($n = 1{,}107$):\\$\bullet$ Upstream-only / generic IDS work\\$\bullet$ No direct downstream focus\\$\bullet$ Tutorial/non-eligible records};

    \node (elig) [block, below=of screen] {\textbf{Eligibility}\\Full-text reviewed ($n = 284$)};
    \node (ex_elig) [excl, right=0.8cm of elig] {Excluded ($n = 192$):\\$\bullet$ No empirical downstream evaluation\\$\bullet$ Insufficient methodological detail\\$\bullet$ Redundant methodology / narrow scope};

    \node (snow) [block, below=of elig] {\textbf{Snowballing}\\Forward/backward additions ($n = 27$)};

    \node (incl) [final, below=of snow] {\textbf{Final retained corpus}\\Total included ($n = 119$)};
    \node (core) [core, below=0.45cm of incl] {\textbf{Core synthesis set}\\Mapped to taxonomy ($n = 87$)};

    % Arrows
    \draw [excarrow] (id) -- (dup);
    \draw [arrow] (id) -- (screen);
    \draw [excarrow] (screen) -- (ex_screen);
    \draw [arrow] (screen) -- (elig);
    \draw [excarrow] (elig) -- (ex_elig);
    \draw [arrow] (elig) -- (snow);
    \draw [arrow] (snow) -- (incl);
    \draw [arrow, double, double distance=1.4pt] (incl) -- (core);
\end{tikzpicture}
\caption{PRISMA-2020 selection flow.}
\Description{A PRISMA-2020 flow diagram showing four solid blocks (Identification, Screening, Eligibility, Snowballing) connected by solid arrows, with dashed side-blocks showing the records excluded at each stage. A filled block at the bottom represents the final retained corpus of 119 records, with a further block below representing the core synthesis set of 87 records.}
\label{fig:prisma}
\end{figure}

% Updated unified methodology covering 2015--2026
Our literature selection follows PRISMA-2020 (Preferred Reporting Items for Systematic Reviews and Meta-Analyses)~\cite{ref7}. We searched the major bibliographic indices (IEEE Xplore, ACM Digital Library, USENIX, arXiv cs.CR/cs.LG, Semantic Scholar, Google Scholar) over the 2015--2026 window, supplemented by hand-search of leading security and SOC venues; the full venue list, search strings, and a targeted provenance/audit-log recovery query are reproduced in Supplementary Table~S0 and \texttt{search\_queries.md}. The same screening pipeline included a bibliography audit of the directly competing surveys~\cite{T1,T2}, contributing seven snowballed additions; the remaining twenty snowballed records came from citation chaining on retained Tier~1 and Tier~2 studies and from seminal pre-2017 works re-introduced for problem framing. Table~\ref{tab:prisma-pipeline} and Figure~\ref{fig:prisma} summarize the pipeline.
\begin{table}[!tbp]
\centering
%\scriptsize
%\setlength{\tabcolsep}{3pt}
\renewcommand{\arraystretch}{1.15}
\caption{PRISMA-2020 literature selection pipeline.}
\label{tab:prisma-pipeline}
\begin{adjustbox}{max width=\linewidth}
\begin{tabular}{@{}L{0.31\linewidth}L{0.51\linewidth}L{0.12\linewidth}@{}}
\toprule
PRISMA Stage & Criterion & Count \\
\midrule
Records identified (database search) & Title/abstract relevance & n = 1,842 \\
After duplicate removal & N/A & n = 1,391 \\
After title/abstract screen & Direct SOC/IDS + AI relevance & n = 284 \\
After full-text eligibility screen & Core-synthesis eligibility and methodological transparency & n = 92 \\
Added via snowball sampling & Citation chaining, seminal pre-2017 works, and competitor-survey bibliography audit & n = 27 \\
Final included records & N/A & n = 119 \\
\bottomrule
\end{tabular}
\end{adjustbox}
\end{table}

Inclusion in the 87-paper core synthesis required (a) direct application of AI/ML to alert management, FP reduction, alert correlation, or analyst workflow augmentation; (b) empirical evaluation with reported quantitative results; (c) sufficient methodological detail for the dataset/evaluation setting and metrics; and (d) publication at a peer-reviewed venue or, for arXiv manuscripts, at least 10 Semantic Scholar citations as of the 31 March--5 April 2026 window. A bounded exception retains three 2025 preprints (Learning to Defer with Human Feedback (L2DHF)~\cite{ref27}, Singh et al.~\cite{ref43}, and the blue-team LLM benchmark~\cite{ref61}) that both screeners judged methodologically transparent and directly relevant to downstream screening; post-lock corroborating entries are treated separately in Section~\ref{sec:validity-reproducibility}. At title/abstract screening, Cohen's $\kappa = 0.82$; Section~\ref{sec:threats-validity} gives the rater-agreement details. Full-text eligibility used the same rubric. Deduplication matched DOIs and normalized titles, with manual reconciliation of ambiguous duplicates. Records that improved upstream intrusion detection without an explicit downstream focus were excluded from the core synthesis. Of the 119 retained records, 87 are core; the remainder are competitor surveys, foundational background references, or contextual studies for problem framing and evaluation discussion.

\paragraph{Evidence-Quality Scheme.} Each core study is assigned an evidence tier used in the per-category tables. \textbf{Tier~1} (production/field) splits into \textbf{T1a} = peer-reviewed intervention with analyst-in-the-loop validation (e.g., Ban et al.~\cite{ref63,ref18}, Aminanto et al.~\cite{ref69}, DeepCASE~\cite{ref62}; full T1a list in Section~\ref{sec:cross-cat-synthesis}); \textbf{T1b} = vendor or preprint production evidence (Microsoft Security Copilot randomized controlled trial (RCT)~\cite{ref47}, Singh et al.~\cite{ref43}); \textbf{FM} = field measurement on real SOC telemetry without an intervention (Yang et al.~\cite{ref120}). \textbf{Tier~2} = public benchmarks (DARPA TC, CICIDS, CERT) or simulation with explicit threat scenarios. \textbf{Tier~3} = limited/synthetic data, simplified simulation, or no quantitative evaluation. Preprint status is orthogonal to tier. The T1/T2/T3 markers appear in the `Notable Finding / Limitation' column with T1 sub-markers (a/b) and FM noted inline. All assignments are released in \texttt{core\_studies.csv}; self-authored core studies were tier-coded by the non-conflicted coauthors. Tier and Automation--Augmentation--Collaboration (A2C) assignments are author-coded rubric-guided descriptors, not separately inter-rater coded; Table~\ref{tab:evidence-landscape} should be read accordingly.

\subsection{Paper Organization}

The remainder is organized as: background and taxonomy (Section~\ref{sec:background-taxonomy}), workflow review (Section~\ref{sec:workflow-taxonomy}), cross-cutting themes (Section~\ref{sec:cross-cutting}), datasets (Section~\ref{sec:datasets}), cross-category synthesis (Section~\ref{sec:cross-cat-synthesis}), open challenges (Section~\ref{sec:open-challenges}), research agenda (Section~\ref{sec:research-agenda}), validity and ethics (Section~\ref{sec:validity-reproducibility}), and conclusion (Section~\ref{sec:conclusion}).

\section{Security Alert Screening: Background, Prior Surveys, and Taxonomy}
\label{sec:background-taxonomy}

This section grounds the survey: the modern SOC, the comparator audit, the evidence-quality rubric, the formal downstream task, and the four-stage taxonomy used by Sections~\ref{sec:workflow-taxonomy}--\ref{sec:cross-cutting}.

\subsection{The Modern Security Operations Center}

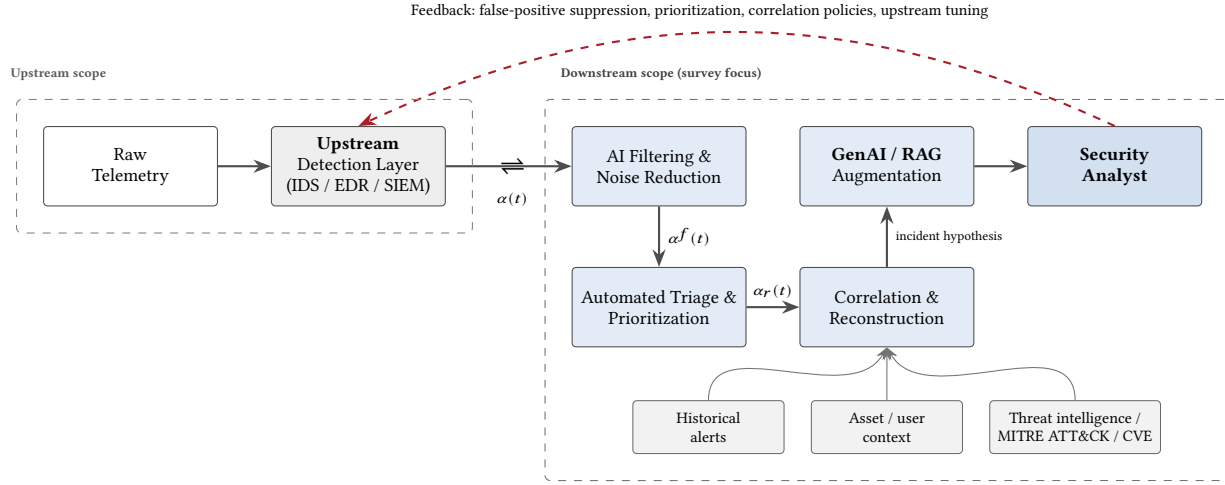
\begin{figure}[!tbp]
\centering
\definecolor{feedbackRed}{RGB}{170, 30, 45}
\definecolor{upstreamGray}{RGB}{238, 238, 238}
\definecolor{downstreamBlue}{RGB}{226, 235, 246}
\definecolor{analystAccent}{RGB}{210, 224, 240}

\begin{tikzpicture}[
    node distance=0.8cm and 0.7cm,
    box/.style={draw=black!70, line width=0.4pt, minimum width=2.3cm, minimum height=1.05cm, align=center, font=\footnotesize, fill=white, rounded corners=1.5pt},
    smallbox/.style={draw=black!60, line width=0.3pt, minimum width=2.0cm, minimum height=0.7cm, align=center, font=\scriptsize, fill=white!95!black, rounded corners=1.5pt},
    upbox/.style={box, fill=upstreamGray},
    downbox/.style={box, fill=downstreamBlue},
    analystbox/.style={box, fill=analystAccent},
    arrow/.style={-Stealth, thick, draw=black!70},
    sidearrow/.style={-Stealth, thin, draw=black!55},
    fb/.style={-Stealth, thick, dashed, draw=feedbackRed}
]
    % Top row (upstream)
    \node (telemetry) [box] {Raw\\Telemetry};
    \node (upstream) [upbox, right=of telemetry] {\textbf{Upstream}\\Detection Layer\\(IDS / EDR / SIEM)};

    % Alert queue glyph
    \node (queue) [inner sep=3pt, right=0.6cm of upstream, font=\large] {$\rightleftharpoons$};
    \node[below=0.02cm of queue, font=\tiny] {$\alpha(t)$};

    % Downstream main pipeline (linear, left to right)
    \node (filtering) [downbox, right=0.5cm of queue] {AI Filtering \&\\Noise Reduction};
    \node (triage) [downbox, below=of filtering] {Automated Triage \&\\Prioritization};
    \node (corr) [downbox, right=of triage] {Correlation \&\\Reconstruction};
    \node (genai) [downbox, above=of corr] {\textbf{GenAI / RAG}\\Augmentation};

    % Side inputs to Correlation (anchored directly below the Correlation node so arrows are unambiguous)
    \node (ctx) [smallbox, below=0.7cm of corr] {Asset / user\\context};
    \node (hist) [smallbox, left=0.35cm of ctx] {Historical\\alerts};
    \node (cti) [smallbox, right=0.35cm of ctx] {Threat intelligence /\\MITRE ATT\&CK / CVE};

    % Analyst
    \node (analyst) [analystbox, right=0.7cm of genai] {\textbf{Security}\\\textbf{Analyst}};

    % Main flow connectors with intermediate queue labels α_f(t), α_r(t), and hypothesis
    \draw [arrow] (telemetry) -- (upstream);
    \draw [arrow] (upstream) -- (filtering);
    \draw [arrow] (filtering) -- node[midway, right, font=\tiny] {$\alpha^{f}(t)$} (triage);
    \draw [arrow] (triage) -- node[midway, above, font=\tiny] {$\alpha_r(t)$} (corr);
    \draw [arrow] (corr) -- node[midway, right, font=\tiny] {incident hypothesis} (genai);
    \draw [arrow] (genai) -- (analyst);

    % Side inputs into Correlation, all three explicitly terminate on corr.south
    \draw [sidearrow] (hist.north) to[out=90, in=240] (corr.south);
    \draw [sidearrow] (ctx.north)  -- (corr.south);
    \draw [sidearrow] (cti.north)  to[out=90, in=300] (corr.south);

    % Feedback loop
    \path (analyst.north) edge [fb, bend right=25] node[above, font=\scriptsize, black, pos=0.55, yshift=2pt] {Feedback: false-positive suppression, prioritization, correlation policies, upstream tuning} (upstream.north);

    % Scope boundaries
    \node[draw=black!50, dashed, line width=0.4pt, inner sep=10pt, fit=(telemetry) (upstream), rounded corners=2pt] (upb) {};
    \node[anchor=south, draw=none, font=\tiny\bfseries, xshift=-70, yshift=4pt, color=black!65] at (upb.north) {Upstream scope};

    \node[draw=black!55, dashed, line width=0.4pt, inner sep=10pt, fit=(filtering) (corr) (genai) (analyst) (hist) (cti), rounded corners=2pt] (downb) {};
    \node[anchor=south, draw=none, font=\tiny\bfseries, xshift=-85, yshift=4pt, color=black!75] at (downb.north) {Downstream scope (survey focus)};
\end{tikzpicture}
\caption{Architecture of the downstream SOC alert-screening workflow.}
\Description{A two-layer architecture diagram with two dashed scope boxes. The upstream scope contains a Raw Telemetry block feeding an Upstream Detection Layer block (IDS/EDR/SIEM), which then passes alerts through an alert-queue glyph into the downstream scope. The downstream scope contains four blocks (AI Filtering and Noise Reduction, Correlation and Reconstruction, Automated Triage and Prioritization, GenAI/RAG Augmentation) that feed a Security Analyst block on the right. A dashed red feedback arrow runs from the analyst back to the upstream layer to indicate analyst-validated outcomes feeding upstream tuning.}
\label{fig:soc-architecture}
\end{figure}

A SOC integrates people, processes, and technology for continuous monitoring, detection, analysis, and response~\cite{ref8}. Tier 1 analysts perform initial triage, Tier 2 incident investigation and escalation, Tier 3 threat hunting and forensics, with documented analyst--manager disagreement on priorities and tool effectiveness~\cite{ref96}. The technology stack centers on the SIEM as aggregation and correlation layer~\cite{ref97}, augmented by EDR, network detection and response (NDR), user and entity behavior analytics (UEBA), threat intelligence platforms (TIPs), and security orchestration, automation, and response (SOAR). Figure~\ref{fig:soc-architecture} schematizes the upstream/downstream boundary that is this survey's focus: the upstream layer (IDS, network security monitoring (NSM), EDR engines, SIEM correlation rules) generates alerts from raw telemetry, and the downstream layer applies four workflow stages (filtering, prioritization, correlation, and analyst augmentation) to surface the subset requiring human investigation. Downstream quality is bounded by, but analytically distinct from, upstream quality.

Each layer generates its own alert stream. Vielberth et al. \cite{ref98} provide a systematic study of SOC building blocks and open challenges across deployment models. Practitioner reports describe large enterprise SIEM deployments ingesting tens to hundreds of billions of events per day; detection rule sets number in the hundreds to thousands and produce a comparable volume of alerts per shift~\cite{ref2}. Alert fatigue arises from the tension between high-recall detection rules, which reduce missed threats, and finite analyst capacity.

%\FloatBarrier
\subsection{Comparison With Directly Competing Surveys}
\label{sec:competing-surveys}

Seven directly relevant surveys or survey-adjacent synthesis papers appeared between 2024 and 2025, including three in ACM Computing Surveys. Table~\ref{tab:competing-surveys} enumerates them and clarifies the comparative niche of the present manuscript. The Habibzadeh et al.\ LLM-for-SOC survey~\cite{S5}, frozen post-lock, is discussed in Section~\ref{sec:validity-reproducibility}.

\begin{table*}[!htbp]
\centering
%\scriptsize
%\setlength{\tabcolsep}{3pt}
\renewcommand{\arraystretch}{1.15}
\caption{Comparison with directly competing surveys (2024--2025).}
\label{tab:competing-surveys}
\begin{adjustbox}{max width=\textwidth}
\small
\begin{tabular}{@{}L{0.14\linewidth}L{0.05\linewidth}L{0.20\linewidth}L{0.10\linewidth}L{0.10\linewidth}L{0.11\linewidth}L{0.11\linewidth}L{0.12\linewidth}L{0.07\linewidth}@{}}
\toprule
Survey & Year & Venue & Alert Screening & Prioritization & Correlation & LLM / GenAI & Adv.\ Robust.\ / Federated & Per-Paper Tables \\
\midrule
\cite{T1} Tariq et al. & 2025 & ACM CSUR & Brief & Brief & Brief & Brief & None & No \\
\cite{T2} Jalalvand et al. & 2024 & ACM CSUR & None & Primary & None & None & None & No \\
\cite{T3} Chhetri et al. & 2024 & ACM TOIT & None & Brief (HAT) & None & Brief & None & No \\
\cite{T4} Srinivas et al. & 2025 & J. Cybersecur.\ Priv. & None & Brief & None & Primary & None & No \\
\cite{T5} Binbeshr et al. & 2025 & IEEE Open J.\ Comput.\ Soc. & None & Brief & Brief & Brief & None & No \\
\cite{T6} Khayat et al. & 2025 & IEEE Access & Brief & Brief & Brief & Brief & None & No \\
\cite{T7} Karunasingha et al. & 2025 & EuroUSEC & None & None & None & Brief (analyst situation awareness) & None & No \\
Present survey$^{*}$ & & & Dedicated (Ban et al.\ screening, DeepCASE, adversarial evasion) & Dedicated & Dedicated (GNNs, provenance, DeepCASE bridge) & Dedicated (RAG, foundation models, prompt injection, hallucination) & Both (Sections~\ref{sec:cross-cat-synthesis} and~\ref{sec:research-agenda}) & Yes (all 4) \\
\multicolumn{9}{@{}l@{}}{\footnotesize Scope labels: \emph{Primary}, \emph{Dedicated}, \emph{Brief}, \emph{None}. IDS-only surveys~\cite{S1,S2} are excluded. $^{*}$Author self-assessment of the present survey, not a prior publication.} \\
\bottomrule
\end{tabular}
\end{adjustbox}
\end{table*}

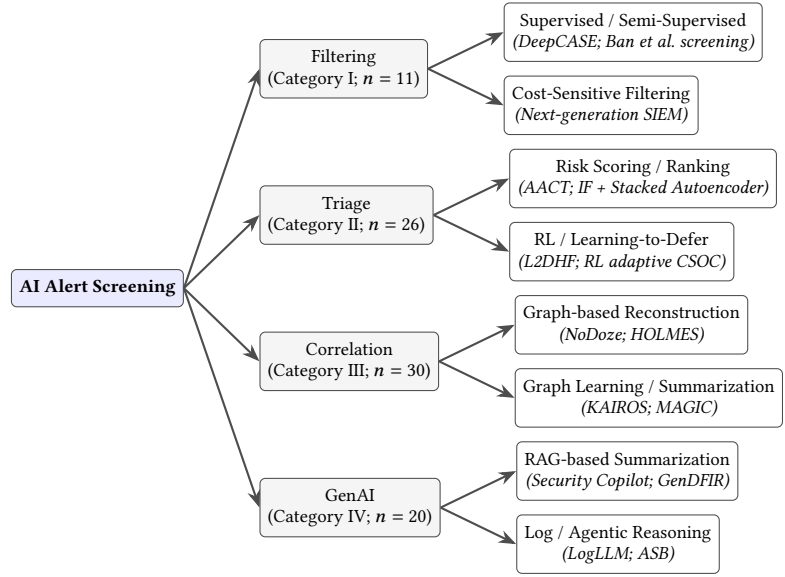
\begin{figure}[!tbp]
\centering
\begin{forest}
  for tree={
    draw=black!70, line width=0.4pt, font=\footnotesize, rounded corners=2pt,
    edge={-Stealth, thick, draw=black!70}, grow'=0, l sep=1.0cm, s sep=0.18cm,
    child anchor=west, parent anchor=east, anchor=west, align=center, fill=white,
    inner sep=3pt
  }
  [{\textbf{AI Alert Screening}}, fill=blue!8
    [{Filtering\\(Category I; $n=11$)}, fill=gray!8
      [{Supervised / Semi-Supervised\\\textit{(DeepCASE; Ban et al.\ screening)}}]
      [{Cost-Sensitive Filtering\\\textit{(Next-generation SIEM)}}]
    ]
    [{Triage\\(Category II; $n=26$)}, fill=gray!8
      [{Risk Scoring / Ranking\\\textit{(AACT; IF + Stacked Autoencoder)}}]
      [{RL / Learning-to-Defer\\\textit{(L2DHF; RL adaptive CSOC)}}]
    ]
    [{Correlation\\(Category III; $n=30$)}, fill=gray!8
      [{Graph-based Reconstruction\\\textit{(NoDoze; HOLMES)}}]
      [{Graph Learning / Summarization\\\textit{(KAIROS; MAGIC)}}]
    ]
    [{GenAI\\(Category IV; $n=20$)}, fill=gray!8
      [{RAG-based Summarization\\\textit{(Security Copilot; GenDFIR)}}]
      [{Log / Agentic Reasoning\\\textit{(LogLLM; ASB)}}]
    ]
  ]
\end{forest}
\caption{Workflow-aligned four-dimensional taxonomy of AI-driven alert screening.}
\Description{A horizontal tree diagram rooted at AI Alert Screening with four child branches: Filtering (Category I, 11 studies), Triage (Category II, 26 studies), Correlation (Category III, 30 studies), and GenAI (Category IV, 20 studies). Each branch expands into two technique families with two example study names.}
\label{fig:taxonomy}
\end{figure}

Figure~\ref{fig:taxonomy} maps the workflow-aligned taxonomy as an interdependent pipeline: filtering shapes the queue, prioritization changes what reaches correlation, correlation feeds GenAI augmentation, and analyst feedback from any stage propagates into future filtering and upstream tuning.

The survey by Tariq et al.~\cite{T1} is closest in scope and foundational for Section~\ref{sec:cross-cutting}, but it organizes the literature around the A2C framework rather than the downstream pipeline. The survey by Jalalvand et al.~\cite{T2} is most focused on prioritization; we extend its treatment with RLHF, optimization-based Cybersecurity Operations Center (CSOC) formulations, FL, and adversarial analysis. Chhetri et al.~\cite{T3} introduce the A2C framing we adopt. Srinivas et al.~\cite{T4}, Binbeshr et al.~\cite{T5}, Khayat et al.~\cite{T6}, and Karunasingha et al.~\cite{T7} cover LLMs, cognitive-SOC framing, broader SOC AI, and analyst situation awareness (SA), but none combines workflow-aligned decomposition, upstream/downstream boundary-setting, per-paper tables across all four stages, and integrated treatment of deployment, robustness, privacy, and human-workflow concerns.

This survey integrates these perspectives into a downstream workflow taxonomy and weights claims by evidence context.

\subsection{Evidence-Quality Validation Rubric}
The evidence-tier scheme introduced in Section~\ref{sec:methodology} (T1a/T1b, FM, T2, T3) is applied consistently across the per-category comparison tables (Tables~\ref{tab:filtering-studies}, \ref{tab:prioritization-studies}, \ref{tab:correlation-studies}, and \ref{tab:genai-studies}). Reproducibility and deployment realism are treated as separate dimensions: a production paper may have weaker artifact transparency than a benchmark paper, but it still provides stronger evidence about operational validity; conversely, a fully reproducible conceptual paper may be methodologically valuable without establishing deployment readiness.

This rubric therefore measures evidence context rather than intrinsic scientific merit. In the analytic synthesis reported later in the manuscript, only a small subset of core studies are Tier~1, whereas the representative Category~III studies summarized in Table~\ref{tab:correlation-studies} are overwhelmingly Tier~2. The resulting imbalance is not a minor reporting artifact but a persistent deployment-evaluation gap: the field produces many technically sophisticated systems, yet comparatively little evidence that those systems remain effective under the temporal, organizational, and data-distribution pressures of real SOCs.

\subsection{Formal Characterization of the Security Alert Screening Task}
\label{sec:formal-task}

We formalize the security alert screening task and its failure mode as follows. Let $\alpha(t) = \{a_1, a_2, \ldots, a_n\}$ denote the discrete alert stream arriving at rate $\lambda$ (alerts per unit time). Each alert $a_i$ has a ground-truth label $y_i \in \{\text{malicious}, \text{benign}\}$ and an analyst-assigned label $\hat{y}_i$. We define:

\begin{itemize}[leftmargin=*]
\item \textbf{False Positive Rate:} $\text{FPR} = |\{a_i : \hat{y}_i = \text{malicious},\; y_i = \text{benign}\}| \;/\; |\{a_i : y_i = \text{benign}\}|$
\item \textbf{Uninvestigated Rate:} $\text{UIR} = |\{a_i : \hat{y}_i = \emptyset \;(\text{no decision rendered})\}| \;/\; |\alpha(t)|$
\item \textbf{Decision Quality Degradation:} $\Delta Q(t) = Q(t{=}0) - Q(t)$ where $Q(t)$ denotes analyst classification accuracy at elapsed shift time t.
\end{itemize}

Alert fatigue emerges when one or more of these quantities exceed site-specific tolerances: $\text{UIR} > \tau_{\text{UIR}}$, $\text{FPR} > \tau_{\text{FPR}}$, or $\Delta Q(t) > \tau_{\text{DQD}}$; severe alert fatigue often involves all three. Alahmadi et al.~\cite{ref9} report qualitative analyst perceptions of decision-quality degradation under sustained FP conditions; practitioner reports~\cite{ref14} describe similar dynamics. Axelsson~\cite{ref114} formalized the base-rate fallacy as a fundamental statistical obstacle to alert triage: even highly accurate classifiers produce unacceptable FP volumes when the base rate of genuine attacks is low, a constraint that remains a primary driver of alert fatigue two decades later. Recent SOC telemetry measurement in a single large operational environment reports the same base-rate skew in practice (only \textasciitilde{}0.01\% of daily alerts linked to true attacks), consistent with this constraint under modern operational conditions~\cite{ref120}. This establishes both the magnitude of the problem and the specific measurable objectives for AI intervention. Shah et al.~\cite{ref101} formalize the throughput--quality--cost tradeoff in distributed CSOC alert analysis (distinct from the \emph{Cognitive SOC} framing of Section~\ref{sec:research-agenda}), providing complementary analytical grounding for the metrics defined above.

\section{Workflow Taxonomy of AI-Driven Alert Screening}
\label{sec:workflow-taxonomy}

We organize AI approaches along four primary workflow-aligned dimensions. Unlike technique-based taxonomies (which group all RL methods together regardless of SOC application), our taxonomy enables direct operational mapping: a SOC architect facing a triage bottleneck can navigate directly to Category~II, while one addressing multi-stage incident investigation consults Category~III. Cross-cutting concerns (HAT, explainability, privacy) are treated separately in Section~\ref{sec:cross-cutting} as they apply across all four categories. Filtering reduces the queue before analyst exposure; prioritization orders surviving alerts; correlation reconstructs multi-alert incidents; and GenAI augmentation supports analyst interpretation across all stages. Each category is treated as a subsection with a paper-level comparison table. Table~\ref{tab:taxonomy} summarizes the workflow mapping and the study counts used throughout the remainder of the manuscript; Table~\ref{tab:metrics-by-category} records the per-category evaluation metrics and protocols that this taxonomy is meant to support.
\begin{table*}[!ht]
\centering
%\scriptsize
%\setlength{\tabcolsep}{3pt}
\renewcommand{\arraystretch}{1.15}
\caption{Taxonomy of the security alert-screening pipeline by workflow stage.}
\label{tab:taxonomy}
\begin{adjustbox}{max width=\textwidth}
\small
\begin{tabular}{@{}L{0.18\linewidth}L{0.28\linewidth}L{0.23\linewidth}L{0.15\linewidth}L{0.10\linewidth}@{}}
\toprule
Category & Primary Objective & Representative Techniques & Screening Stage & Papers Surveyed \\
\midrule
I. Filtering & Suppress FPs before analyst queue & Supervised classifiers, cost-sensitive learning, anomaly detection, rule learning & Pre-queue noise reduction & n = 11 \\
II. Triage \& Prioritization & Rank and route alerts to optimize analyst attention allocation & Risk scoring, RL, RLHF, L2D, multi-armed bandits & Triage & n = 26 \\
III. Correlation \& Reconstruction & Aggregate related alerts into coherent multi-stage attack narratives & GNNs, provenance analysis, temporal sequence models, threat intelligence fusion & Analysis & n = 30 \\
IV. GenAI \& LLM Augmentation & Natural-language interfaces, automated playbooks, log analysis, report generation & LLMs, RAG, multi-agent systems, fine-tuning, prompt engineering & All stages & n = 20 \\
\multicolumn{5}{@{}l@{}}{\footnotesize Supplementary Table~S1 provides the complete study-to-category mapping.} \\
\bottomrule
\end{tabular}
\end{adjustbox}
\end{table*}

\begin{table*}[!htbp]
\centering
\renewcommand{\arraystretch}{1.15}
\caption{Recommended evaluation metrics and protocols by taxonomic category.}
\label{tab:metrics-by-category}
\begin{adjustbox}{max width=\textwidth}
\small
\begin{tabular}{@{}L{0.16\linewidth}L{0.30\linewidth}L{0.50\linewidth}@{}}
\toprule
Category & Primary metric(s) & Required protocol \\
\midrule
I.\ Filtering & QRR at fixed FNR; AUC-ROC/AUC-PR & Chronological split (e.g., 70/30 forward-chained); conservative operating point selection; report degradation over time \\
II.\ Triage \& Prioritization & nDCG@$k$, MRR, time-to-investigate & Analyst-capacity $k$ matched to daily caseload; severity-only baseline; per-shift longitudinal evaluation \\
III.\ Correlation \& Reconstruction & Incident reconstruction precision/recall; graph reduction ratio; latency and memory footprint & Attack-scenario ground truth (DARPA TC-style or simulator); simple alert-grouping baseline; report enterprise-scale runtime \\
IV.\ GenAI \& LLM Augmentation & Hallucination rate; investigation time delta; prompt-injection robustness; NASA-TLX & IRB-approved human-subject protocol; tool-permission boundary audit; deployment-risk checklist (\texttt{LLM\_Risk\_Checklist.md}) \\
\multicolumn{3}{@{}l@{}}{\footnotesize Metric and protocol definitions are given in Section~\ref{sec:eval-metrics}.}\\
\bottomrule
\end{tabular}
\end{adjustbox}
\end{table*}

\FloatBarrier

\subsection{Category I: Alert Filtering and Noise Reduction (Screening Stage: Pre-Queue)}

\subsubsection{Problem Formulation}

Alert filtering is the first stage of the downstream screening pipeline: it receives the raw alert stream $\alpha(t)$ from the upstream detection layer and produces a filtered queue $\alpha^{f}(t) \subseteq \alpha(t)$ by identifying and removing alerts highly likely to be benign before they consume analyst attention. This is distinct from upstream IDS tuning (which reduces the alert generation rate at the source by adjusting detection thresholds): filtering operates on already-generated alerts and makes no modifications to the detection layer itself. The formal objective is to learn a function $f: \mathbb{R}^d \to \{\text{pass}, \text{suppress}\}$, mapping an alert's $d$-dimensional feature vector to a binary disposition; we write $\hat{w}$ for the filter's disposition (distinct from the analyst label $\hat{y}$ of Section~\ref{sec:formal-task}). The training objective is then: (i) minimize the $\text{False Negative Rate (FNR)} = P(\hat{w} = \text{suppress} \mid y = \text{malicious})$, subject to (ii) a $\text{Suppression Rate (SR)} = P(\hat{w} = \text{suppress}) \geq \tau_{\text{target}}$. The asymmetric cost structure (where a suppressed true positive (missed threat) incurs unbounded organizational damage while a passed FP merely consumes analyst time) drives the optimization toward extreme conservatism at the decision threshold.

This cost asymmetry makes alert filtering qualitatively different from standard binary classification and demands cost-sensitive formulations, uncertainty quantification, and conservative operating point selection. The design question is not ‘maximize accuracy’ but ‘maximize SR subject to a tolerable FNR budget’, typically FNR < 2\% in production systems \cite{ref16}.

\subsubsection{Supervised Classification}

In the Category~I studies surveyed here, a common paradigm trains supervised classifiers on labeled analyst triage decisions. Features typically encode static alert attributes (type, source, destination, severity), temporal features (inter-arrival time, hourly/daily frequencies), historical analyst decision patterns, asset context (business criticality, vulnerability state), and threat intelligence correlation scores.

Random Forests and gradient-boosted trees (XGBoost, LightGBM) are common choices for tabular alert data because they are robust to feature noise, train efficiently, and produce relatively interpretable output. In the surveyed structured-alert settings, deep neural networks offer a limited advantage over ensemble methods unless the task includes sequence or graph structure. Reported performance is typically summarized by the area under the receiver-operating-characteristic curve (AUC-ROC), which we abbreviate AUC throughout the per-paper tables. Representative filtering studies are summarized in Table~\ref{tab:filtering-studies}.

\begingroup
\small
\setlength{\tabcolsep}{3.5pt}
\renewcommand{\arraystretch}{1.06}
\setlength{\LTpre}{4pt}
\setlength{\LTpost}{4pt}
\begin{longtable}{@{}L{0.14\linewidth}L{0.05\linewidth}L{0.24\linewidth}L{0.16\linewidth}L{0.19\linewidth}L{0.22\linewidth}@{}}
\caption{Representative studies in alert filtering and noise reduction.}
\label{tab:filtering-studies} \\
\toprule
Study & Year & Method & Dataset & Metric & Finding / limitation \\
\midrule
\endfirsthead
\caption[]{Representative studies in alert filtering and noise reduction (continued)} \\
\toprule
Study & Year & Method & Dataset & Metric & Finding / limitation \\
\midrule
\endhead
\midrule
\multicolumn{6}{r@{}}{Continued on next page} \\
\endfoot
\bottomrule
\endlastfoot
\cite{ref18} Ban et al. & 2023 & Next-gen SIEM: cost-sensitive IWSVM + segmentation + SOAR & Enterprise logs & FP count $-53\%$ & [T1a] Multi-IDS coverage gain; no longitudinal evaluation. \\
\cite{ref63} Ban et al. & 2021 & AI-assisted screening: ML + visualization & Enterprise SIEM SOC & Recall 99.598\% / FPR 0.001\% on highly-critical-alert class & [T1a] Explicit downstream-screening formulation; single-org, non-public features. \\
\cite{ref69} Aminanto et al. & 2019 & Online IF for streaming alert fatigue & Enterprise NIDS logs & AUC 0.91; <1\,ms update & [T1a] Early online alert-fatigue model; hyperparameter sensitivity. \\
\cite{ref65} Ndichu et al. & 2023 & Ensemble oversampling + subsampling for class imbalance & Enterprise + CICIDS & F1 $+22$\,pp vs.\ SMOTE & [T2] Class-balance gain; precision trade-off unresolved. \\
\cite{ref62} van Ede et al.\ (DeepCASE) & 2022 & Semi-supervised LSTM context builder + attention-based clustering & Real enterprise + synthetic & Workload $-90.53\%$; risk under-est.\ $<0.001\%$ & [T1a] Dual-use screener with implicit event correlation; counted under Cat~I; single-feature-per-event SIEM model. \\
\cite{ref71} He et al.\ (Drain) & 2017 & Fixed-depth parse tree for streaming log parsing & HDFS, BGL, Zookeeper, 13 systems & Parsing acc.\ 0.96 & [Contextual, not core] Widely deployed log parser; brittle to template evolution. \\
\cite{ref73} Guo et al.\ (LogBERT) & 2021 & BERT pre-training on log key sequences & HDFS, BGL & F1 $+4.7$\,pp vs.\ DeepLog & [Contextual, not core] Transformer log model; high compute, slow inference. \\
\cite{ref74} Yang et al.\ (PLELog) & 2021 & Semi-supervised log anomaly detection via probabilistic label estimation & HDFS, BGL & F1 $+6.3$\,pp under noisy labels & [Contextual, not core] Addresses label noise; EM convergence sensitivity. \\
\end{longtable}
\endgroup

SwissLog~\cite{ref72} complements these contextual infrastructure rows by emphasizing robust log anomaly detection across diverse system faults.

Several Category~I studies report evaluation on real enterprise alert data: Ban et al.~\cite{ref63,ref18} (AI-assisted screening and next-generation SIEM), Aminanto et al.~\cite{ref69} (online alert fatigue), and Ndichu et al.~\cite{ref65} (imbalance-aware screening on an enterprise alert dataset and UNSW-NB15). Variants~\cite{ref64,ref66,ref67} are contextual. Goud et al.~\cite{refGoudEvasionAware} extend the direction with a semi-supervised, evasion-aware framework coupling concept-drift handling with adversarial considerations. A comparable operations-research line by Shah et al.~\cite{ref101} addresses capacity and cost modeling explicitly; the full set (including~\cite{ref102,ref130,ref131,ref132}) is discussed under Category~II below.

\subsubsection{DeepCASE and Cost-Sensitive Approaches}

DeepCASE~\cite{ref62} combines contextual sequence modeling, clustering, and analyst cluster labeling: a Context Builder (Long Short-Term Memory (LSTM) with multi-head attention) summarizes preceding events; an Interpreter clusters sequences with similar attention vectors; analysts label clusters rather than individual events. DeepCASE is counted under Category~I (primary contribution: pre-queue workload reduction) and cross-referenced in Section~\ref{sec:correlation-cat} for its correlation role. The system reports analyst workload reduction above 90\% under the reported evaluation setting (Table~\ref{tab:filtering-studies}); limitations include cold-start and theoretical vulnerability to adversarial event injection. When false-negative (FN) and FP costs differ, standard binary cross-entropy is inappropriate; cost-sensitive losses (asymmetric focal loss~\cite{ref19}, class-weighted cross-entropy, calibrated thresholding) remain standard, while Monte Carlo Dropout, deep ensembles, and conformal prediction add calibrated uncertainty for variable-threshold policies (suppress at high confidence, escalate at high uncertainty, otherwise pass to the analyst queue).

\subsubsection{Unsupervised, Semi-Supervised, and Adversarial Robustness}

Label scarcity is pervasive: analyst decisions are expensive at scale and historical labels may be stale. Unsupervised approaches (Isolation Forest (IF), Local Outlier Factor~\cite{ref20}, DBSCAN (Density-Based Spatial Clustering of Applications with Noise), Variational Autoencoders) flag structural deviants without labels but cannot distinguish novel benign events from novel threats; hybrid semi-supervised approaches that combine these with sparse labeled supervision on confirmed threats partially address this. Self- and semi-supervised representation learning (DeepCASE~\cite{ref62}, self-supervised graph learning for provenance~\cite{ref81}) further reduce analyst effort, although direct evidence for alert-stream contrastive pre-training in production remains limited. Adversarial evasion is a critical and underexplored vulnerability: attackers who model the defense can craft semantically valid attacks that look benign to the classifier. Pierazzi et al.~\cite{ref94} and the IDS-pipeline survey of Alotaibi and Rassam~\cite{ref95} document analogous vulnerabilities (both retained as contextual references since their primary contribution is upstream detection robustness). Common adversarial-ML mitigations, such as randomized smoothing, adversarial training with realistic models, and ensemble diversity, are seldom reported in production alert-filtering studies of the surveyed corpus, leaving a security gap.

\subsection{Category II: Automated Triage and Prioritization (Screening Stage: Queue Ordering)}

\subsubsection{From Filtering to Prioritization}

After queue reduction, the operational problem shifts from suppression to ordering: given the filtered queue $\alpha^{f}(t)$ from Category~I (or the raw queue when no filtering is deployed), the triage model ranks alerts so analyst time is spent first on the highest-value investigations. The formal objective is to learn a ranking $r: \mathcal{A} \to \mathbb{R}$ that maximizes expected threat coverage within the analyst time budget; ideally evaluated by ranking metrics (normalized discounted cumulative gain (nDCG); mean reciprocal rank (MRR)), although Table~\ref{tab:prioritization-studies} shows that most surveyed studies still report classification metrics due to dataset constraints. Earlier IDS-alert prioritization and response-support systems~\cite{ref154,ref155}, analyst-trace retrieval and data-triage learning~\cite{ref156,ref157}, and context-aware prioritization and representation learning~\cite{ref158,ref159,ref160} together establish prioritization as also a queue-management and investigation-support problem.

\subsubsection{Multi-Feature Risk Scoring}

A common production approach combines multiple evidence signals into a composite risk score. Features used in published scoring models include alert-level signals (severity, confidence, type, source/destination reputation), asset context (host criticality, vulnerability state, network exposure), user and entity context from UEBA (behavioral anomaly scores, privilege level, recent activity), threat-intelligence correlation (matches against indicators of compromise (IOCs), MITRE ATT\&CK technique relevance, threat-actor profile), and temporal features (recent-alert clustering, attack-timing proximity).

Peer-reviewed SOC surveys argue that severity-only ranking is inadequate for modern SOC prioritization: Jalalvand et al.~\cite{T2}, Kokulu et al.~\cite{ref96}, and Vielberth et al.~\cite{ref98} document this gap empirically, and practitioner-consensus accounts in vendor documentation~\cite{ref2,ref85,ref86} converge on the same design pattern of combining alert severity with entity context, asset criticality, historical behavior, and threat-intelligence enrichment. The implication is practical rather than merely architectural: organizations that lack asset inventory, entity context, or usable threat-intelligence pipelines are unlikely to realize the benefits of advanced prioritization models.

AACT~\cite{ref143} further illustrates analyst-action-based triage learning in a real-SOC deployment; Section~\ref{sec:validity-reproducibility} explains its post-lock treatment. Representative prioritization studies are summarized in Table~\ref{tab:prioritization-studies}; the full Category~II core list of 26 studies (including human-factors and operations-research entries discussed in Section~\ref{sec:cross-cutting}) is given in Supplementary Table~S1.

\begingroup
\small
\setlength{\tabcolsep}{3.5pt}
\renewcommand{\arraystretch}{1.06}
\setlength{\LTpre}{4pt}
\setlength{\LTpost}{4pt}
\begin{longtable}{@{}L{0.14\linewidth}L{0.05\linewidth}L{0.24\linewidth}L{0.16\linewidth}L{0.19\linewidth}L{0.22\linewidth}@{}}
\caption{Representative studies in automated triage and prioritization.}
\label{tab:prioritization-studies} \\
\toprule
Study & Year & Method & Dataset & Metric & Finding / limitation \\
\midrule
\endfirsthead
\caption[]{Representative studies in automated triage and prioritization (continued)} \\
\toprule
Study & Year & Method & Dataset & Metric & Finding / limitation \\
\midrule
\endhead
\midrule
\multicolumn{6}{r@{}}{Continued on next page} \\
\endfoot
\bottomrule
\endlastfoot
\cite{ref27} Jalalvand et al. & 2025 & L2DHF for adaptive prioritization & UNSW-NB15; CICIDS2017 & ``AP accuracy'' $+13$--$16\%$ / $+60$--$67\%$; deferrals $-37\%$ \emph{(source-reported)} & [T3] Simulated benchmark feedback; nonstandard ``AP accuracy'' and undisclosed baselines limit comparability. \\
\cite{ref16} Aminanto et al. & 2020 & IF + stacked autoencoder; day-forward-chaining temporal analysis & Enterprise network-IDS alerts & AUC 0.93; FPR $-31\%$ & [T1a] Day-forward-chaining preserves chronology; fixed window, no feedback. \\
\cite{ref102} Shah et al. & 2020 & RL-based adaptive alert management for distributed CSOCs & Simulation (multi-CSOC) & Optimization objective; no single headline metric reported & [T3] RL formulation for CSOC alert routing; simplified analyst model. \\
\cite{ref130} Shah et al. & 2019 & Two-step optimal alert selection under analyst capacity & Simulation (CSOC queue) & Optimization objective; no single headline metric reported & [T3] Operations-research framing with capacity constraints; static threat model. \\
\cite{ref132} Ghadermazi et al. & 2024 & Joint ML + optimization for CSOC alert management & Real SOC alert logs & Queue efficiency improvement & [T1a] ML--optimization bridge in production CSOC; single-org. \\
\cite{ref119} Kim \& D\'{a}n & 2022 & Active learning for dynamic prioritization and real-time SA & Simulation (alert stream) & Label efficiency $+3\times$ & [T3] Targets analyst feedback on decision-critical alerts; simulated only. \\
\end{longtable}
\endgroup

\subsubsection{Reinforcement Learning, Risk-Based Alerting (RBA), and Concept Drift}

Static scorers cannot adapt to changing threats, organizational context, or analyst preference drift. RL provides a principled adaptive framework: Shah et al.~\cite{ref102} formulate distributed CSOC alert management as an RL problem balancing throughput and detection quality, with state $s$ encoding alert and queue features, action $a$ assigning a queue position, and reward $r$ derived from analyst outcomes; active-learning extensions target analyst feedback to decision-critical alerts~\cite{ref119}, and decision-aware trust signal alignment (calibrated confidence, uncertainty cues, and cost-sensitive thresholds) has been proposed as an analyst-facing triage layer evaluated on UNSW-NB15~\cite{refTrustSignal}.

The L2DHF framework~\cite{ref27} combines deep RLHF by adding (i)~a learned reward model from analyst preferences, (ii)~an explicit L2D mechanism, and (iii)~an Analyst-Validated Alert Repository for periodic reward updates. The source reports gains on critical-alert classification over its own baselines together with a deferral-rate reduction; the raw percentages and the metric-definition caveat (``Average Precision (AP) accuracy'', baselines not disclosed) are recorded in Table~\ref{tab:prioritization-studies}. Adjacent operations-research formulations model queueing, staffing, and capacity constraints explicitly~\cite{ref101,ref130,ref131,ref132}.

RBA, as described in Splunk-centered operational documentation, replaces binary alert semantics with entity-level risk accumulation; analyst-facing alerts fire only after a risk threshold is crossed. Operational documentation reports substantial volume reduction~\cite{ref85,ref86}, but peer-reviewed evaluation of commercial RBA remains limited; the broader risk-aware management problem is receiving systematic academic treatment~\cite{ref130,ref131,ref132}.

Finally, alert distributions are non-stationary, so trained models drift; explicit drift-detection methods (adaptive windowing (ADWIN), Page-Hinkley, drift detection method (DDM)) monitor prediction-stream statistics and trigger retraining. Pendlebury et al.~\cite{ref103} show that ignoring temporal ordering in splits inflates reported performance; the broader concept-drift literature surveyed by Lu et al.~\cite{ref87} catalogs ADWIN, Page-Hinkley, and DDM as commonly used drift detectors, with targeted feature-distribution monitoring used to scope retraining to drifted dimensions.

\subsubsection{Emerging Deployment Contexts}

Three forward-looking contexts deserve mention. Cloud-native alerts (Amazon Web Services (AWS) GuardDuty, Defender for Cloud, Google Cloud Security Command Center (SCC)) are Application Programming Interface (API)-centric and include identity and access management (IAM) escalation patterns, with multi-tenant managed security service provider (MSSP) workflows documented in operational sources~\cite{ref88,ref89}. Operational technology and industrial control system (OT/ICS) environments add supervisory control and data acquisition (SCADA) protocols (Modbus, DNP3), deterministic process baselines, and safety-critical FN tolerances, illustrated by decision-tree screening~\cite{ref90}. Workload- and shift-aware routing recognizes that analyst fatigue peaks at handoffs, building on context-loss findings~\cite{ref100}.

\subsection{Category III: Alert Correlation and Incident Reconstruction (Screening Stage: Grouping and Narrative)}
\label{sec:correlation-cat}

\subsubsection{The Multi-Stage Attack Challenge}

Sophisticated attacks (Advanced Persistent Threats (APTs), ransomware campaigns, supply chain intrusions) generate distributed signal patterns across multiple tools, hosts, and time windows that may span hours, days, or weeks. A typical chain proceeds from a spear-phishing email (email/antivirus alert) through credential theft, lateral movement (authentication/NDR alert), privilege escalation (EDR/UEBA alert), data staging (data loss prevention (DLP) alert), and exfiltration (proxy/firewall alert). Each individual alert, viewed in isolation, may be unremarkable; their temporal and causal relationship constitutes the attack narrative. Alert correlation (the automated grouping of semantically and temporally related alerts into coherent incident hypotheses) is therefore both the primary mechanism for attack narrative reconstruction and a potent alert volume reducer (collapsing hundreds of related alerts into a single enriched incident).

Several adjacent systems sit directly on the Category~II/III boundary by coupling prioritization with post-correlation analysis or context-aware alert investigation \cite{ref153,ref159}, illustrating that ranking and reconstruction are operationally intertwined even when individual papers foreground one stage over the other.

\subsubsection{Graph-Based Correlation}

Alert correlation predates graph learning. Early rule-based, probabilistic, and adaptive approaches established the clustering, inference, and cooperative-correlation foundations on which later graph methods build~\cite{Debar2001,Valdes2001,Cuppens2002,Pietraszek2004,ref23,ref115}. Contemporary systems instead model alerts as nodes in a property graph connected by semantic, causal, and temporal relations, then recover attack hypotheses via clustering, path analysis, or learned graph representations. Under the Section~\ref{sec:formal-task} boundary, we include provenance systems when their evaluated output supports downstream investigation, triage, or reconstruction.

Graph Neural Networks (GNNs) have become a prominent learning family for correlation. Heterogeneous graph models capture alert, host, user, and process entities within a unified message-passing framework, and provenance-graph representation learning has reported strong results on DARPA TC-style attack benchmarks \cite{ref79,ref81,ref110,ref136}. Recent work also targets investigation quality and analyst effort through dependency-impact back-propagation \cite{ref124}, comparative provenance-IDS analysis \cite{ref122}, attribution optimization in ORTHRUS \cite{ref123}, and adversarially robust root-cause-aware detection in R-CAID \cite{ref136}. The evidence-weighted conclusion is more cautious than the raw benchmark scores suggest: GNNs are extensively benchmarked for high-fidelity correlation, but most evaluations remain tied to DARPA TC-style provenance graphs, and only a minority of papers report latency, memory overhead, or analyst-time savings under realistic enterprise loads. The literature therefore supports GNNs as the most extensively benchmarked correlation family in the surveyed corpus, not yet as universally deployment-ready SOC components. Representative studies are summarized in Table~\ref{tab:correlation-studies}.

Beyond GNN-based methods, hybrid rule-plus-learning approaches are also emerging for provenance-based intrusion detection. The CAPTAIN system \cite{ref148} augments traditional rule-based provenance monitoring with gradient-based learning to adapt rules and reduce FPs without incurring high overhead, achieving improved detection accuracy and lower latency relative to static rule engines.

\begingroup
\small
\setlength{\tabcolsep}{3.5pt}
\renewcommand{\arraystretch}{1.06}
\setlength{\LTpre}{4pt}
\setlength{\LTpost}{4pt}
\begin{longtable}{@{}L{0.14\linewidth}L{0.05\linewidth}L{0.24\linewidth}L{0.16\linewidth}L{0.19\linewidth}L{0.22\linewidth}@{}}
\caption{Representative studies in alert correlation and incident reconstruction.}
\label{tab:correlation-studies} \\
\toprule
Study & Year & Method & Dataset & Metric & Finding / limitation \\
\midrule
\endfirsthead

\caption[]{Representative studies in alert correlation and incident reconstruction (continued)}\\
\toprule
Study & Year & Method & Dataset & Metric & Finding / limitation \\
\midrule
\endhead

\midrule
\multicolumn{6}{r@{}}{Continued on next page}\\
\endfoot

\bottomrule
\endlastfoot

\cite{ref34} Hassan et al.\ (NoDoze) & 2019 & Provenance triage + dependency explosion mitigation & Synthetic APT scenarios & Graph red.\ $>99\%$; Prec.\ 0.91 & [T2] Provenance triage at scale; tuned for Linux. \\

\cite{ref35} Zeng et al.\ (WATSON) & 2021 & Audit-log abstraction via aggregation of contextual semantics & DARPA ENGAGE & DR 0.96; FPR 0.02 & [T2] Enterprise-scale provenance analysis; OS-level audit logging required. \\

\cite{ref36} Han et al.\ (UNICORN) & 2020 & Streaming provenance graph sketching + anomaly detection & TC: CADETS/TRACE & APT F1 0.85 & [T2] Streaming sketches; high memory; FPR at scale inadequately evaluated. \\

\cite{ref75} Zeng et al.\ (SHADEWATCHER) & 2022 & Graph recommendation + GNN for provenance-based detection & TC & DR 0.96; FPR 0.001 & [T2] Recommendation framing enables scalable provenance GNN; full graph; high memory. \\

\cite{ref76} Wang et al.\ (threaTrace) & 2022 & Node-level provenance-graph learning for host-based threat detection & TC; StreamSpot & F1 0.978 node-level & [T2] Fine-grained node classification; per-host, no cross-host correlation. \\

\cite{ref77} Hossain et al.\ (SLEUTH) & 2017 & Causal analysis: tag propagation + backward/forward tracing & TC; synthetic APT & Foundational baseline & [T2] Causal provenance analysis baseline; tag propagation causes dependency explosion without pruning. \\

\cite{ref79} Cheng et al.\ (KAIROS) & 2024 & Whole-system provenance with GNN + hierarchical graph construction & TC: CADETS/THEIA & AUC 0.99 & [T2] Strong DARPA TC results; high computational cost limits real-time use. \\

\cite{ref80} Xu et al.\ (DEPCOMM) & 2022 & Graph summarization via process community detection & TC; enterprise scenarios & Graph $-97.5\%$; DR maintained & [T2] Provenance graph compression; summarization can obscure rare attack paths. \\

\cite{ref81} Jia et al.\ (MAGIC) & 2024 & Masked graph representation learning for APT detection & TC; StreamSpot & F1 0.963; label-efficient & [T2] Self-supervised pre-training reduces label requirement; masking strategy sensitive. \\

\cite{ref136} Goyal et al.\ (R-CAID) & 2024 & GNN + root cause embedding for adversarially robust provenance detection & TC: THEIA/TRACE; StreamSpot; ATLASv2 & AUC 0.99 passive; 0.94 adversarial & [T2] Adversarially evaluated provenance GNN; root-node overhead scales with graph size. \\

\cite{ref82} Alsaheel et al.\ (ATLAS) & 2021 & Alert-driven attack scenario reconstruction from system logs via NLP + causal analysis & TC; enterprise logs & Scenario F1 0.91 & [T2] Bridges correlation and narrative reconstruction for analyst reporting. Limitation: depends on upstream alert quality. \\

\end{longtable}
\endgroup

\subsubsection{Provenance Analysis, Sequence Models, and Threat Intelligence Fusion}

Provenance analysis reconstructs attack causality from OS-level events such as process creation, file access, network connections, and registry changes. Its central challenge is dependency explosion. The systems cataloged in Table~\ref{tab:correlation-studies}, together with RapSheet~\cite{ref113}, HOLMES~\cite{ref78}, POIROT~\cite{ref106}, PrioTracker~\cite{ref116}, and OmegaLog~\cite{ref117}, extend the line toward scalable streaming analysis, graph learning, robustness, and analyst-facing narratives. The provenance-IDS surveys of Inam et al.~\cite{ref107} and Zipperle et al.~\cite{ref108} synthesize this literature; NODLINK~\cite{ref109} targets online fine-grained APT detection. Deployment remains constrained by telemetry coverage, OS/application variability, and runtime overhead (e.g., WATSON's 7--8\% per-host CPU~\cite{ref35}). As a complementary route, LSTM, gated recurrent unit (GRU), and transformer architectures~\cite{ref37}, adapted to security contexts, are used to learn temporal attack-progression patterns from alert sequences; this can be more data-efficient where incident labels are absent, though susceptible to adversaries who disrupt temporal regularity.

Correlation quality further improves when alerts are enriched with threat-intelligence sources: STIX/TAXII (Structured Threat Information eXpression and Trusted Automated eXchange of Indicator Information) feeds, MITRE ATT\&CK technique mappings, and Common Vulnerabilities and Exposures (CVE) / National Vulnerability Database (NVD) records. POIROT~\cite{ref106} aligns cyber threat-intelligence (CTI) behavior graphs with kernel audit provenance for fine-grained threat hunting; LOCALINTEL~\cite{ref39} extracts organization-specific intelligence from heterogeneous sources so correlation models can account for the organization's technology stack and historical exposure; AttacKG~\cite{ref84} adds structured ATT\&CK-oriented knowledge-graph (KG) construction. The critical challenge is timeliness and quality: STIX feeds often lag active campaigns and indicator quality varies considerably across sources.

\subsection{Category IV: GenAI and LLM Augmentation (Screening Stage: Analyst Cognitive Bandwidth)}

\subsubsection{LLMs in Security Operations}

The fourth pipeline stage addresses a constraint that filtering, prioritization, and correlation cannot fully resolve: the cognitive bandwidth of human analysts. GenAI and LLMs~\cite{ref42} augment that bandwidth across all earlier stages, explaining filter decisions, enriching prioritization context, narrating correlated incident hypotheses, and automating routine investigation steps. Unlike classical ML, LLMs offer multi-task capability through natural language and may lower some data-labeling barriers for smaller SOCs that lack large labeled training pipelines, although rigorous cost-benefit evidence is still limited; a recent LLM-for-SOC survey documents the rapid expansion of this sub-area~\cite{S5}. The evidence quality is correspondingly less mature than in Categories~I--III, a pattern visible in publication-type composition (Table~\ref{tab:evidence-landscape}) and revisited in the cross-category synthesis (Section~\ref{sec:cross-cat-synthesis}).

An empirical field study~\cite{ref43} analyzed 3{,}090 GPT-4 queries from 45 SOC analysts over a 10-month deployment. The dominant task themes were Command Understanding, Analysis \& Generation (31\% of queries) and Text Generation \& Editing (22\%); 41\% of analyst interactions were one-off conversations, while a subset of analysts showed sustained adoption (single-organization, preprint). These adoption metrics do not directly measure investigation quality or accuracy; the qualitative finding is that LLM use centered on sensemaking and context-building rather than autonomous high-stakes decisions. The Microsoft Security Copilot RCT~\cite{ref47} reports speed and accuracy gains in the same direction (Table~\ref{tab:genai-studies}), but as an unblinded vendor self-evaluation those figures are not interchangeable with peer-reviewed evidence.

\subsubsection{Log Analysis and Anomaly Detection}

Log analysis is cognitively demanding because log semantics are heterogeneous and context-dependent. Drain~\cite{ref71}, SwissLog~\cite{ref72}, LogBERT~\cite{ref73}, and PLELog~\cite{ref74} are enabling infrastructure rather than alert-disposition systems and are cataloged here only for that role. DeepLog~\cite{ref68} established the pre-LLM deep-learning baseline by framing log anomaly detection as language modeling (stacked LSTM next-log-key prediction plus a parameter-level anomaly LSTM); on HDFS and OpenStack, DeepLog reaches precision/recall above 0.95 with online incremental updates and bridges naturally to the correlation problem of Section~\ref{sec:correlation-cat}. The trajectory from DeepLog through LogBERT to LLM-based LogPrompt~\cite{ref44} and LogLLM~\cite{ref145} progressively replaces binary anomaly flags with natural-language explanations. LogPrompt pairs zero-shot LLM log parsing (average F1 $0.797$ across eight log-parsing datasets) with a six-expert interpretability study rating usefulness and readability at $4.19$--$4.50$/5 (\emph{High-Interpretability Percentage}, HIP, of $73$--$98\%$, i.e., the fraction of samples scoring $>4$); on the separate anomaly-detection task (BGL and Spirit), online F1 remains low ($\leq 0.45$). LogParser-LLM~\cite{ref144} adapts parsers across formats; LogLLM combines semantic representations with sequence classification on four public datasets. Representative studies are summarized in Table~\ref{tab:genai-studies}.

\begingroup
\small
\setlength{\tabcolsep}{3.5pt}
\renewcommand{\arraystretch}{1.06}
\setlength{\LTpre}{4pt}
\setlength{\LTpost}{4pt}
\begin{longtable}{@{}L{0.14\linewidth}L{0.05\linewidth}L{0.24\linewidth}L{0.16\linewidth}L{0.19\linewidth}L{0.22\linewidth}@{}}
\caption{Representative studies in GenAI and LLM-based SOC augmentation.}
\label{tab:genai-studies} \\
\toprule
Study & Year & Method & Setting & Metric & Finding / limitation \\
\midrule
\endfirsthead
\caption[]{Representative studies in GenAI and LLM-based SOC augmentation (continued)} \\
\toprule
Study & Year & Method & Setting & Metric & Finding / limitation \\
\midrule
\endhead
\midrule
\multicolumn{6}{r@{}}{Continued on next page} \\
\endfoot
\bottomrule
\endlastfoot
\cite{ref43} Singh et al. & 2025 & Empirical study: GPT-4 in live SOC & 45 analysts; 3,090 queries; 10 months & Top task themes: Command Understanding 31\%, Text Generation \& Editing 22\%; one-off interactions 41\%; sustained adoption by a subset of analysts (single-org) & [T1b] Preprint field study; pre-replication; single organization. \\
\cite{ref47} Edelman et al.\ (Copilot RCT) & 2023 & GPT-4 integrated SOC assistant (Microsoft Security Copilot; RCT) & 149 novice analysts; industry trial (preprint) & \emph{Vendor self-evaluation, unblinded, not peer-reviewed:} novices $+25.9\%$ overall task speed (holding accuracy constant), $+7\%$ accuracy, $+46.2\%$ on Incident Summarization subtask & [T1b] Vendor evaluating its own commercial product; proprietary, not independently replicated. \\
\cite{ref44} Liu et al.\ (LogPrompt) & 2024 & Zero-shot LLM log analysis with natural-language explanations & 8 log datasets; 6-expert interpretability study & Log-parsing F1 0.797 avg (range 0.65--0.89 across datasets); 6-expert interpretability rated 4.19--4.50/5 (HIP 73--98\%) & [T2] Natural-language explanation value; online anomaly-detection F1 still low ($\leq 0.45$). \\
\cite{ref68} Du et al.\ (DeepLog) & 2017 & LSTM log-key sequence model + parameter-value anomaly detection & HDFS; OpenStack & HDFS: precision 0.95, recall 0.96, F1 0.96; OpenStack: precision 0.94, recall 0.99, F1 0.97 & [T2] Pre-LLM log-anomaly baseline; fixed vocabulary, struggles with novel templates. \\
\cite{ref45} Song et al.\ (Audit-LLM) & 2024 & Multi-agent LLM for insider threat detection & CERT Insider Threat r6.2 & FP $-40\%$ vs.\ single agent & [T2] Multi-agent specialization; coordination overhead, prompt-injection risk. \\
\cite{ref46} Ali \& Kostakos (HuntGPT) & 2023 & ML anomaly detection + XAI + LLM interface & CICIDS2017; analyst study & Investigation time $-37\%$ & [T2] NL query interface lowers analyst barrier; limited evaluation scale. \\
\cite{ref48} Loumachi et al.\ (GenDFIR) & 2024 & RAG + LLM for incident timeline analysis & DFIR case studies & Report time $-58\%$ & [T3] RAG grounds responses; retrieval quality sensitive to knowledge-base curation. \\
\end{longtable}
\endgroup

\subsubsection{Retrieval-Augmented Generation, Benchmarks, and Knowledge Graphs}

Retrieval-Augmented Generation (RAG) couples the LLM with a real-time retrieval system over a continuously updated knowledge base, addressing knowledge-staleness in frozen-parameter models. RAG-based SOC assistants report investigation-time gains (e.g., GenDFIR's $-58\%$ report time, Table~\ref{tab:genai-studies}; see also~\cite{ref48,ref49}); production RAG still requires knowledge-base freshness, retrieval precision, context-window management, and verifiable citation quality. Standardized SOC-specific evaluation remains a gap: SecBench~\cite{ref59}, CyberSecEval~\cite{ref83}, Agent Security Bench (ASB)~\cite{ref50}, and the blue-team embodied benchmark~\cite{ref61} cover question answering (QA), code, prompt-injection, and adversarial dimensions but none directly measures triage accuracy on real alerts, investigation speed, hallucination rates, or analyst-tool injection robustness (R3). KGs are a complementary signal: AttacKG~\cite{ref84} builds a technique KG from threat-intelligence reports, and KG+RAG combines structured traversal with natural-language synthesis.

\subsubsection{Security Foundation Models and Fine-Tuning}

General LLMs perform credibly on security tasks, but domain-specific pre-training and fine-tuning yield measurable gains: SecureFalcon~\cite{ref58} fine-tunes Falcon on security code, CyberPal.AI~\cite{ref57} instruction-tunes on expert corpora, and SecBench~\cite{ref59} provides multi-dimensional evaluation. The optimal strategy among instruction tuning, RLHF with expert preferences, and domain-adaptive pre-training is not yet systematically evaluated. The RAG-vs-fine-tune choice is also open: RAG~\cite{ref48,ref49} offers freshness, grounding, and citation trails but inherits retrieval-quality and context-truncation failure modes, while fine-tuned models~\cite{ref57,ref58,ref60} reduce jargon mismatch and latency for narrow tasks but stale faster. No surveyed study compares them head-to-head on the same SOC task.

\subsubsection{Agentic AI and LLM-Specific Risks}

LLM agents with tool access (SIEM queries, threat-intelligence APIs, sandbox detonation, network utilities) execute multi-step investigations: alert-context retrieval, host/user history queries, artifact detonation, TI correlation, and report synthesis. PentestGPT~\cite{ref111} demonstrates agent capability in sequential security tasks; CORTEX~\cite{ref162} extends this to collaborative LLM-based alert triage with specialized agents. Recent agentic-SOC studies extend this line with tooling-gap, policy-compliance, narrative-generation, and end-to-end SOC-agent architectures~\cite{ref150,ref151,refAgentSOC,refCognitiveSOC,refAutonomousSOC,refSOCpilot}. The shift from augmentation to autonomous operation introduces hallucination and prompt-injection risks not central to classical ML; mitigations include RAG grounding with mandatory citations, confidence-elicitation prompting, multi-agent cross-checking, and mandatory human review. Fang et al.~\cite{ref112} show LLM agents can autonomously exploit certain known web vulnerabilities (specific CVEs), and Agent Security Bench~\cite{ref50} benchmarks injection, plan-of-thought backdoors, and memory poisoning with candidate defenses.

\section{Cross-Cutting Themes}
\label{sec:cross-cutting}

This section reviews themes that cut across the four workflow stages: HAT, explainability, RLHF, privacy-preserving learning, platform integration, and closed-loop SOC operation.

\subsection{Human--AI Teaming (HAT)}

The surveyed evidence does not yet support fully autonomous AI decision-making for security-critical SOC operations. Analysts contribute contextual judgment~\cite{ref96}, creative adversarial reasoning, ethical accountability, and recognition of novel threats outside the AI's training distribution; AI systems contribute high-volume structured-data processing, anomaly detection, and threat-intelligence synthesis. HAT frameworks combine these through deliberate interface design~\cite{ref52}. Autonomy-tiered collaboration maps SOC task criticality to trust thresholds and oversight roles~\cite{ref161}, and cognitive-hierarchy-driven RL extends this to analyst-attacker interaction in cloud security~\cite{refCHTDQN}. We use A2C in the sense developed by Chhetri et al.~\cite{T3} and Tariq et al.~\cite{T1,ref53}: \emph{Automation} runs autonomously; \emph{Augmentation} enriches the analyst workflow while leaving the decision to a human; and \emph{Collaboration} implements bidirectional adaptation. Our coding (Table~\ref{tab:evidence-landscape}b) shows the literature concentrated in Automation and Augmentation, with Collaboration largely unrealized. Embedded production-SOC fieldwork (Kersten et al.~\cite{ref127}), the SAIBERSOC conference~\cite{ref125} and journal~\cite{ref126} papers, and the USEC 2025 workflow study~\cite{ref146} converge on the finding that explicit investigation structure and modest process changes measurably improve analyst consistency, usability, and outcome quality. AI support for analyst SA deserves treatment distinct from generic automation~\cite{T7}, with information fusion, contextual visibility, and workflow continuity as core design concerns~\cite{ref135}.

\subsection{Explainable AI (XAI)}

The black-box problem is a persistent adoption barrier: opaque suppression, escalation, or prioritization invites either over-trust (automation bias) or under-trust (automation skepticism). Post-hoc methods (SHapley Additive exPlanations (SHAP) and TreeSHAP~\cite{ref54}, Local Interpretable Model-agnostic Explanations (LIME), attention visualization, counterfactual explanations) provide feature-level attribution; counterfactuals are particularly suited to SOC contexts because they suggest actionable investigation pathways. Recent survey work argues that XAI value depends less on standalone widgets than on integration into existing analyst workflows~\cite{ref55}; the most persistent barriers are weak tool integration, explanation formats that do not map onto operational security concepts, and live-investigation performance overhead. LLMs offer a natural bridge because they can translate feature attributions into analyst-facing language. The PAIEL framework of Kaneko et al.~\cite{ref152} demonstrates this approach by injecting protocol metadata and contextual signals into LLM prompts to produce explanations for low-level network events that analysts judge more accurate and actionable than baseline model outputs.

\subsection{Reinforcement Learning from Human Feedback (RLHF)}

RLHF aligns model behavior with human preferences and is increasingly applied to SOC prioritization. The pipeline collects analyst preferences over candidate prioritizations, trains a reward model that learns the organizational value function, optimizes a policy via RL (typically Proximal Policy Optimization, PPO), and iterates. The L2DHF framework~\cite{ref27} is the most complete SOC instantiation in the surveyed corpus, combining preference-based reward learning with an explicit L2D mechanism. Open questions include feedback-collection interfaces (explicit comparisons vs.\ implicit decision signals), preference inconsistency across analysts, and reward hacking against proxies that do not reflect detection quality.

\subsection{Privacy-Preserving and Federated Learning}

Alert data and analyst decisions are sensitive, regulated (e.g., General Data Protection Regulation (GDPR) Art.~6, California Consumer Privacy Act (CCPA) Section~1798.140), and rarely shared, leaving individual organizations without enough data for robust models. FL lets each SOC train locally and share only parameter updates; differential privacy (DP) provides formal guarantees against reconstruction of individual records, and Byzantine-robust aggregation (e.g., Krum, FLTrust) handles compromised participants. \emph{No core study in the surveyed corpus implements FL specifically for SOC alert screening}; adjacent IDS literature shows that FL and federated-transfer learning can approach centralized performance across heterogeneous environments~\cite{ref93}. Open questions for SOC alert models include non-IID (i.e., not independent and identically distributed) alert distributions, incentive mechanisms for participation, and system heterogeneity.

\subsection{Deployment Integration: SIEM, SOAR, and XDR}

How AI components integrate into operational architectures is underexamined. SIEM platforms (Splunk ML Toolkit, Sentinel notebooks) are the primary integration surface; SOAR platforms (XSOAR~\cite{ref25}, Splunk SOAR) couple AI-driven triage to playbook automation with confidence-tiered auto-close/escalation/auto-containment branches; and Extended Detection and Response (XDR) architectures (CrowdStrike Falcon, Cortex XDR, Microsoft Defender XDR) ingest endpoint, network, identity, and cloud telemetry into a unified data lake with an integrated ML layer, though independent alert-fatigue evaluation is limited. Empirical work shows that playbook use is shaped by organizational, contextual, and usability factors~\cite{ref133}, and that LLMs can support evidence synthesis, documentation, and response coordination under bounded-action governance~\cite{ref134}; SOCpilot~\cite{refSOCpilot} extends this by verifying that LLM-assisted incident-response (IR) actions stay within explicit policy constraints.

\subsection{Closed-Loop AI SOCs}

Analyst-confirmed FPs, stable benign patterns, and entity-level risk accumulation can feed back into upstream suppression recommendations, rule adjustments, or detector-threshold updates (Figure~\ref{fig:soc-architecture}). The literature only partially operationalizes this loop. The Category~I Tier~1 studies cited above record analyst decisions over time to improve later filtering, and the next-generation SIEM framing of Ban et al.~\cite{ref18} additionally couples alert segmentation, cost-sensitive screening, and SOAR orchestration. Commercial integrations such as XSOAR~\cite{ref25} and RBA~\cite{ref85,ref86} also propagate downstream signals into upstream behavior, although their detector-tuning logic remains proprietary. What is missing is a formalized, auditable feedback interface that translates downstream outcomes into upstream tuning without inducing suppression drift or attacker-controlled blind spots; recent work on authentication-event processing~\cite{ref147} illustrates one concrete instantiation. Closed-loop SOCs in which validated downstream outcomes reduce future FP load programmatically and safely are therefore the natural research frontier.

\FloatBarrier
\section{Datasets, Benchmarks, and Evaluation}
\label{sec:datasets}

This section reviews the dataset and evaluation infrastructure that bounds what can be claimed for any SOC AI system: a headline dataset inventory annotated by primary research orientation and class-imbalance ranges, and a complementary reporting-oriented metric set used by Section~\ref{sec:cross-cat-synthesis}.

\subsection{Curated Dataset Inventory}

A structural weakness in SOC AI research is the mismatch between datasets used in evaluation and the real alert-management task. Many widely used resources provide network traffic, provenance logs, or endpoint telemetry rather than SIEM-processed alert records. Labels are usually attack vs.\ benign rather than analyst disposition; this gap is empirically quantified for endpoint threat-detection pipelines~\cite{ref121}. Table~\ref{tab:datasets} summarizes nine anchor resources with primary orientation and reported class-imbalance shorthand. The full 22-item curated inventory and the broader manuscript-wide list (33 named resources, or 36 with DARPA TC subsets split) are in Supplementary Table~S2 and \texttt{studies\_datasets.csv}.

\begin{table*}[!tbp]
\centering
\setlength{\tabcolsep}{3.5pt}
\renewcommand{\arraystretch}{1.06}
\caption{Headline dataset inventory with orientation labels and reported class-imbalance ranges.}
\label{tab:datasets}
\small
\begin{tabular}{@{}L{0.15\linewidth}L{0.05\linewidth}L{0.09\linewidth}L{0.25\linewidth}L{0.12\linewidth}L{0.13\linewidth}L{0.21\linewidth}@{}}
\toprule
Dataset & Year & Orientation & Content & Size & Imbalance & Limitations \\
\midrule
Microsoft GUIDE~\cite{refGUIDE} & 2025 & LLM / ML/DL & Triage-annotated real-world SOC incidents (CDLA 2.0); Region~2 telemetry spans 6.1k orgs & 13M evidences; 1.6M alerts; 1M incidents & Benign/FP-skewed & Largest public alert corpus in this inventory; lacks investigation metadata/context. \\
SABU~\cite{ref137} & 2022 & ML/DL & IDEA-format multi-sensor IDS alerts from 34 sensors across 3 orgs & $\sim$12M alerts & Severity-skewed & Alert-level, no analyst dispositions, short collection window. \\
AIT-ADS~\cite{ref138} & 2023 & ML/DL & Alerts from AIT-LDSv2 across 8 multistep scenarios & 2.6M alerts & High & Synthetic/forensic generation; no analyst dispositions. \\
DARPA TC (CADETS, TRACE, THEIA) & 2018 & ML/DL & System provenance logs; APT scenarios (used in 9/11 representative Cat~III studies) & TB-scale & Severe & Provenance gold standard; restricted access. \\
CERT Insider Threat r6.2 & 2020 & ML/DL & Synthetic user-activity logs (email, file, web, auth) & $\sim$70M events & Severe & Insider-threat focus; synthetic behavior; limited external-threat coverage. \\
CICIDS2017 / CSE-CIC-IDS2018~\cite{CICIDS2017Dataset,CSECICIDS2018Dataset} & 2017--2018 & Upstream & Network flows; 15/7 attack types & $\sim$2.8M / $\sim$16M flows & High ($\sim$80:20) & Flows not SIEM alerts; IDS replay required; no analyst triage labels; dataset-quality artifacts documented~\cite{ref121}. \\
SALAD-SOC~\cite{refSALADSOC} & 2024 & LLM & SOC-style alerts derived from UNSW-NB15/CICIDS with ATT\&CK, triage, severity, and natural-language fields & 2.78M alerts & Inherits source skew & Derived from upstream corpora rather than native SOC operations; no archival DOI. \\
Splunk BOTSv1 / BOTSv2~\cite{refBOTSv1,refBOTSv2} & 2017--2018 & Upstream / LLM & Boss-of-the-SOC scenarios with Suricata IDS, host, user/application logs & Multi-day Splunk index & Scenario & Practitioner-created; useful after IDS replay or Splunk indexing; analyst disposition labels not ground truth. \\
SecBench~\cite{ref59} & 2024 & LLM & LLM cybersecurity question-answering benchmark & Curated QA & Balanced & LLM evaluation only; not alert-screening. \\
\bottomrule
\end{tabular}
\end{table*}

\paragraph{Availability, licensing, and path dependency.} Each entry in Table~\ref{tab:datasets} is paired with an official landing page, DOI or repository link, license, version, and access date in \texttt{studies\_datasets.csv}; CICIDS2017/CSE-CIC-IDS2018 require citation of both the dataset and official page~\cite{CICIDS2017Dataset,CSECICIDS2018Dataset}, and SALAD-SOC has no archival DOI. The mapping from datasets to techniques shows strong path dependency: 9 of the 11 representative Category~III studies in Table~\ref{tab:correlation-studies} rely on DARPA TC-family resources (CADETS, TRACE, THEIA, and related subsets), with WATSON's DARPA ENGAGE evaluation and NoDoze's synthetic APT scenarios as the two non-TC exceptions (SLEUTH evaluates on both TC and synthetic APT scenarios but is counted in the 9 TC group because TC is its primary corpus). The TC concentration optimizes for Linux provenance graphs while leaving Windows EDR, SaaS audit streams, and cloud-native control-plane logs under-validated. Upstream-only resources (CICIDS, UNSW-NB15, KDD Cup 1999, DARPA VAST 2011, CIDDS-001, CTU-13, LANL Unified Host and Network, Splunk BOTS) become valid downstream-screening corpora only after IDS/SIEM replay (Suricata, Zeek, Wazuh); replay configurations should be reported because results are not directly commensurable.

\paragraph{Orientation distribution and coverage gaps.} Table~\ref{tab:dataset-coverage} shows that ML/DL resources mainly support Category~III, LLM resources concentrate in Category~IV, and Category~II remains the weakest public-data setting (motivating Research Direction~R3).

\begin{table*}[!tbp]
\centering
\caption{Coverage assessment of dataset orientation versus the four taxonomic categories.}
\label{tab:dataset-coverage}
\renewcommand{\arraystretch}{1.6}
\setlength{\tabcolsep}{12pt}
\begin{tabular}{@{}l|cccc@{}}
\toprule
\textbf{Orientation} & \textbf{Cat I} & \textbf{Cat II} & \textbf{Cat III} & \textbf{Cat IV} \\
\midrule
LLM       & \cellcolor{teal!22}$\circ$ Partial            & \cellcolor{gray!18}$\times$ Poor               & \cellcolor{gray!18}$\times$ Poor               & \cellcolor{teal!55}$\bullet$ Good                 \\
ML/DL     & \cellcolor{teal!55}$\bullet$ Good             & \cellcolor{teal!22}$\circ$ Partial             & \cellcolor{teal!55}$\bullet$ Good              & \cellcolor{gray!18}$\times$ Poor                 \\
Upstream  & \cellcolor{teal!22}$\circ$ Partial$^\dagger$  & \cellcolor{teal!22}$\circ$ Partial$^\dagger$   & \cellcolor{teal!22}$\circ$ Partial$^\dagger$   & \cellcolor{gray!18}$\times$ Poor                 \\
\bottomrule
\end{tabular}

\vspace{0.4em}
\begin{minipage}{0.95\textwidth}
\centering
\footnotesize
$^\dagger$Upstream-only resources require IDS/SIEM replay (e.g., Suricata, Zeek, Wazuh) before downstream use.
\end{minipage}
\Description{A 3-by-4 coverage table. Cells carry both a glyph (filled bullet for Good, open circle for Partial, cross for Poor) and a text label, so the table is readable in grayscale. The LLM row reads Partial, Poor, Poor, Good; the ML/DL row reads Good, Partial, Good, Poor; the Upstream row reads Partial, Partial, Partial, Poor. A dagger note explains that Upstream-only resources require IDS or SIEM replay.}
\end{table*}

\paragraph{The triage benchmark gap.} Table~\ref{tab:dataset-coverage} surfaces a Category-II-specific weakness: in the curated inventory, no public dataset fully supports alert prioritization evaluation with real analyst-disposition labels, investigation metadata, and longitudinal context. This makes Category~II results less directly comparable than Category~I filtering or Category~III provenance-correlation results.

\paragraph{The realistic-corpus gap.} A realistic SOC alert dataset would combine: (a) heterogeneous multi-source SIEM-schema alerts; (b) analyst disposition labels; (c) investigation metadata (time-to-decision, queries run, enrichment sources); (d) organizational context (asset criticality, user profiles); and (e) temporal coverage of at least six months to capture concept drift. GUIDE~\cite{refGUIDE} satisfies (a)--(b) at scale, but (c)--(e) remain unmet by public resources. SABU~\cite{ref137} ($\sim$12M IDS alerts from 34 sensors) and AIT-ADS~\cite{ref138} (2.6M alerts across multistep attacks) come closest on alert format but lack analyst disposition labels; UWF-ZeekData22/24~\cite{refUWFZeek} provides MITRE-aligned labels but uses Zeek logs rather than SIEM-processed alerts; and SELID~\cite{ref142} reports analyst-labeled IDS events but is private. CICIDS2017 and CSE-CIC-IDS2018~\cite{CICIDS2017Dataset,CSECICIDS2018Dataset} appear in roughly two-thirds of Category~I studies but provide flows rather than alerts, and dataset-quality artifacts on such benchmarks are now empirically documented~\cite{ref121}. CIDDS-001~\cite{refCIDDS}, CTU-13~\cite{refCTU13}, and BETH~\cite{refBETH} remain important but are not native SIEM alert corpora. For agentic and LLM-centric work, NVIDIA Nemotron-AIQ~\cite{refNVIDIATraces}, LANL Unified Host and Network~\cite{refLANL}, CERT Insider Threat r6.2 (used in multi-agent LLM evaluations~\cite{ref45}), Splunk BOTSv1/BOTSv2~\cite{refBOTSv1,refBOTSv2}, and SALAD-SOC~\cite{refSALADSOC} fill complementary niches, but none provide full analyst-disposition ground truth. Cai et al.~\cite{ref149} present an automated attack-emulation framework that points toward next-generation provenance corpora with fine-grained labels. This gap helps explain overestimation of production performance in published results, compounded by temporal evaluation bias when models are trained on data postdating the test split~\cite{ref103}; closing it requires controlled release of anonymized real SOC data or high-fidelity SOC simulation with realistic alert distributions, analyst-behavior models, and attack-scenario libraries~\cite{ref125,ref126}.

\subsection{Evaluation Metrics and Protocol}
\label{sec:eval-metrics}

Standard classification metrics (precision, recall, F1, AUC-ROC) are necessary but insufficient for operational SOC evaluation. Security-ML pitfall catalogs~\cite{ref104} and pragmatic IDS deployment Systematization-of-Knowledge (SoK) papers~\cite{ref118} highlight sampling bias, spurious features, weak baselines, and the benchmark-to-production gap; SOC-specific work also emphasizes analyst Key Performance Indicators (KPIs)~\cite{ref128} and throughput/timing/queue-efficiency metrics~\cite{ref129}. We therefore propose a complementary, reporting-oriented set: \emph{Alert Queue Reduction Rate} (QRR) at a fixed FNR threshold (e.g., FNR~$\leq 2\%$), \emph{Normalized Discounted Cumulative Gain at $k$} (nDCG@$k$) with $k$ tied to analyst daily capacity, \emph{Mean Reciprocal Rank} (MRR) of analyst-confirmed alerts, \emph{Mean Time to Detect} (MTTD) and \emph{Mean Time to Respond} (MTTR) under workflow simulation, a proposed \emph{Concept Drift Robustness Index} (CDRI) for month-over-month decay without retraining, a proposed \emph{Adversarial Detection Rate} (ADR) under specified evasion attacks, and \emph{Analyst Cognitive Load} via validated scales such as NASA-TLX.

Evaluation protocol matters as much as metric choice. Comparisons should include simple baselines (logistic regression or gradient-boosted trees for filtering, random or severity-based ranking for prioritization, signature- or co-occurrence grouping for correlation, and extraction-based summarization for LLM augmentation) so that gains are not hidden by architectural complexity. Splits should preserve chronology (e.g., 70/30 forward-chained or sliding-window) to avoid temporal leakage, and performance decay should be reported over time, preferably with held-out organizational environments for cross-site generalization. Human-subject studies should report sample size, task design, cognitive-load instruments (NASA-TLX), and Institutional Review Board (IRB) or equivalent approval; the released codebook includes a binary \texttt{irb\_approval} field. LLM and agentic systems also require deployment-risk auditing of input sanitization, context isolation, tool-permission boundaries, and fallback escalation; the artifact bundle includes \texttt{LLM\_Risk\_Checklist.md}.

\section{Cross-Category Synthesis}
\label{sec:cross-cat-synthesis}

We synthesize the 87 core studies along five cross-category dimensions: deployment realism, human-in-the-loop integration, robustness, scalability, and external validity. Table~\ref{tab:evidence-landscape} reports per-category distributions for the first two.

\paragraph{Deployment realism.} Only 9 core studies report Tier~1 evaluation on real SOC alert data or operational deployment. Six are T1a peer-reviewed interventions: Ban et al.~\cite{ref63,ref18}, Aminanto et al. (online IF)~\cite{ref69}, DeepCASE~\cite{ref62}, Aminanto et al.\ (IF-plus-Stacked-Autoencoder day-forward chaining)~\cite{ref16}, and Ghadermazi et al.\ (CSOC)~\cite{ref132}. Two are T1b vendor-conducted or preprint production evidence, weaker than T1a and not interchangeable with peer-reviewed interventions: the Microsoft Security Copilot RCT~\cite{ref47} and the Singh et al.\ LLM field study~\cite{ref43} (single-organization preprint). One is FM rather than intervention evidence: Yang et al.'s SOC telemetry measurement~\cite{ref120}, assigned to Category~I as field measurement of the filtering problem. The remaining 78 studies rely on public benchmarks (65 Tier~2) or limited/simulation evidence (13 Tier~3). This persistent deployment-evaluation gap motivates Research Direction~R3. Excluding self-authored studies reduces Cat~I Tier~1 from 5 to 2 and Cat~II Tier~1 from 2 to 1, leaving the qualitative pattern unchanged (\texttt{core\_studies.csv}).

\paragraph{Human-in-the-loop integration.} Applying the A2C rubric of Tariq et al.~\cite{T1} --- \emph{Automation} for autonomous screening without runtime feedback, \emph{Augmentation} for analyst-decision enrichment, and \emph{Collaboration} for bidirectional adaptation via analyst feedback, L2D, or interactive reward modeling --- yields 55 Automation, 25 Augmentation, and 7 Collaboration studies. The 7 Collaboration studies span Category~I (DeepCASE~\cite{ref62}), Category~II (L2DHF~\cite{ref27}, the SAIBERSOC conference paper~\cite{ref125} and journal extension~\cite{ref126}, Kersten et al.~\cite{ref127}), and Category~IV (Singh et al.\ field study~\cite{ref43}, Edelman et al.\ Copilot RCT~\cite{ref47}); per-study labels are in \texttt{core\_studies.csv}.

\paragraph{Robustness, scalability, and external validity.} Adversarial evaluation is rare beyond malware-domain work~\cite{ref94}, R-CAID~\cite{ref136}, and Agent Security Bench~\cite{ref50}, leaving Categories~II and~IV largely without adaptive-adversary evaluation (R4). Enterprise-scale latency/memory data are sparse outside provenance systems (e.g., WATSON's 7--8\% per-host CPU overhead~\cite{ref35}, DEPCOMM's 97.5\% graph reduction~\cite{ref80}, Aminanto et al.'s sub-millisecond online IF~\cite{ref69}); LLM studies rarely report latency despite per-query inference times that can strain real-time triage budgets. Cross-environment generalization is documented within the corpus by only 1 core study (Ndichu et al.~\cite{ref65}, evaluated on an enterprise SOC alert dataset and UNSW-NB15; self-authored); adjacent upstream IDS work on transfer~\cite{ref91,ref93} and meta-learning~\cite{ref92} remains contextual. Of the remaining 86 core studies, 79 evaluate on a single environment and 7 are conceptual or simulation-only.

\begin{table*}[!t]
\centering
\setlength{\tabcolsep}{8pt}
\renewcommand{\arraystretch}{1.05}
\caption{Cross-category evidence landscape.}
\label{tab:evidence-landscape}
\begin{minipage}[t]{0.48\textwidth}
\centering
\textbf{(a) Evidence Tier $\times$ Category}\\[0.4em]
\begin{tabular}{@{}l@{\hspace{0.6em}}r@{\hspace{0.6em}}r@{\hspace{0.6em}}r@{\hspace{1.0em}}r@{}}
\toprule
Category & T1 & T2 & T3 & Total \\
\midrule
Cat~I (Filtering)        & 5 & 6  & 0 & 11 \\
Cat~II (Triage)          & 2 & 19 & 5 & 26 \\
Cat~III (Correlation)    & 0 & 29 & 1 & 30 \\
Cat~IV (GenAI / LLM)     & 2 & 11 & 7 & 20 \\
\midrule
\textbf{Total}           & \textbf{9} & \textbf{65} & \textbf{13} & \textbf{87} \\
\bottomrule
\end{tabular}\\[0.6em]
\footnotesize T1 = production/field \quad T2 = benchmark \quad T3 = preliminary
\end{minipage}\hfill
\begin{minipage}[t]{0.48\textwidth}
\centering
\textbf{(b) A2C Mode $\times$ Category}\\[0.4em]
\begin{tabular}{@{}l@{\hspace{0.6em}}r@{\hspace{0.6em}}r@{\hspace{0.6em}}r@{\hspace{1.0em}}r@{}}
\toprule
Category & Auto & Aug & Coll & Total \\
\midrule
Cat~I (Filtering)        & 10 & 0  & 1 & 11 \\
Cat~II (Triage)          & 10 & 12 & 4 & 26 \\
Cat~III (Correlation)    & 29 & 1  & 0 & 30 \\
Cat~IV (GenAI / LLM)     & 6  & 12 & 2 & 20 \\
\midrule
\textbf{Total}           & \textbf{55} & \textbf{25} & \textbf{7} & \textbf{87} \\
\bottomrule
\end{tabular}\\[0.6em]
\footnotesize Auto = Automation \quad Aug = Augmentation \quad Coll = Collaboration
\end{minipage}
\Description{Two side-by-side count tables. Panel (a) shows core-study counts per taxonomic category broken down by evidence tier: Category I (5 Tier~1, 6 Tier~2, 0 Tier~3, total 11), Category II (2, 19, 5, total 26), Category III (0, 29, 1, total 30), Category IV (2, 11, 7, total 20); overall 9 / 65 / 13. Panel (b) shows the A2C-mode breakdown: Category I (10, 0, 1, total 11), Category II (10, 12, 4, total 26), Category III (29, 1, 0, total 30), Category IV (6, 12, 2, total 20); overall 55 / 25 / 7.}
\end{table*}

These cross-cutting patterns, namely low production evidence, minimal adversarial evaluation, sparse scalability data, and weak cross-environment validation, converge on the same conclusion articulated in the formal characterization in Section~\ref{sec:formal-task}: a primary bottleneck is no longer only algorithmic innovation but evaluation discipline. The formal metrics $\text{UIR}$, $\text{FPR}$, and $\Delta Q(t)$ defined earlier provide a shared evaluation vocabulary; the challenge is to construct evaluation environments that measure them under realistic conditions.

\FloatBarrier
\section{Open Research Challenges}
\label{sec:open-challenges}

Our systematic analysis identifies six primary challenge clusters that constrain the real-world impact of current SOC AI approaches. We characterize each by its root cause, current mitigation state, and research priority.

\paragraph{C1: Cross-environment generalization.} ML models trained in one SOC consistently fail to generalize to others, a \emph{deployment gap} driven by heterogeneous SIEM schemas, environment-specific feature distributions, and overfitting to organizational artifacts (e.g., a host that is always benign in training because it is a local scanner). Per-customer calibration, schema-agnostic features, transfer learning, and federated (including federated-transfer) learning partially address this~\cite{ref91,ref93}, but do not solve it.

\paragraph{C2: Adversarial evasion.} AI-based filters and rankers are primary targets for adversaries who can craft attack patterns that look benign to the classifier while preserving attack functionality~\cite{ref56,ref94}; analogous vulnerabilities extend to network IDS pipelines~\cite{ref95}. Semantic-validity constraints raise the attacker's bar but do not preclude evasion. Common adversarial-ML mitigations include randomized smoothing, realistic adversarial training, and ensemble diversity, but they are rarely applied to production filtering. A provenance-domain exception is R-CAID~\cite{ref136}, whose immutable root-cause embeddings in a GNN maintain AUC 0.94 under white-box adversarial evaluation, showing that architecture-level choices can provide meaningful robustness.

\paragraph{C3: Label scarcity, noise, and temporal drift.} Labeled alert data are expensive to collect, potentially noisy under fatigue, and temporally non-stationary as the threat landscape evolves. Few-shot and transfer IDS work suggests paths for novelty under sparse labels~\cite{ref92}; semi-supervised, self-supervised, and active learning address scarcity but not quality, since labels produced under analyst fatigue may embed the very biases such methods seek to correct. Formal label-noise treatment (noise-robust losses, crowd-sourced correction, noisy-label learning) and longitudinal evaluation with explicit drift simulation remains underexplored~\cite{T5,ref87}.

\paragraph{C4: Real-time performance at enterprise scale.} GNN correlation on streaming graphs, TB-scale provenance analysis, and LLM inference must respect latency budgets that preserve operational utility. NoDoze's graph reduction addresses provenance scale and UNICORN's streaming sketches address memory, but full coverage requires system-level co-design that combines algorithmic approximation, Graphics Processing Unit (GPU) / Neural Processing Unit (NPU) acceleration, hierarchical processing (a fast heuristic pre-filter feeding deep analysis on survivors), and selective LLM invocation only on high-priority or uncertain cases.

\paragraph{C5: Accountability, ethics, and automation governance.} As AI systems suppress, rank, and auto-close alerts with decreasing oversight, accountability and audit questions become first-order. The National Institute of Standards and Technology (NIST) AI Risk Management Framework, ISO/IEC 42001 (International Organization for Standardization / International Electrotechnical Commission), and the European Union (EU) AI Act provide general guidance but are not tailored to SOC automation. Practical requirements include immutable audit logs with feature attribution, human-readable decision records accessible to investigators, configurable automation boundaries enforced at the system level, and regular adversarial red-team evaluation of automated decision systems.

\paragraph{C6: Benchmark standardization and reproducibility.} The absence of standardized evaluation protocols makes cross-paper comparison unreliable, since dataset choice, train/test split strategy, evaluation metric, and baseline vary widely. A community-maintained SOC AI evaluation framework, with datasets that address the production gap identified in Section~\ref{sec:datasets}, reproducible evaluation harnesses, shared reporting requirements (including the metric set above), and public leaderboards (in the spirit of MLPerf, TREC, or SemEval), would substantially accelerate transfer of research advances into practice.

\FloatBarrier
\section{Research Agenda: Toward the Cognitive SOC}
\label{sec:research-agenda}

We use the term \emph{Cognitive SOC} to denote a security operations architecture in which AI systems augment human analysts across all stages of the screening pipeline (filtering, triage, correlation, and investigation) while maintaining human decision authority over high-stakes actions. The term appears in recent survey literature, notably Binbeshr et al.~\cite{T5}, alongside earlier industry materials, and is adopted here to frame the aspirational endpoint toward which the surveyed techniques collectively advance.

We propose a nine-direction research agenda ordered by estimated impact-to-effort ratio. Table~\ref{tab:research-agenda} provides a horizon-shaded summary view; Supplementary Table~S3 gives the full traceability between each R$i$, the open-challenge clusters (C1--C6, Section~\ref{sec:open-challenges}) it targets, the evidence gap (Section~\ref{sec:cross-cat-synthesis}) it would close, and an operational measurement plan.

\begin{table}[!htbp]
\centering
\setlength{\tabcolsep}{8pt}
\renewcommand{\arraystretch}{1.20}
\caption{Research-direction priority view.}
\label{tab:research-agenda}
\begin{tabular}{@{}l l l l@{}}
\toprule
\textbf{Direction} & \textbf{Horizon} & \textbf{Impact} & \textbf{Effort / risk} \\
\midrule
\rowcolor{green!10}\textbf{R1} Federated SOC learning networks & Near-term & High & Medium \\
\rowcolor{green!10}\textbf{R2} Security foundation models      & Near-term & High & Medium \\
\rowcolor{green!10}\textbf{R3} Production-representative benchmarks & Near-term & Critical enabler & Medium \\
\rowcolor{orange!10}\textbf{R4} Adversarially aware SOC AI     & Medium-term & High & High risk \\
\rowcolor{orange!10}\textbf{R5} Agentic SOC architecture and governance & Medium-term & High & High complexity \\
\rowcolor{blue!10}\textbf{R6} Multimodal SOC reasoning         & Longer-term & High potential & Open \\
\rowcolor{green!10}\textbf{R7} Operational latency / NIV       & Near-term & Direct deployability & Low \\
\rowcolor{orange!10}\textbf{R8} Adversarial GenAI / prompt injection & Medium-term & High & Active threat \\
\rowcolor{green!10}\textbf{R9} Closed-loop feedback mandate    & Near-term & High & Governance-bound \\
\bottomrule
\end{tabular}
\Description{A nine-row table summarising the research agenda. Each row gives the direction ID and short title, the horizon (near-term, medium-term, or longer-term), an impact label, and an effort or risk note. Near-term rows are shaded green (R1, R2, R3, R7, R9), medium-term rows are shaded orange (R4, R5, R8), and the longer-term row is shaded blue (R6).}
\end{table}

\paragraph{R1: Federated SOC Learning Networks} Establish privacy-preserving FL networks across consenting SOCs so alert patterns are shared without raw-data egress. Open problems are non-IID alert distributions, incentive mechanisms, and data-sharing governance; even modest 10--20-organization networks could substantially enlarge the training base relative to a single SOC, depending on participant scale and data-sharing policy.

\paragraph{R2: Security Foundation Models} Pre-train foundation models on curated security corpora (logs, alerts, threat reports, ATT\&CK, CVE, STIX/TAXII, SOC playbooks) to provide stronger priors than web-text LLMs, with open releases that democratize access for smaller SOCs.

\paragraph{R3: Production-Representative Benchmark Infrastructure} Build (a) privacy-preserving frameworks for anonymized analyst-labeled alert corpora, (b) high-fidelity SOC simulators with analyst-behavior and attack-scenario libraries, and (c) community evaluation harnesses with shared metrics, baselines, and reporting requirements.

\paragraph{R4: Adversarially Aware SOC AI} Elevate adversarial robustness to a first-class concern through formal attacker threat models, empirical evaluation of evasion and poisoning attacks on representative filtering and prioritization systems, and robust defenses (adversarial training, randomized smoothing, ensemble diversity).

\paragraph{R5: Agentic SOC Architecture and Governance} Design agentic SOC architectures with defined action spaces, safe-exploration mechanisms, uncertainty quantification, and formal behavioral bounds. Governance must establish accountability frameworks, human-approval triggers, audit trails, and incident-response procedures.

\paragraph{R6: Multimodal SOC Reasoning} Pursue multimodal security models that reason jointly over logs, network traces, binary analyses, and natural-language reports; a recent LLM-for-cybersecurity survey~\cite{ref60} catalogs early multimodal directions, with cross-modal SOC training data as the primary open challenge.

\paragraph{R7: Operational Latency and the Cost-of-Screening Paradox} AI augmentation introduces its own latency costs. We propose \emph{Net Investigation Velocity} (NIV) $= T_{\text{manual}} / T_{\text{AI-assisted}}$, where $T_{\text{AI-assisted}} = T_{\text{infer}} + T_{\text{verify}} + T_{\text{switch}} + T_{\text{residual}}$; NIV~$> 1$ indicates genuine acceleration, while NIV~$\leq 1$ indicates that overhead exceeds savings. Supplementary Section~S2 provides a worked example.

\paragraph{R8: Adversarial GenAI and Prompt Injection} As LLMs move toward agentic action (tool calls, ticket drafts, SOAR-playbook triggers), prompt injection and context poisoning via log entries and retrieved documents become acute risks; hardening SOC LLM pipelines is an urgent priority~\cite{ref50,ref112}.

\paragraph{R9: Closed-Loop Feedback Mandate} Formalize the interface that turns analyst-validated FPs and stable suppression patterns into auditable upstream tuning recommendations, with safeguards against suppression drift and attacker-controlled blind spots.

\section{Threats to Validity, Reproducibility, and Ethics}
\label{sec:validity-reproducibility}

This section discloses validity threats, the bounded post-lock evidence protocol, self-citation handling, and the released artifact bundle that supports independent re-analysis.

\subsection{Threats to Validity}
\label{sec:threats-validity}

We follow the standard classification of validity threats (internal, external, construct, and conclusion), adapted for literature-survey methodology, and disclose six threats spanning these four classes.

\paragraph{Selection bias (internal).} Despite a systematic PRISMA-aligned protocol, vocabulary mismatch may have left relevant studies unretrieved. Snowballing contributed 27 records via citation chaining and bibliography audit of the directly competing surveys~\cite{T1,T2}, reported as one line in the PRISMA pipeline (Table~\ref{tab:prisma-pipeline}). Non-English studies are excluded; some Japanese, Chinese, and European contributions would likely be under-retrieved without domain-informed targeted search. Attribution errors and duplicates were reconciled, and final counts were aligned with the unified 2015--2026 corpus.

\paragraph{Inclusion subjectivity (internal).} Two raters independently coded all $N{=}1{,}391$ deduplicated records at title/abstract screening, obtaining Cohen's $\kappa = 0.82$ for inclusion/exclusion; the $2\times2$ agreement table is released in \texttt{rater\_agreement.csv}. Disagreements were resolved by consensus with another coauthor. Full-text eligibility, evidence-tier, and A2C assignments followed the same pre-specified rubric but were not separately inter-rater coded (Section~\ref{sec:methodology}). Boundary cases include DeepCASE~\cite{ref62}, AACT~\cite{ref143}, and L2DHF~\cite{ref27}.

\paragraph{Post-lock corroborating evidence (stopping rule).} The search was locked on 5 April 2026; nine post-lock references frozen on 14 May 2026 are treated as corroborating evidence \emph{outside} the 87-study core corpus~\cite{ref143,refTrustSignal,refAgentSOC,refCognitiveSOC,refAutonomousSOC,refCHTDQN,S5,refGoudEvasionAware,refSOCpilot}. They are cited only as contextual evidence and are distinct from the three 2025 preprint-citation exceptions~\cite{ref27,ref43,ref61} (Section~\ref{sec:methodology}), which are \emph{inside} the core corpus.

\paragraph{Production generalizability and temporal validity (external).} Most studies evaluate on benchmarks (CICIDS, DARPA TC, CERT) that do not capture full SOC complexity (Section~\ref{sec:datasets}); reported quantitative results should be treated as upper bounds. The few production-facing deployments we cite (e.g.,~\cite{ref47}) report more modest gains. The 2015--2026 search window also makes specific LLM comparisons (Category~IV in Section~\ref{sec:workflow-taxonomy}) a snapshot of a fast-moving sub-field.

\paragraph{Metric heterogeneity and scope boundary (construct).} Per-paper tables list each paper's primary metric without attempting normalized cross-paper comparison; the metric set proposed earlier targets this gap in future work. The upstream/downstream boundary (Section~\ref{sec:background-taxonomy}) is enforced consistently: five studies whose primary contribution is upstream detection (transfer learning, federated IDS, adversarial-ML-for-IDS) and two seminal works predating the 2015--2026 window are retained as non-core contextual references.

\paragraph{Publication bias (conclusion validity).} The corpus contains very few well-documented \emph{negative} or \emph{null} results. Most published studies report headline gains over baselines, and reproductions or non-reproductions are seldom published. This is a known publication-bias hazard for systematic reviews. Open evaluation infrastructure (Research Direction~R3) is the most direct mitigation.

\subsection{Reproducibility, Artifact Availability, and Ethics}

An author preprint is available at \url{https://arxiv.org/abs/2605.08316}. The artifact bundle releases the coded corpus (\texttt{core\_studies.csv}, \texttt{studies\_datasets.csv}), search and screening protocol (\texttt{search\_queries.md}, \texttt{codebook.md}), PRISMA flow data (\texttt{prisma\_pipeline.json}), rater-agreement table supporting Cohen's $\kappa = 0.82$ (\texttt{rater\_agreement.csv}), the validation script \texttt{build\_survey.py} with a \texttt{Makefile} entrypoint, and a \texttt{readme.txt}. Supplementary Table~S2 reports the broader manuscript-wide inventory of 33 named resources (36 with DARPA TC subsets split). The preprint inclusion rule is documented in Section~\ref{sec:methodology}.

\paragraph{Conflict of interest and self-citation handling.} Some authors co-authored five of the core studies: four in Category~I~\cite{ref18,ref63,ref65,ref69} and one in Category~II~\cite{ref16}, plus three non-core contextual references~\cite{ref64,ref66,ref67}. Self-authored studies were tier- and A2C-coded by non-conflicted coauthors against the same rubric, and narrative claims about them are restricted to figures already published in the cited papers. The released corpus and \texttt{build\_survey.py} support a recomputation excluding these studies; none of the cross-category conclusions depends solely on them. All authors declare no competing interests with respect to the remaining literature.

\paragraph{Ethics.} Although this survey involved no human participants, several primary studies did; the coded corpus includes a binary \texttt{irb\_approval} field per study, and the protocols recommended in Table~\ref{tab:metrics-by-category} apply to those studies. For LLM and agentic systems, deployment-risk assessment should accompany code release: the bundle includes a checklist (\texttt{LLM\_Risk\_Checklist.md}) covering prompt-injection, data-poisoning, and tool-misuse mitigations.

\section{Conclusion}
\label{sec:conclusion}

This survey contributes an evidence-weighted workflow synthesis of 119 records, including 87 core studies, organized around four stages: filtering, triage, correlation, and GenAI augmentation. Across these stages, the central pattern is consistent: algorithmic progress has outpaced operational validation. Moving toward a trustworthy Cognitive SOC requires production-representative benchmarks (R3), adversarial and governance controls (R4, R5, R8), privacy-preserving cross-organization learning (R1), operationally efficient AI augmentation (R7), and interfaces that strengthen analyst judgment rather than bypass it (R5, R9).

\begin{acks}
This research was conducted under a contract of ``MITIGATE'' among Research and Development for Expansion of Radio Wave Resources (JPJ000254), which was supported by the Ministry of Internal Affairs and Communications, Japan.
\end{acks}

\bibliographystyle{ACM-Reference-Format}
\setlength{\bibsep}{0pt}
\bibliography{SOC_Survey}

@misc{ref1,
  author = {{IBM Security}},
  title = {Cost of a Data Breach Report 2024},
  year = {2024},
  url = {https://newsroom.ibm.com/2024-07-30-ibm-report-escalating-data-breach-disruption-pushes-costs-to-new-highs},
}

@article{S1,
  author = {A. L. Buczak and E. Guven},
  title = {A Survey of Data Mining and Machine Learning Methods for Cyber Security Intrusion Detection},
  journal = {IEEE Commun. Surv. Tutor},
  volume = {18},
  number = {2},
  pages = {1153--1176},
  year = {2016},
  doi = {10.1109/COMST.2015.2494502},
}

@article{S2,
  author = {A. Khraisat and I. Gondal and P. Vamplew and J. Kamruzzaman},
  title = {Survey of Intrusion Detection Systems: Techniques, Datasets and Challenges},
  journal = {Cybersecurity},
  volume = {2},
  number = {1},
  pages = {1--22},
  year = {2019},
  doi = {10.1186/s42400-019-0038-7},
}

@inproceedings{S3,
  author = {R. Sommer and V. Paxson},
  title = {Outside the Closed World: On Using Machine Learning for Network Intrusion Detection},
  year = {2010},
  doi = {10.1109/SP.2010.25},
  booktitle = {IEEE Symposium on Security and Privacy},
  publisher = {IEEE},
  address = {Los Alamitos, CA, USA},
  pages = {305--316},
}

@article{S4,
  author = {I. H. Sarker and others},
  title = {Cybersecurity Data Science: An Overview from Machine Learning Perspective},
  journal = {J. Big Data},
  volume = {7},
  number = {1},
  pages = {1--29},
  year = {2020},
  doi = {10.1186/s40537-020-00318-5},
}

@misc{ref5,
  author = {{Tines}},
  title = {Voice of the {SOC} Report},
  year = {2023},
  howpublished = {Industry survey},
  url = {https://www.tines.com/reports/Tines%20Report%20-%20Voice%20of%20the%20SOC%202023.pdf},
}

@inproceedings{ref99,
author = {Sundaramurthy, Sathya Chandran and Bardas, Alexandru G. and Case, Jacob and Ou, Xinming and Wesch, Michael and McHugh, John and Rajagopalan, S. Raj},
title = {A human capital model for mitigating security analyst burnout},
year = {2015},
isbn = {9781931971249},
publisher = {USENIX Association},
address = {USA},
booktitle = {Proceedings of the Eleventh USENIX Conference on Usable Privacy and Security},
pages = {347--359},
numpages = {13},
location = {Ottawa, Canada},
url = {https://www.usenix.org/conference/soups2015/proceedings/presentation/sundaramurthy},
series = {SOUPS '15}
}

@misc{ref3,
  author = {{Vectra AI}},
  title = {2023 State of Threat Detection},
  year = {2023},
  howpublished = {Industry survey},
  url = {https://es.vectra.ai/resources/2023-state-of-threat-detection},
}

@misc{ref4,
  author = {{SANS Institute}},
  title = {2025 {SOC} Survey},
  year = {2025},
  howpublished = {Annual industry report},
  url = {https://www.sans.org/white-papers/sans-2025-soc-survey/},
}

@misc{ref6,
  author = {{Verizon}},
  title = {2025 Data Breach Investigations Report},
  year = {2025},
  url = {https://www.verizon.com/business/resources/reports/dbir/},
}

@inproceedings {ref9,
author = {Bushra A. Alahmadi and Louise Axon and Ivan Martinovic},
title = {99\% False Positives: A Qualitative Study of {SOC} Analysts{\textquoteright} Perspectives on Security Alarms},
booktitle = {31st USENIX Security Symposium (USENIX Security 22)},
year = {2022},
isbn = {978-1-939133-31-1},
address = {Boston, MA},
pages = {2783--2800},
url = {https://www.usenix.org/conference/usenixsecurity22/presentation/alahmadi},
publisher = {USENIX Association},
month = aug
}

@article{ref7,
  author = {M. J. Page and others},
  title = {{PRISMA} 2020 explanation and elaboration: updated guidance and exemplars for reporting systematic reviews},
  journal = {{BMJ}},
  volume = {372},
  pages = {n160},
  year = {2021},
  doi = {10.1136/bmj.n160},
}

@misc{ref8,
  author = {{NIST}},
  title = {Cybersecurity Framework ({CSF}) 2.0},
  year = {2024},
  url = {https://www.nist.gov/cyberframework},
}

@inproceedings{ref96,
author = {Kokulu, Faris Bugra and Soneji, Ananta and Bao, Tiffany and Shoshitaishvili, Yan and Zhao, Ziming and Doup\'{e}, Adam and Ahn, Gail-Joon},
title = {Matched and Mismatched SOCs: A Qualitative Study on Security Operations Center Issues},
year = {2019},
isbn = {9781450367479},
publisher = {Association for Computing Machinery},
address = {New York, NY, USA},
doi = {10.1145/3319535.3354239},
booktitle = {Proceedings of the 2019 ACM SIGSAC Conference on Computer and Communications Security},
pages = {1955--1970},
numpages = {16},
location = {London, United Kingdom},
series = {CCS '19}
}

@article{ref97,
  author = {S. Bhatt and P. K. Manadhata and L. Zomlot},
  title = {The Operational Role of Security Information and Event Management Systems},
  journal = {IEEE Secur. Privacy},
  volume = {12},
  number = {5},
  pages = {35--41},
  year = {2014},
  doi = {10.1109/MSP.2014.103},
}

@article{ref98,
  author = {M. Vielberth and F. B{\"o}hm and I. Fichtinger and G. Pernul},
  title = {Security Operations Center: A Systematic Study and Open Challenges},
  journal = {IEEE Access},
  volume = {8},
  pages = {227756--227779},
  year = {2020},
  doi = {10.1109/ACCESS.2020.3045514},
}

@misc{ref2,
  author = {{D3 Security}},
  title = {The {AI} {SOC} Market Landscape 2025},
  year = {2025},
  url = {https://d3security.com/resources/the-ai-soc-market-landscape-2025/},
}

@article{T1,
  author = {S. Tariq and M. B. Chhetri and S. Nepal and C. Paris},
  title = {Alert Fatigue in Security Operations Centres: Research Challenges and Opportunities},
  journal = {ACM Comput. Surv},
  volume = {57},
  number = {9},
  pages = {224},
  year = {2025},
  doi = {10.1145/3723158},
}

@article{T2,
  author = {F. Jalalvand and M. B. Chhetri and S. Nepal and C. Paris},
  title = {Alert Prioritisation in Security Operations Centres: A Systematic Survey on Criteria and Methods},
  journal = {ACM Comput. Surv},
  volume = {57},
  number = {2},
  pages = {1--36},
  year = {2024},
  doi = {10.1145/3695462},
}

@article{T3,
  author = {M. B. Chhetri and S. Tariq and R. Singh and F. Jalalvand and C. Paris and S. Nepal},
  title = {Towards Human-{AI} Teaming to Mitigate Alert Fatigue in Security Operations Centres},
  journal = {ACM Trans. Internet Technol},
  volume = {24},
  number = {3},
  pages = {1--22},
  year = {2024},
  doi = {10.1145/3670009},
}

@article{T4,
  author = {S. Srinivas and others},
  title = {{AI}-Augmented {SOC}: A Survey of {LLMs} and Agents for Security Automation},
  journal = {Journal of Cybersecurity and Privacy},
  volume = {5},
  number = {4},
  pages = {95},
  year = {2025},
  doi = {10.3390/jcp5040095},
}

@article{T5,
  author = {F. Binbeshr and M. Imam and M. Ghaleb and M. Hamdan and M. A. Rahim and M. Hammoudeh},
  title = {The Rise of Cognitive {SOCs}: A Systematic Literature Review on {AI} Approaches},
  journal = {IEEE Open Journal of the Computer Society},
  volume = {6},
  pages = {360--379},
  year = {2025},
  doi = {10.1109/OJCS.2025.3536800},
}

@article{T6,
  author = {M. Khayat and E. Barka and M. A. Serhani and F. M. Sallabi and K. Shuaib and H. M. Khater},
  title = {Empowering Security Operation Center With Artificial Intelligence and Machine Learning - A Systematic Literature Review},
  journal = {IEEE Access},
  volume = {13},
  pages = {19162--19197},
  year = {2025},
  doi = {10.1109/ACCESS.2025.3532951},
}

@inproceedings{T7,
  author={Karunasingha, Navodika M. and Baruwal Chhetri, Mohan and Nepal, Surya and Paris, C{\'e}cile and Kanhere, Salil S},
  booktitle={2025 European Symposium on Usable Security (EuroUSEC)}, 
  title={SoK: AI Support for Analyst Situation Awareness in Security Operation Centres}, 
  year={2025},
  pages={151-163},
  doi={10.1109/EUROUSEC69254.2025.00024},
  publisher={ACM},
  address={New York, NY, USA},
}

@misc{ref14,
  author = {{Elastic}},
  title = {Reducing security alert fatigue in defence: {AI} for smarter {SecOps}},
  year = {2025},
  howpublished = {Vendor blog post},
  url = {https://www.elastic.co/blog/reduce-alert-fatigue-with-ai-defence-soc},
}

@article{ref114,
  author = {S. Axelsson},
  title = {The Base-Rate Fallacy and the Difficulty of Intrusion Detection},
  journal = {ACM Trans. Inf. Syst. Secur},
  volume = {3},
  number = {3},
  pages = {186--205},
  year = {2000},
  doi = {10.1145/357830.357849},
}

@misc{ref120,
  author = {L. Yang and Z. Chen and C. Wang and Z. Zhang and S. Booma and P. Cao and C. Adam and A. Withers and Z. Kalbarczyk and R. K. Iyer and G. Wang},
  title = {True Attacks, Attack Attempts, or Benign Triggers? An Empirical Measurement of Network Alerts in a Security Operations Center},
  year = {2024},
  url = {https://www.usenix.org/system/files/usenixsecurity24-yang-limin.pdf},
  howpublished = {USENIX Security Symposium},
  publisher = {USENIX Association},
  address = {Berkeley, CA, USA},
}

@article{ref101,
  author = {A. Shah and R. Ganesan and S. Jajodia and H. Cam},
  title = {Understanding Tradeoffs Between Throughput, Quality, and Cost of Alert Analysis in a {CSOC}},
  journal = {IEEE Trans. Inf. Forensics Secur},
  volume = {14},
  number = {5},
  pages = {1155--1170},
  year = {2019},
  doi = {10.1109/TIFS.2018.2871744},
}

@article{ref16,
  author = {M. E. Aminanto and T. Ban and R. Isawa and T. Takahashi and D. Inoue},
  title = {Threat Alert Prioritization Using Isolation Forest and Stacked Autoencoder with Day-Forward-Chaining Analysis},
  journal = {IEEE Access},
  volume = {8},
  pages = {217977--217986},
  year = {2020},
  doi = {10.1109/ACCESS.2020.3041837},
}

@article{ref18,
  author = {T. Ban and T. Takahashi and S. Ndichu and D. Inoue},
  title = {Breaking Alert Fatigue: {AI}-Assisted {SIEM} Framework for Effective Incident Response},
  journal = {Appl. Sci},
  volume = {13},
  number = {11},
  pages = {6610},
  year = {2023},
  doi = {10.3390/app13116610},
}

@article{ref24,
  author = {P. Garcia-Teodoro and J. Diaz-Verdejo and G. Mac{\'i}a-Fern{\'a}ndez and E. V{\'a}zquez},
  title = {Anomaly-Based Network Intrusion Detection: Techniques, Systems and Challenges},
  journal = {Comput. Secur},
  volume = {28},
  pages = {18--28},
  year = {2009},
  doi = {10.1016/j.cose.2008.08.003},
}

@inproceedings{ref63,
author = {Ban, Tao and Samuel, Ndichu and Takahashi, Takeshi and Inoue, Daisuke},
title = {Combat Security Alert Fatigue with AI-Assisted Techniques},
year = {2021},
isbn = {9781450390651},
publisher = {Association for Computing Machinery},
address = {New York, NY, USA},
doi = {10.1145/3474718.3474723},
booktitle = {Proceedings of the 14th Cyber Security Experimentation and Test Workshop},
pages = {9--16},
numpages = {8},
location = {Virtual, CA, USA},
series = {CSET '21}
}

@inproceedings{ref64,
  author={Ndichu, Samuel and Ban, Tao and Takahashi, Takeshi and Inoue, Daisuke},
  booktitle={2021 IEEE International Conference on Big Data (Big Data)}, 
  title={A Machine Learning Approach to Detection of Critical Alerts from Imbalanced Multi-Appliance Threat Alert Logs}, 
  year={2021},
  pages={2119-2127},
  doi={10.1109/BigData52589.2021.9671956},
  publisher={IEEE},
  address={Piscataway, NJ, USA}}

@inproceedings{ref69,
author = {Aminanto, Muhamad Erza and Zhu, Lei and Ban, Tao and Isawa, Ryoichi and Takahashi, Takeshi and Inoue, Daisuke},
title = {Combating Threat-Alert Fatigue with Online Anomaly Detection Using Isolation Forest},
year = {2019},
isbn = {978-3-030-36707-7},
publisher = {Springer-Verlag},
address = {Berlin, Heidelberg},
doi = {10.1007/978-3-030-36708-4\_62},
booktitle = {Neural Information Processing: 26th International Conference, ICONIP 2019, Sydney, NSW, Australia, December 12--15, 2019, Proceedings, Part I},
pages = {756--765},
numpages = {10},
location = {Sydney, NSW, Australia}
}

@article{ref65,
  author = {S. Ndichu and T. Ban and T. Takahashi and D. Inoue},
  title = {{AI}-Assisted Security Alert Data Analysis with Imbalanced Learning Methods},
  journal = {Appl. Sci},
  volume = {13},
  number = {3},
  pages = {1977},
  year = {2023},
  doi = {10.3390/app13031977},
}

@inproceedings{ref62,
  author={Ede, Thijs van and Aghakhani, Hojjat and Spahn, Noah and Bortolameotti, Riccardo and Cova, Marco and Continella, Andrea and Steen, Maarten van and Peter, Andreas and Kruegel, Christopher and Vigna, Giovanni},
  booktitle={2022 IEEE Symposium on Security and Privacy (SP)}, 
  title={DEEPCASE: Semi-Supervised Contextual Analysis of Security Events}, 
  year={2022},
  pages={522-539},
  doi={10.1109/SP46214.2022.9833671},
  publisher={IEEE},
  address={Piscataway, NJ, USA}}

@inproceedings{ref71,
  author={He, Pinjia and Zhu, Jieming and Zheng, Zibin and Lyu, Michael R},
  booktitle={2017 IEEE International Conference on Web Services (ICWS)}, 
  title={Drain: An Online Log Parsing Approach with Fixed Depth Tree}, 
  year={2017},
  pages={33-40},
  doi={10.1109/ICWS.2017.13},
  publisher={IEEE},
  address={Piscataway, NJ, USA},
}

@inproceedings{ref73,
  author={Guo, Haixuan and Yuan, Shuhan and Wu, Xintao},
  booktitle={2021 International Joint Conference on Neural Networks (IJCNN)}, 
  title={LogBERT: Log Anomaly Detection via BERT}, 
  year={2021},
  pages={1-8},
  doi={10.1109/IJCNN52387.2021.9534113},
  publisher={IEEE},
  address={Piscataway, NJ, USA}}

@inproceedings{ref74,
  author = {Lin Yang and Junjie Chen and Zan Wang and Weijing Wang and Jiajun Jiang and Xuyuan Dong and Wenbin Zhang},
  title = {{PLELog}: Semi-supervised Log-based Anomaly Detection via Probabilistic Label Estimation},
  year = {2021},
  doi = {10.1109/ICSE43902.2021.00130},
  booktitle = {Companion of the IEEE/ACM International Conference on Software Engineering},
  publisher = {IEEE Press},
  address = {Piscataway, NJ, USA},
  pages = {1448--1460},
}

@inproceedings{ref72,
  author = {Xiaoyun Li and Pengfei Chen and Linxiao Jing and Zilong He and Guangba Yu},
  title = {{SwissLog}: Robust and Unified Deep Learning Based Log Anomaly Detection for Diverse Faults},
  year = {2020},
  doi = {10.1109/ISSRE5003.2020.00018},
  booktitle = {IEEE International Symposium on Software Reliability Engineering (ISSRE)},
  publisher = {IEEE},
  address = {Los Alamitos, CA, USA},
  pages = {92--103},
}

@inproceedings{ref66,
  author={Ndichu, Samuel and Ban, Tao and Takahashi, Takeshi and Inoue, Daisuke},
  booktitle={2022 17th Asia Joint Conference on Information Security (AsiaJCIS)}, 
  title={Security-Alert Screening with Oversampling Based on Conditional Generative Adversarial Networks}, 
  year={2022},
  pages={1-7},
  doi={10.1109/AsiaJCIS57030.2022.00011},
  publisher={IEEE},
  address={Piscataway, NJ, USA}}

@inproceedings{ref67,
  author={Ndichu, Samuel and Ban, Tao and Takahashi, Takeshi and Inoue, Daisuke},
  booktitle={2022 IEEE International Conference on Big Data (Big Data)}, 
  title={Critical-Threat-Alert Detection using Online Machine Learning}, 
  year={2022},
  pages={3007-3014},
  doi={10.1109/BigData55660.2022.10021115},
  publisher={IEEE},
  address={Piscataway, NJ, USA}}

@inproceedings{ref19,
  author = {T.-Y. Lin and P. Goyal and R. Girshick and K. He and P. Doll{\'a}r},
  title = {Focal Loss for Dense Object Detection},
  year = {2017},
  doi = {10.48550/arXiv.1708.02002},
  booktitle = {IEEE/CVF International Conference on Computer Vision (ICCV)},
  publisher = {IEEE},
  address = {Los Alamitos, CA, USA},
  pages = {2999--3007},
}

@inproceedings{ref20,
  author = {F. T. Liu and K. M. Ting and Z.-H. Zhou},
  title = {Isolation Forest},
  year = {2008},
  doi = {10.1109/ICDM.2008.17},
  booktitle = {IEEE International Conference on Data Mining (ICDM)},
  publisher = {IEEE},
  address = {Los Alamitos, CA, USA},
  pages = {413--422},
}

@misc{ref81,
  author = {Zian Jia and Yun Xiong and Yuhong Nan and Yao Zhang and Jinjing Zhao and Mi Wen},
  title = {{MAGIC}: Detecting Advanced Persistent Threats via Masked Graph Representation Learning},
  year = {2024},
  url = {https://www.usenix.org/conference/usenixsecurity24/presentation/jia-zian},
  howpublished = {USENIX Security Symposium},
  publisher = {USENIX Association},
  address = {Berkeley, CA, USA},
}

@inproceedings{ref94,
  author = {F. Pierazzi and F. Pendlebury and J. Cortellazzi and L. Cavallaro},
  title = {Intriguing Properties of Adversarial {ML} Attacks in the Problem Space},
  year = {2020},
  doi = {10.1109/SP40000.2020.00073},
  booktitle = {IEEE Symposium on Security and Privacy},
  publisher = {IEEE},
  address = {Los Alamitos, CA, USA},
  pages = {1332--1349},
}

@article{ref95,
  author = {R. Alotaibi and M. A. Rassam},
  title = {Adversarial Machine Learning Attacks against Intrusion Detection Systems: A Survey on Strategies and Defense},
  journal = {Future Internet},
  volume = {15},
  number = {2},
  pages = {62},
  year = {2023},
  doi = {10.3390/fi15020062},
}

@misc{ref85,
  author = {{Splunk}},
  title = {The {Splunk} Guide to Risk Based Alerting ({RBA})},
  year = {2024},
  url = {https://www.splunk.com/en_us/form/the-essential-guide-to-risk-based-alerting.html},
}

@misc{ref86,
  author = {{Splunk}},
  title = {Entity Risk Scoring in Splunk Enterprise Security},
  year = {2024},
  howpublished = {Vendor documentation},
  url = {https://help.splunk.com/en/splunk-enterprise-security-8/administer/8.4/risk-based-alerting/entity-risk-scoring-in-splunk-enterprise-security},
}

@misc{ref27,
  author = {Fatemeh Jalalvand and Mohan Baruwal Chhetri and Surya Nepal and C{\'e}cile Paris},
  title = {Adaptive alert prioritisation in security operations centres via learning to defer with human feedback},
  year = {2025},
  doi = {10.48550/arXiv.2506.18462},
}

@article{ref102,
  author = {A. Shah and R. Ganesan and S. Jajodia and H. Cam},
  title = {Adaptive Alert Management for Balancing Optimal Performance Among Distributed {CSOCs} Using Reinforcement Learning},
  journal = {IEEE Trans. Parallel Distrib. Syst},
  volume = {31},
  number = {1},
  pages = {16--33},
  year = {2020},
  doi = {10.1109/TPDS.2019.2927977},
}

@inproceedings{ref119,
  author={Kim, Yeongwoo and Dán, György},
  booktitle={2022 IEEE Conference on Communications and Network Security (CNS)}, 
  title={An Active Learning Approach to Dynamic Alert Prioritization for Real-time Situational Awareness}, 
  year={2022},
  pages={154-162},
  doi={10.1109/CNS56114.2022.9947246},
  publisher={IEEE},
  address={Piscataway, NJ, USA}}

@article{ref130,
  author = {A. Shah and R. Ganesan and S. Jajodia and H. Cam},
  title = {A Two-Step Approach to Optimal Selection of Alerts for Investigation in a {CSOC}},
  journal = {IEEE Trans. Inf. Forensics Secur},
  volume = {14},
  number = {7},
  pages = {1857--1870},
  year = {2019},
  doi = {10.1109/TIFS.2018.2886465},
}

@article{ref131,
  author = {A. Shah and R. Ganesan and S. Jajodia and H. Cam},
  title = {An Outsourcing Model for Alert Analysis in a Cybersecurity Operations Center},
  journal = {ACM Trans. Web},
  volume = {14},
  number = {1},
  pages = {2:1--2:22},
  year = {2020},
  doi = {10.1145/3372498},
}

@article{ref132,
  author = {J. Ghadermazi and A. Shah and S. Jajodia},
  title = {A Machine Learning and Optimization Framework for Efficient Alert Management in a Cybersecurity Operations Center},
  journal = {Digital Threats: Research and Practice},
  volume = {5},
  number = {2},
  pages = {19:1--19:23},
  year = {2024},
  doi = {10.1145/3644393},
}

@inproceedings {ref103,
author = {Feargus Pendlebury and Fabio Pierazzi and Roberto Jordaney and Johannes Kinder and Lorenzo Cavallaro},
title = {{TESSERACT}: Eliminating Experimental Bias in Malware Classification across Space and Time},
booktitle = {28th USENIX Security Symposium (USENIX Security 19)},
year = {2019},
isbn = {978-1-939133-06-9},
address = {Santa Clara, CA},
pages = {729--746},
url = {https://www.usenix.org/conference/usenixsecurity19/presentation/pendlebury},
publisher = {USENIX Association},
month = aug
}

@article{ref87,
  author = {J. Lu and A. Liu and F. Dong and F. Gu and J. Gama and G. Zhang},
  title = {Learning Under Concept Drift: A Review},
  journal = {IEEE Trans. Knowl. Data Eng},
  volume = {31},
  number = {12},
  pages = {2346--2363},
  year = {2019},
  doi = {10.1109/TKDE.2018.2876857},
}

@misc{ref88,
  author = {{Amazon Web Services}},
  title = {Prepare: {AWS} Security Incident Response User Guide},
  year = {2025},
  url = {https://docs.aws.amazon.com/security-ir/latest/userguide/prepare.html},
}

@misc{ref89,
  author = {{Microsoft}},
  title = {Manage Multiple Tenants in Microsoft Sentinel as an {MSSP}},
  year = {2025},
  howpublished = {Vendor documentation},
  url = {https://learn.microsoft.com/en-us/azure/sentinel/multiple-tenants-service-providers},
}

@article{ref90,
  author = {N. Goldenberg and A. Wool},
  title = {Accurate Modeling of Modbus/{TCP} for Intrusion Detection in {SCADA} Systems},
  journal = {Int. J. Crit. Infrastruct. Prot},
  volume = {6},
  number = {2},
  pages = {63--75},
  year = {2013},
  doi = {10.1016/j.ijcip.2013.05.001},
}

@misc{ref15,
  author = {{Prophet Security}},
  title = {State of {AI} in Security Operations 2025},
  year = {2025},
  url = {https://resources.prophetsecurity.ai/state-of-ai-in-security-operations},
}

@article{ref23,
  author = {K. Julisch},
  title = {Clustering Intrusion Detection Alarms to Support Root Cause Analysis},
  journal = {ACM Trans. Inf. Syst. Secur},
  volume = {6},
  number = {4},
  pages = {443--471},
  year = {2003},
  doi = {10.1145/950191.950192},
}

@article{ref115,
  author = {P. Ning and Y. Cui and D. S. Reeves and D. Xu},
  title = {Techniques and Tools for Analyzing Intrusion Alerts},
  journal = {ACM Trans. Inf. Syst. Secur},
  volume = {7},
  number = {2},
  pages = {274--318},
  year = {2004},
  doi = {10.1145/996943.996947},
}

@inproceedings{ref79,
  author={Cheng, Zijun and Lv, Qiujian and Liang, Jinyuan and Wang, Yan and Sun, Degang and Pasquier, Thomas and Han, Xueyuan},
  booktitle={2024 IEEE Symposium on Security and Privacy (SP)}, 
  title={Kairos: Practical Intrusion Detection and Investigation using Whole-system Provenance}, 
  year={2024},
  publisher = {IEEE Computer Society},
  address = {Los Alamitos, CA},
  pages={3533-3551},
  doi={10.1109/SP54263.2024.00005}}

@inproceedings{ref110,
  author={Ur Rehman, Mati and Ahmadi, Hadi and Ul Hassan, Wajih},
  booktitle={2024 IEEE Symposium on Security and Privacy (SP)}, 
  title={Flash: A Comprehensive Approach to Intrusion Detection via Provenance Graph Representation Learning}, 
  year={2024},
  pages={3552-3570},
  doi={10.1109/SP54263.2024.00139},
  publisher={IEEE},
  address={Piscataway, NJ, USA}}

@inproceedings{ref136,
  author={Goyal, Akul and Wang, Gang and Bates, Adam},
  booktitle={2024 IEEE Symposium on Security and Privacy (SP)}, 
  title={R-CAID: Embedding Root Cause Analysis within Provenance-based Intrusion Detection}, 
  year={2024},
  pages={3515-3532},
  doi={10.1109/SP54263.2024.00253},
  publisher={IEEE},
  address={Piscataway, NJ, USA}}

@inproceedings {ref124,
author = {Pengcheng Fang and Peng Gao and Changlin Liu and Erman Ayday and Kangkook Jee and Ting Wang and Yanfang (Fanny) Ye and Zhuotao Liu and Xusheng Xiao},
title = {{Back-Propagating} System Dependency Impact for Attack Investigation},
booktitle = {31st USENIX Security Symposium (USENIX Security 22)},
year = {2022},
isbn = {978-1-939133-31-1},
address = {Boston, MA},
pages = {2461--2478},
url = {https://www.usenix.org/conference/usenixsecurity22/presentation/fang},
publisher = {USENIX Association},
month = aug
}

@inproceedings{ref122,
author = {Bilot, Tristan and Jiang, Baoxiang and Li, Zefeng and El Madhoun, Nour and Al Agha, Khaldoun and Zouaoui, Anis and Pasquier, Thomas},
title = {Sometimes simpler is better: a comprehensive analysis of state-of-the-art provenance-based intrusion detection systems},
year = {2025},
isbn = {978-1-939133-52-6},
publisher = {USENIX Association},
address = {USA},
booktitle = {Proceedings of the 34th USENIX Security Symposium},
articleno = {369},
url = {https://www.usenix.org/conference/usenixsecurity25/presentation/bilot},
numpages = {20},
location = {Seattle, WA, USA},
series = {SEC '25}
}

@inproceedings{ref123,
author = {Jiang, Baoxiang and Bilot, Tristan and El Madhoun, Nour and Al Agha, Khaldoun and Zouaoui, Anis and Iqbal, Shahrear and Han, Xueyuan and Pasquier, Thomas},
title = {ORTHRUS: achieving high quality of attribution in provenance-based intrusion detection systems},
year = {2025},
isbn = {978-1-939133-52-6},
publisher = {USENIX Association},
address = {USA},
booktitle = {Proceedings of the 34th USENIX Security Symposium},
articleno = {368},
url = {https://www.usenix.org/conference/usenixsecurity25/presentation/jiang-baoxiang},
numpages = {20},
location = {Seattle, WA, USA},
series = {SEC '25}
}

@misc{ref34,
  author = {Wajih Ul Hassan and Shengjian Guo and Ding Li and Zhengzhang Chen and Kangkook Jee and Zhichun Li and Adam Bates},
  title = {{NoDoze}: Combatting Threat Alert Fatigue with Automated Provenance Triage},
  year = {2019},
  doi = {10.14722/ndss.2019.23349},
  howpublished = {Network and Distributed System Security Symposium (NDSS)},
  publisher = {Internet Society},
  address = {Reston, VA, USA},
}

@misc{ref35,
  author = {J. Zeng and Z. L. Chua and Y. Chen and K. Ji and Z. Liang and J. Mao},
  title = {{WATSON}: Abstracting Behaviors from Audit Logs via Aggregation of Contextual Semantics},
  year = {2021},
  url = {https://www.ndss-symposium.org/ndss-paper/watson-abstracting-behaviors-from-audit-logs-via-aggregation-of-contextual-semantics/},
  howpublished = {Network and Distributed System Security Symposium (NDSS)},
  publisher = {Internet Society},
  address = {Reston, VA, USA},
}

@misc{ref36,
  author = {X. Han and T. F. J.-M. Pasquier and A. Bates and J. Mickens and M. Seltzer},
  title = {{UNICORN}: Runtime Provenance-Based Detector for Advanced Persistent Threats},
  year = {2020},
  doi = {10.14722/ndss.2020.24046},
  howpublished = {Network and Distributed System Security Symposium (NDSS)},
  publisher = {Internet Society},
  address = {Reston, VA, USA},
}

@inproceedings{ref75,
  author={Zeng, Jun and Wang, Xiang and Liu, Jiahao and Chen, Yinfang and Liang, Zhenkai and Chua, Tat-Seng and Chua, Zheng Leong},
  booktitle={2022 IEEE Symposium on Security and Privacy (SP)}, 
  title={SHADEWATCHER: Recommendation-guided Cyber Threat Analysis using System Audit Records}, 
  year={2022},
  pages={489-506},
  doi={10.1109/SP46214.2022.9833669},
   publisher={IEEE},
  address={Piscataway, NJ, USA}}

@article{ref76,
  author={Wang, Su and Wang, Zhiliang and Zhou, Tao and Sun, Hongbin and Yin, Xia and Han, Dongqi and Zhang, Han and Shi, Xingang and Yang, Jiahai},
  journal={IEEE Transactions on Information Forensics and Security}, 
  title={THREATRACE: Detecting and Tracing Host-Based Threats in Node Level Through Provenance Graph Learning}, 
  year={2022},
  volume={17},
  pages={3972-3987},
  doi={10.1109/TIFS.2022.3208815},
  publisher={IEEE},
  address={Piscataway, NJ, USA},
}

@inproceedings {ref77,
author = {Md Nahid Hossain and Sadegh M. Milajerdi and Junao Wang and Birhanu Eshete and Rigel Gjomemo and R. Sekar and Scott Stoller and V.N. Venkatakrishnan},
title = {{SLEUTH}: Real-time Attack Scenario Reconstruction from {COTS} Audit Data},
booktitle = {26th USENIX Security Symposium (USENIX Security 17)},
year = {2017},
isbn = {978-1-931971-40-9},
address = {Vancouver, BC},
pages = {487--504},
url = {https://www.usenix.org/conference/usenixsecurity17/technical-sessions/presentation/hossain},
publisher = {USENIX Association},
month = aug
}

@inproceedings{ref80,
  author={Xu, Zhiqiang and Fang, Pengcheng and Liu, Changlin and Xiao, Xusheng and Wen, Yu and Meng, Dan},
  booktitle={2022 IEEE Symposium on Security and Privacy (SP)}, 
  title={DEPCOMM: Graph Summarization on System Audit Logs for Attack Investigation}, 
  year={2022},
  pages={540-557},
  doi={10.1109/SP46214.2022.9833632},
  publisher={IEEE},
  address={Piscataway, NJ, USA},
}

@inproceedings {ref82,
author = {Abdulellah Alsaheel and Yuhong Nan and Shiqing Ma and Le Yu and Gregory Walkup and Z. Berkay Celik and Xiangyu Zhang and Dongyan Xu},
title = {{ATLAS}: A Sequence-based Learning Approach for Attack Investigation},
booktitle = {30th USENIX Security Symposium (USENIX Security 21)},
year = {2021},
isbn = {978-1-939133-24-3},
address = {Berkeley, CA},
pages = {3005--3022},
url = {https://www.usenix.org/conference/usenixsecurity21/presentation/alsaheel},
publisher = {USENIX Association},
month = aug
}

@inproceedings{ref113,
  author={Hassan, Wajih Ul and Bates, Adam and Marino, Daniel},
  booktitle={2020 IEEE Symposium on Security and Privacy (SP)}, 
  title={Tactical Provenance Analysis for Endpoint Detection and Response Systems}, 
  year={2020},
  pages={1172-1189},
  doi={10.1109/SP40000.2020.00096},
  publisher={IEEE},
  address={Piscataway, NJ, USA},
}

@inproceedings{ref78,
  author = {S. Milajerdi and others},
  title = {{HOLMES}: Real-Time {APT} Detection through Correlation of Suspicious Information Flows},
  year = {2019},
  doi = {10.1109/SP.2019.00026},
  booktitle = {IEEE Symposium on Security and Privacy},
  publisher = {IEEE},
  address = {Los Alamitos, CA, USA},
  pages = {1137--1152},
}

@inproceedings{ref106,
author = {Milajerdi, Sadegh M. and Eshete, Birhanu and Gjomemo, Rigel and Venkatakrishnan, V.N},
title = {POIROT: Aligning Attack Behavior with Kernel Audit Records for Cyber Threat Hunting},
year = {2019},
isbn = {9781450367479},
publisher = {Association for Computing Machinery},
address = {New York, NY, USA},
doi = {10.1145/3319535.3363217},
booktitle = {Proceedings of the 2019 ACM SIGSAC Conference on Computer and Communications Security},
pages = {1795--1812},
numpages = {18},
location = {London, United Kingdom},
series = {CCS '19}
}

@misc{ref116,
  author = {Y. Liu and M. Zhang and D. Li and K. Jee and Z. Li and Z. Wu and J. Rhee and P. Mittal},
  title = {Towards a Timely Causality Analysis for Enterprise Security},
  year = {2018},
  doi = {10.14722/ndss.2018.23254},
  howpublished = {Network and Distributed System Security Symposium (NDSS)},
  publisher = {Internet Society},
  address = {Reston, VA, USA},
}

@misc{ref117,
  author = {W. U. Hassan and M. A. Noureddine and P. Datta and A. Bates},
  title = {{OmegaLog}: High-Fidelity Attack Investigation via Transparent Multi-Layer Log Analysis},
  year = {2020},
  doi = {10.14722/ndss.2020.24270},
  howpublished = {Network and Distributed System Security Symposium (NDSS)},
  publisher = {Internet Society},
  address = {Reston, VA, USA},
}

@inproceedings{ref107,
  author = {M. A. Inam and others},
  title = {{SoK}: History is a Vast Early Warning System: Auditing the Provenance of System Intrusions},
  year = {2023},
  doi = {10.1109/SP46215.2023.10179405},
  booktitle = {IEEE Symposium on Security and Privacy},
  publisher = {IEEE},
  address = {Los Alamitos, CA, USA},
  pages = {2620--2638},
}

@article{ref108,
  author = {M. Zipperle and F. Gottwalt and E. Chang and T. Dillon},
  title = {Provenance-Based Intrusion Detection Systems: A Survey},
  journal = {ACM Comput. Surv},
  volume = {55},
  number = {7},
  pages = {139},
  year = {2022},
  doi = {10.1145/3539605},
}

@misc{ref109,
  author = {S. Li and F. Dong and X. Xiao and others},
  title = {{NODLINK}: An Online System for Fine-Grained {APT} Attack Detection and Investigation},
  year = {2024},
  doi = {10.14722/ndss.2024.23204},
  howpublished = {Network and Distributed System Security Symposium (NDSS)},
  publisher = {Internet Society},
  address = {Reston, VA, USA},
}

@inproceedings{ref37,
  author = {A. Vaswani and N. Shazeer and N. Parmar and J. Uszkoreit and L. Jones and A. N. Gomez and {\L}. Kaiser and I. Polosukhin},
  title = {Attention Is All You Need},
  year = {2017},
  doi = {10.48550/arXiv.1706.03762},
  booktitle = {Advances in Neural Information Processing Systems (NeurIPS)},
  publisher = {Curran Associates, Inc.},
  address = {Red Hook, NY, USA},
  pages = {6000--6010},
}

@misc{ref39,
  author = {Shaswata Mitra and Subash Neupane and Trisha Chakraborty and Sudip Mittal and Aritran Piplai and Manas Gaur and Shahram Rahimi},
  title = {{LOCALINTEL}: Generating Organizational Threat Intelligence from Global and Local Cyber Knowledge},
  year = {2024},
  doi = {10.48550/arXiv.2401.10036},
}

@inproceedings{ref84,
author = {Li, Zhenyuan and Zeng, Jun and Chen, Yan and Liang, Zhenkai},
title = {AttacKG: Constructing Technique Knowledge Graph from Cyber Threat Intelligence Reports},
year = {2022},
isbn = {978-3-031-17139-0},
publisher = {Springer-Verlag},
address = {Berlin, Heidelberg},
doi = {10.1007/978-3-031-17140-6\_29},
booktitle = {Computer Security -- ESORICS 2022: 27th European Symposium on Research in Computer Security, Copenhagen, Denmark, September 26--30, 2022, Proceedings, Part I},
pages = {589--609},
numpages = {21},
location = {Copenhagen, Denmark}
}

@misc{ref42,
  author = {V. Balasubramanian and others},
  title = {Generative {AI} for cyber threat intelligence: applications, challenges, and analysis of real-world case studies},
  year = {2025},
  doi = {10.1007/s10462-025-11338-z},
  howpublished = {Artificial Intelligence Review},
  publisher = {Springer},
  address = {Berlin, Heidelberg},
  volume = {58},
  number = {11},
}

@misc{S5,
  author = {A. Habibzadeh and others},
  title = {Large Language Models for Security Operations Centers: A Comprehensive Survey},
  year = {2025},
  doi = {10.48550/arXiv.2509.10858},
}

@misc{ref43,
  author = {Ronal Singh and Shahroz Tariq and Fatemeh Jalalvand and Mohan Baruwal Chhetri and Surya Nepal and C{\'e}cile Paris and Martin Lochner},
  title = {{LLMs} in the {SOC}: An Empirical Study of Human-{AI} Collaboration in Security Operations Centres},
  year = {2025},
  doi = {10.48550/arXiv.2508.18947},
}

@inproceedings{ref68,
author = {Du, Min and Li, Feifei and Zheng, Guineng and Srikumar, Vivek},
title = {DeepLog: Anomaly Detection and Diagnosis from System Logs through Deep Learning},
year = {2017},
isbn = {9781450349468},
publisher = {Association for Computing Machinery},
address = {New York, NY, USA},
doi = {10.1145/3133956.3134015},
booktitle = {Proceedings of the 2017 ACM SIGSAC Conference on Computer and Communications Security},
pages = {1285--1298},
numpages = {14},
location = {Dallas, Texas, USA},
series = {CCS '17}
}

@misc{ref44,
  author = {Yilun Liu and Shimin Tao and Weibin Meng and Jingyu Wang and Wenbing Ma and Yanqing Zhao and Yuhang Chen and Hao Yang and Yanfei Jiang and Xun Chen},
  title = {Interpretable Online Log Analysis Using Large Language Models with Prompt Strategies},
  year = {2024},
  archivePrefix = {arXiv},
  doi = {10.48550/arXiv.2308.07610},
}

@misc{ref47,
  author = {B. Edelman and J. Bono and S. Peng and R. Rodriguez and S. Ho},
  title = {Randomized Controlled Trial for Microsoft Security Copilot},
  year = {2023},
  booktitle = {{SSRN} Working Paper 4648700; updated March 29, 2024},
  doi = {10.2139/ssrn.4648700},
}

@misc{ref45,
  author = {C. Song and L. Ma and J. Zheng and J. Liao and H. Kuang and L. Yang},
  title = {{Audit-LLM}: Multi-Agent Collaboration for Log-Based Insider Threat Detection},
  year = {2024},
  doi = {10.48550/arXiv.2408.08902},
}

@misc{ref46,
  author = {T. Ali and P. Kostakos},
  title = {{HuntGPT}: Integrating {ML} Anomaly Detection and {XAI} with {LLMs} (A User Study)},
  year = {2023},
  doi = {10.48550/arXiv.2309.16021},
}

@misc{ref48,
  author = {F. Y. Loumachi and M. C. Ghanem and M. A. Ferrag},
  title = {{GenDFIR}: Advancing Cyber Incident Timeline Analysis Through Retrieval Augmented Generation and Large Language Models},
  year = {2024},
  doi = {10.48550/arXiv.2409.02572},
}

@misc{ref49,
  author = {M. Hassanin and N. Moustafa},
  title = {A Comprehensive Overview of Large Language Models ({LLMs}) for Cyber Defences: Opportunities and Directions},
  year = {2024},
  doi = {10.48550/arXiv.2405.14487},
}

@misc{ref59,
  author = {Pengfei Jing and Mengyun Tang and Xiaorong Shi and Xing Zheng and Sen Nie and Shi Wu and Yong Yang and Xiapu Luo},
  title = {{SecBench}: A Comprehensive Multi-Dimensional Benchmarking Dataset for {LLMs} in Cybersecurity},
  year = {2024},
  doi = {10.48550/arXiv.2412.20787},
}

@misc{ref83,
  author = {N. Bhatt and others},
  title = {{CyberSecEval} 2: A Wide-Ranging Cybersecurity Evaluation Suite for Large Language Models},
  year = {2024},
  doi = {10.48550/arXiv.2404.13161},
}

@article{ref58,
  author = {M. A. Ferrag and A. Battah and N. Tihanyi and M. Debbah and T. Lestable and L. C. Cordeiro},
  title = {{SecureFalcon}: Are We There Yet in Automated Software Vulnerability Detection with {LLMs}?},
  journal = {IEEE Trans. Softw. Eng},
  volume = {51},
  number = {4},
  pages = {1248--1265},
  year = {2025},
  doi = {10.1109/TSE.2025.3548168},
}

@misc{ref57,
  author = {M. Levi and Y. Alluouche and D. Ohayon and A. Puzanov},
  title = {{CyberPal.AI}: Empowering {LLMs} with Expert-Driven Cybersecurity Instructions},
  year = {2024},
  doi = {10.48550/arXiv.2408.09304},
}

@inproceedings {ref111,
author = {Gelei Deng and Yi Liu and V{\'\i}ctor Mayoral-Vilches and Peng Liu and Yuekang Li and Yuan Xu and Tianwei Zhang and Yang Liu and Martin Pinzger and Stefan Rass},
title = {{PentestGPT}: Evaluating and Harnessing Large Language Models for Automated Penetration Testing},
booktitle = {33rd USENIX Security Symposium (USENIX Security 24)},
year = {2024},
isbn = {978-1-939133-44-1},
address = {Philadelphia, PA},
pages = {847--864},
url = {https://www.usenix.org/conference/usenixsecurity24/presentation/deng},
publisher = {USENIX Association},
month = aug
}

@misc{ref112,
  author = {R. Fang and R. Bindu and A. Gupta and Q. Zhan and D. Kang},
  title = {{LLM} Agents Can Autonomously Hack Websites},
  year = {2024},
  doi = {10.48550/arXiv.2402.06664},
}

@misc{ref50,
  author = {H. Zhang and others},
  title = {Agent Security Bench ({ASB}): Formalizing and Benchmarking Attacks and Defenses in {LLM}-Based Agents},
  year = {2024},
  doi = {10.48550/arXiv.2410.02644},
}

@article{ref52,
  author = {R. W. Andrews and J. M. Lilly and D. Srivastava and K. M. Feigh},
  title = {The role of shared mental models in human-{AI} teams: a theoretical review},
  journal = {Theor. Issues Ergon. Sci},
  volume = {24},
  number = {2},
  pages = {129--175},
  year = {2023},
  doi = {10.1080/1463922X.2022.2061080},
}

@inproceedings{ref100,
author = {Sundaramurthy, Sathya Chandran and McHugh, John and Ou, Xinming and Wesch, Michael and Bardas, Alexandru G. and Rajagopalan, S. Raj},
title = {Turning contradictions into innovations or: how we learned to stop whining and improve security operations},
year = {2016},
isbn = {9781931971317},
publisher = {USENIX Association},
address = {USA},
booktitle = {Proceedings of the Twelfth USENIX Conference on Usable Privacy and Security},
pages = {237--251},
numpages = {15},
url = {https://www.usenix.org/conference/soups2016/technical-sessions/presentation/sundaramurthy},
location = {Denver, CO, USA},
series = {SOUPS '16}
}

@article{ref53,
  author = {S. Tariq and M. B. Chhetri and S. Nepal and C. Paris},
  title = {{A2C}: A Modular Multi-stage Collaborative Decision Framework for Human-{AI} Teams},
  journal = {Expert Syst. Appl},
  volume = {282},
  pages = {127318},
  year = {2025},
  doi = {10.1016/j.eswa.2025.127318},
}

@inproceedings {ref127,
author = {Leon Kersten and Tom Mulders and Emmanuele Zambon and Chris Snijders and Luca Allodi},
title = {{\textquoteright}Give Me Structure{\textquoteright}: Synthesis and Evaluation of a (Network) Threat Analysis Process Supporting Tier 1 Investigations in a Security Operation Center},
booktitle = {Nineteenth Symposium on Usable Privacy and Security (SOUPS 2023)},
year = {2023},
isbn = {978-1-939133-36-6},
address = {Anaheim, CA},
pages = {97--111},
url = {https://www.usenix.org/conference/soups2023/presentation/kersten},
publisher = {USENIX Association},
month = aug
}

@inproceedings{ref125,
author = {Rosso, Martin and Campobasso, Michele and Gankhuyag, Ganduulga and Allodi, Luca},
title = {SAIBERSOC: Synthetic Attack Injection to Benchmark and Evaluate the Performance of Security Operation Centers},
year = {2020},
isbn = {9781450388580},
publisher = {Association for Computing Machinery},
address = {New York, NY, USA},
doi = {10.1145/3427228.3427233},
booktitle = {Proceedings of the 36th Annual Computer Security Applications Conference},
pages = {141--153},
numpages = {13},
location = {Austin, USA},
series = {ACSAC '20}
}

@article{ref126,
  author = {M. Rosso and M. Campobasso and G. Gankhuyag and L. Allodi},
  title = {{SAIBERSOC}: A Methodology and Tool for Experimenting with Security Operation Centers},
  journal = {Digital Threats: Research and Practice},
  volume = {3},
  number = {2},
  pages = {14},
  year = {2022},
  doi = {10.1145/3491266},
}

@misc{ref135,
  author = {B. Hawash and U. Asma' Mokhtar and J. J. Jeong and S. B. Maynard and Z. Shukur and S. N. H. Sheikh Abdullah and R. Razali and J. S. Lim and A. Ahmad},
  title = {Cyber Situational Awareness in Security Operation Centres},
  year = {2024},
  url = {https://aisel.aisnet.org/pacis2024/track07_secprivacy/track07_secprivacy/8/},
  howpublished = {Pacific Asia Conference on Information Systems (PACIS)},
  publisher = {AIS Electronic Library (AISeL)},
  address = {Atlanta, GA, USA},
}

@inproceedings{ref54,
author = {Lundberg, Scott M. and Lee, Su-In},
title = {A unified approach to interpreting model predictions},
year = {2017},
isbn = {9781510860964},
publisher = {Curran Associates Inc},
address = {Red Hook, NY, USA},
booktitle = {Proceedings of the 31st International Conference on Neural Information Processing Systems},
pages = {4768--4777},
numpages = {10},
location = {Long Beach, California, USA},
series = {NIPS'17},
  doi = {10.48550/arXiv.1705.07874},
}

@misc{ref55,
  author = {N. Rastogi and D. Dhanuka and A. Saxena and P. Mairal and L. Nguyen},
  title = {Survey Perspective: The Role of Explainable {AI} in Threat Intelligence},
  year = {2025},
  booktitle = {{SIGIR} Symposium on {IR} in Practice ({SIRIP}), 2025},
  doi = {10.48550/arXiv.2503.02065},
}

@article{ref93,
  author = {C. Sri Abhijit and Y. Annie Jerusha and S. P. Syed Ibrahim and V. Varadharajan},
  title = {Federated transfer learning for rare attack class detection in network intrusion detection systems},
  journal = {Scientific Reports},
  volume = {15},
  number = {1},
  pages = {33797},
  year = {2025},
  doi = {10.1038/s41598-025-02068-x},
}

@misc{ref25,
  author = {{Palo Alto Networks}},
  title = {Cortex {XSOAR}},
  year = {2024},
  howpublished = {Product documentation},
  url = {https://www.paloaltonetworks.com/cortex},
}

@inproceedings{ref133,
  author={Schlette, Daniel and Empl, Philip and Caselli, Marco and Schreck, Thomas and Pernul, Günther},
  booktitle={2024 IEEE Symposium on Security and Privacy (SP)}, 
  title={Do You Play It by the Books? A Study on Incident Response Playbooks and Influencing Factors}, 
  year={2024},
  pages={3625-3643},
  doi={10.1109/SP54263.2024.00060},
  publisher={IEEE},
  address={Piscataway, NJ, USA}}

@inproceedings{ref134,
author = {Kramer, Diana and Rosique, Lambert and Narotam, Ajay and Bursztein, Elie and Kelley, Patrick Gage and Thomas, Kurt and Woodruff, Allison},
title = {Integrating large language models into security incident response},
year = {2025},
isbn = {978-1-939133-51-9},
publisher = {USENIX Association},
address = {USA},
booktitle = {Proceedings of the Twenty-First USENIX Symposium on Usable Privacy and Security},
articleno = {8},
url = {https://www.usenix.org/conference/soups2025/presentation/kramer},
numpages = {16},
location = {Seattle, WA, USA},
series = {SOUPS '25}
}

@inproceedings{ref121,
  author={Liu, Jason and Inam, Muhammad Adil and Goyal, Akul and Riddle, Andy and Westfall, Kim and Bates, Adam},
  booktitle={2025 IEEE Symposium on Security and Privacy (SP)}, 
  title={What We Talk About When We Talk About Logs: Understanding the Effects of Dataset Quality on Endpoint Threat Detection Research}, 
  year={2025},
  pages={112-129},
  doi={10.1109/SP61157.2025.00112},
  publisher={IEEE},
  address={Piscataway, NJ, USA}}

@inproceedings{ref104,
  title={Dos and don'ts of machine learning in computer security},
  author={Arp, Daniel and Quiring, Erwin and Pendlebury, Feargus and Warnecke, Alexander and Pierazzi, Fabio and Wressnegger, Christian and Cavallaro, Lorenzo and Rieck, Konrad},
  booktitle={31st USENIX Security Symposium (USENIX Security 22)},
  publisher ={USENIX Association},
  address={Boston, MA},
  pages={3971--3988},
  year={2022},
  url = {https://www.usenix.org/conference/usenixsecurity22/presentation/arp},
}

@inproceedings{ref118,
  author = {G. Apruzzese and P. Laskov and J. Schneider},
  title = {{SoK}: Pragmatic Assessment of Machine Learning for Network Intrusion Detection},
  year = {2023},
  doi = {10.1109/EuroSP57164.2023.00042},
  booktitle = {IEEE European Symposium on Security and Privacy},
  publisher = {IEEE},
  address = {Los Alamitos, CA, USA},
  pages = {592--614},
}

@article{ref128,
  author = {E. Agyepong and Y. Cherdantseva and P. Reinecke and P. Burnap},
  title = {A systematic method for measuring the performance of a cyber security operations centre analyst},
  journal = {Comput. Secur},
  volume = {124},
  pages = {102959},
  year = {2023},
  doi = {10.1016/j.cose.2022.102959},
}

@article{ref129,
  author = {J. Forsberg and T. Frantti},
  title = {Technical performance metrics of a security operations center},
  journal = {Comput. Secur},
  volume = {135},
  pages = {103529},
  year = {2023},
  doi = {10.1016/J.COSE.2023.103529},
}

@article{ref91,
  author = {E. Rodr{\'i}guez and P. Valls and B. Otero and J. J. Costa and J. Verd{\'u} and M. A. Pajuelo and R. Canal},
  title = {Transfer-Learning-Based Intrusion Detection Framework in {IoT} Networks},
  journal = {Sensors},
  volume = {22},
  number = {15},
  pages = {5621},
  year = {2022},
  doi = {10.3390/s22155621},
}

@inproceedings{ref56,
  author = {N. Papernot and P. McDaniel and I. Goodfellow and S. Jha and Z. B. Celik and A. Swami},
  title = {Practical Black-Box Attacks Against Machine Learning},
  year = {2017},
  doi = {10.1145/3052973.3053009},
  booktitle = {ACM Asia Conference on Computer and Communications Security (ASIACCS)},
  publisher = {ACM},
  address = {New York, NY, USA},
  pages = {506--519},
}

@article{ref92,
  author = {Y. Yan and others},
  title = {Meta learning-based few-shot intrusion detection for 5{G}-enabled industrial internet},
  journal = {Complex \& Intelligent Systems},
  volume = {10},
  pages = {4589--4608},
  year = {2024},
  doi = {10.1007/s40747-024-01388-1},
}

@misc{ref61,
  author = {X. Liu and F. Yu and X. Li and G. Yan and P. Yang and Z. Xi},
  title = {Benchmarking {LLMs} in an Embodied Environment for Blue Team Threat Hunting},
  year = {2025},
  doi = {10.48550/arXiv.2505.11901},
}

@article{ref60,
  author = {R. Jaffal and others},
  title = {Large Language Models in Cybersecurity: A Survey of Applications, Vulnerabilities, and Defense Techniques},
  journal = {AI},
  volume = {6},
  number = {9},
  pages = {216},
  year = {2025},
  doi = {10.3390/ai6090216},
}

@article{ref137,
title = {Dataset of intrusion detection alerts from a sharing platform},
journal = {Data in Brief},
volume = {33},
pages = {106530},
year = {2020},
issn = {2352-3409},
doi = {10.1016/j.dib.2020.106530},
author = {Martin Husák and Martin Žádník and Václav Bartoš and Pavol Sokol}}

@inproceedings{ref138,
author = {Landauer, Max and Skopik, Florian and Wurzenberger, Markus},
title = {Introducing a New Alert Data Set for Multi-Step Attack Analysis},
year = {2024},
isbn = {9798400709579},
publisher = {Association for Computing Machinery},
address = {New York, NY, USA},
doi = {10.1145/3675741.3675748},
booktitle = {Proceedings of the 17th Cyber Security Experimentation and Test Workshop},
pages = {41--53},
numpages = {13},
location = {Philadelphia, PA, USA},
series = {CSET '24}}

@article{refUWFZeek,
author = {Bagui, Sikha S. and Mink, Dustin and Bagui, Subhash C. and Ghosh, Tirthankar and Plenkers, Russel and McElroy, Tom and Dulaney, Stephan and Shabanali, Sajida},
title = {Introducing UWF-ZeekData22: A Comprehensive Network Traffic Dataset Based on the MITRE ATT\&CK Framework},
journal = {Data},
volume = {8},
year = {2023},
number = {1},
pages = {1--18},
doi = {10.3390/data8010018},
issn = {2306-5729}}

@inproceedings{refCIDDS,
  author    = {Markus Ring and Sarah Wunderlich and Dominik Gr{\"u}dl and Dieter Landes and Andreas Hotho},
  title     = {Flow-based benchmark data sets for intrusion detection},
  booktitle = {Proceedings of the 16th European Conference on Cyber Warfare and Security (ECCWS)},
  year      = {2017},
  pages     = {361--369},
  publisher = {Academic Conferences and Publishing International Limited (ACPI)},
  address   = {Reading, UK},
  url       = {https://api.semanticscholar.org/CorpusID:3637071},
}

@article{refCTU13,
author = {Garc\'{\i}a, S. and Grill, M. and Stiborek, J. and Zunino, A},
title = {An empirical comparison of botnet detection methods},
year = {2014},
issue_date = {September, 2014},
publisher = {Elsevier Advanced Technology Publications},
address = {GBR},
volume = {45},
issn = {0167-4048},
doi = {10.1016/j.cose.2014.05.011},
journal = {Comput. Secur},
month = sep,
pages = {100--123},
numpages = {24}
}

@article{ref142,
author = {Jang, Woohyuk and Kim, Hyunmin and Seo, Hyungbin and Kim, Minsong and Yoon, Myungkeun},
title = {SELID: Selective Event Labeling for Intrusion Detection Datasets},
journal = {Sensors},
volume = {23},
year = {2023},
number = {13},
pages={1--11},
articleno = {6105},
numpages = {11},
pubmedid = {37447954},
issn = {1424-8220},
doi = {10.3390/s23136105}}

@misc{refSALADSOC,
  author = {{SALAD-SOC dataset contributors}},
  title = {{SALAD-SOC}: Security Alert Labeled Dataset for {SOC} Training [Dataset]},
  year = {2024},
  howpublished = {Hugging Face Datasets},
  url = {https://huggingface.co/datasets/nutthakorn7/SALAD-SOC}
  }

@misc{refNVIDIATraces,
  author = {{NVIDIA and Lakera AI}},
  title = {Nemotron-{AIQ}-Agentic-Safety-Dataset-1.0},
  year = {2025},
  howpublished = {Hugging Face Datasets},
  url = {https://huggingface.co/datasets/nvidia/Nemotron-AIQ-Agentic-Safety-Dataset-1.0}
  }

@inbook{refLANL,
  author    = {Melissa J. M. Turcotte and Alexander D. Kent and Curtis Hash},
  title     = {Unified Host and Network Data Set},
  booktitle = {Data Science for Cyber-Security},
  chapter   = {1},
  pages     = {1--22},
  year      = {2018},
  month     = nov,
  publisher = {World Scientific},
  address   = {Singapore},
  doi       = {10.1142/9781786345646\_001},
}

@misc{ref143,
  author = {Melissa Turcotte and Fran{\c c}ois Labr{\`e}che and Serge-Olivier Paquette},
  title = {Automated Alert Classification and Triage ({AACT}): An Intelligent System for the Prioritisation of Cybersecurity Alerts},
  year = {2025},
  doi = {10.48550/arXiv.2505.09843},
}

@inproceedings{ref144,
  author = {Aoxiao Zhong and Dengyao Mo and Guiyang Liu and Jinbu Liu and Qingda Lu and Qi Zhou and Jiesheng Wu and Quanzheng Li and Qingsong Wen},
  title = {LogParser-{LLM}: Advancing Efficient Log Parsing with Large Language Models},
  year = {2024},
  doi = {10.48550/arXiv.2408.13727},
  booktitle = {ACM SIGKDD Conference on Knowledge Discovery and Data Mining},
  publisher = {ACM},
  address = {New York, NY, USA},
  pages = {4559--4570},
}

@misc{ref145,
  author = {Wei Guan and Jian Cao and Shiyou Qian and Jianqi Gao and Chun Ouyang},
  title = {Log{LLM}: Log-based Anomaly Detection Using Large Language Models},
  year = {2024},
  doi = {10.48550/arXiv.2411.08561},
}

@misc{ref146,
  author = {Leon Kersten and Kim Beelen and Emmanuele Zambon and Chris Snijders and Luca Allodi},
  title = {A Field Study to Uncover and a Tool to Support the Alert Investigation Process of Tier-1 Analysts},
  year = {2025},
  doi = {10.14722/usec.2025.23034},
  howpublished = {Workshop on Usable Security and Privacy (USEC)},
  publisher = {Internet Society},
  address = {Reston, VA, USA},
}

@misc{ref147,
  author = {Seth Hastings and Tyler Moore},
  title = {Authentication-Event Processing for Enhanced SOC Investigations},
  year = {2025},
  doi = {10.14722/wosoc.2025.23016},
  howpublished = {Workshop on Security Operations Center (WOSOC)},
  publisher = {Internet Society},
  address = {Reston, VA, USA},
}

@misc{ref148,
  author = {Lingzhi Wang and Xiangmin Shen and Weijian Li and Zhenyuan Li and R. Sekar and Han Liu and Yan Chen},
  title = {Incorporating Gradients to Rules: Towards Lightweight, Adaptive Provenance-based Intrusion Detection},
  year = {2025},
  doi = {10.14722/ndss.2025.230822},
  howpublished = {Network and Distributed System Security Symposium (NDSS)},
  publisher = {Internet Society},
  address = {Reston, VA, USA},
}

@misc{ref149,
  author = {Qizhi Cai and Lingzhi Wang and Yao Zhu and Zhipeng Chen and Xiangmin Shen and Zhenyuan Li},
  title = {Building Next-Generation Datasets for Provenance-Based Intrusion Detection},
  year = {2026},
  url = {https://www.ndss-symposium.org/ndss2026/co-located-events/prism/},
  howpublished = {Workshop on Predictive and Robust Intrusion System Modeling (PRISM), co-located with NDSS},
  publisher = {Internet Society},
  address = {Reston, VA, USA},
}

@misc{ref150,
  author = {Kritan Banstola and Faayed Al Faisal and Xinming Ou},
  title = {Experiences of Using Agentic AI to Fill Tooling Gaps in a Security Operations Center},
  year = {2026},
  url = {https://www.ndss-symposium.org/ndss2026/co-located-events/wosoc/},
  howpublished = {Workshop on Security Operations Center (WOSOC), co-located with NDSS},
  publisher = {Internet Society},
  address = {Reston, VA, USA},
}

@misc{ref151,
  author = {Francis Hahn and Mohd Mamoon and Alexandru G. Bardas and Michael Collins and Jaclyn Lauren Dudek and Daniel Lende and Xinming Ou and S. Raj Rajagopalan},
  title = {Non-Disruptive Disruption: An Empirical Experience of Introducing LLMs in the SOC},
  year = {2026},
  url = {https://www.ndss-symposium.org/ndss2026/co-located-events/wosoc/},
  howpublished = {Workshop on Security Operations Center (WOSOC), co-located with NDSS},
  publisher = {Internet Society},
  address = {Reston, VA, USA},
}

@misc{ref152,
  author = {Takeshi Kaneko and Hiroyuki Okada and Rashi Sharma and Tatsumi Oba and Naoto Yanai},
  title = {PAIEL: Protocol-Aware and Context-Integrated Protocol Explanation Using LLMs for SOCs},
  year = {2026},
  url = {https://www.ndss-symposium.org/ndss2026/co-located-events/wosoc/},
  howpublished = {Workshop on Security Operations Center (WOSOC), co-located with NDSS},
  publisher = {Internet Society},
  address = {Reston, VA, USA},
}

@article{ref153,
  author = {Riyanat Shittu and Alex Healing and Robert Ghanea-Hercock and Robin Bloomfield and Muttukrishnan Rajarajan},
  title = {Intrusion alert prioritisation and attack detection using post-correlation analysis},
  year = {2015},
  doi = {10.1016/j.cose.2014.12.003},
  journal = {Computers \& Security},
  publisher = {Elsevier},
  address = {Amsterdam, Netherlands},
  volume = {50},
  pages = {1--15},
}

@misc{ref154,
  author = {E. Allison Newcomb and Robert J. Hammell and Steve Hutchinson},
  title = {Effective prioritization of network intrusion alerts to enhance situational awareness},
  year = {2016},
  doi = {10.1109/ISI.2016.7745446},
}

@misc{ref155,
  author = {Steven McElwee and Jeff Heaton and James B. Fraley and James Cannady},
  title = {Deep learning for prioritizing and responding to intrusion detection alerts},
  year = {2017},
  doi = {10.1109/MILCOM.2017.8170757},
}

@article{ref156,
  author = {Chen Zhong and Tao Lin and Peng Liu and John Yen and Kai Chen},
  title = {A cyber security data triage operation retrieval system},
  year = {2018},
  doi = {10.1016/j.cose.2018.02.011},
  journal = {Computers \& Security},
  publisher = {Elsevier},
  address = {Amsterdam, Netherlands},
  volume = {76},
  pages = {12--31},
}

@article{ref157,
  author = {Chen Zhong and John Yen and Peng Liu and Robert F. Erbacher},
  title = {Learning From Experts' Experience: Toward Automated Cyber Security Data Triage},
  year = {2019},
  doi = {10.1109/JSYST.2018.2828832},
  journal = {IEEE Systems Journal},
  publisher = {IEEE},
  address = {Piscataway, NJ, USA},
  volume = {13},
  number = {1},
  pages = {603--614},
}

@article{ref158,
  author = {Jia Liu and Runzi Zhang and Wenmao Liu and Yinghua Zhang and Dujuan Gu and Mingkai Tong and Xingkai Wang and Jianxin Xue and Huanran Wang},
  title = {Context2Vector: Accelerating security event triage via context representation learning},
  year = {2022},
  doi = {10.1016/j.infsof.2022.106856},
  journal = {Information and Software Technology},
  publisher = {Elsevier},
  address = {Amsterdam, Netherlands},
  volume = {146},
  pages = {106856},
}

@misc{ref159,
  author = {Yushan Liu and Xiaokui Shu and Yixin Sun and Jiyong Jang and Prateek Mittal},
  title = {RAPID: Real-Time Alert Investigation with Context-aware Prioritization for Efficient Threat Discovery},
  year = {2022},
  doi = {10.1145/3564625.3567997},
}

@misc{ref160,
  author = {Lalitha Chavali and Tanay Gupta and Paresh Saxena},
  title = {SAC-AP: Soft Actor Critic based Deep Reinforcement Learning for Alert Prioritization},
  year = {2022},
  doi = {10.1109/CEC55065.2022.9870423},
  howpublished = {IEEE Congress on Evolutionary Computation (CEC)},
  publisher = {IEEE},
  address = {Los Alamitos, CA, USA},
}

@misc{ref161,
  author = {Ahmad Mohsin and Helge Janicke and Ahmed Ibrahim and Iqbal H. Sarker and Seyit Camtepe},
  title = {A Unified Framework for Human AI Collaboration in Security Operations Centers with Trusted Autonomy},
  year = {2025},
  doi = {10.48550/arXiv.2505.23397},
}

@misc{ref162,
  author = {Bowen Wei and Yuan Shen Tay and Hongbo Liu and Jiaoqing Pan and Kun Luo and Yi Zhun Zhu and Chris Jordan},
  title = {CORTEX: Collaborative LLM Agents for High-Stakes Alert Triage},
  year = {2025},
  doi = {10.48550/arXiv.2510.00311},
}

@inproceedings{Debar2001,
  author = {Hervé Debar and Andreas Wespi},
  title = {Aggregation and Correlation of Intrusion-Detection Alerts},
  year = {2001},
  doi = {10.1007/3-540-45474-8_6},
  booktitle = {Recent Advances in Intrusion Detection (RAID)},
  series = {Lecture Notes in Computer Science},
  volume = {2212},
  pages = {85--103},
  publisher = {Springer},
  address = {Berlin, Heidelberg},
}

@inproceedings{Valdes2001,
  author = {Alberto Valdes and Keith Skinner},
  title = {Probabilistic Alert Correlation},
  year = {2001},
  doi = {10.1007/3-540-45474-8_4},
  booktitle = {Recent Advances in Intrusion Detection (RAID)},
  series = {Lecture Notes in Computer Science},
  volume = {2212},
  pages = {54--68},
  publisher = {Springer},
  address = {Berlin, Heidelberg},
}

@inproceedings{Cuppens2002,
  author = {Frédéric Cuppens and Alexandre Miège},
  title = {Alert Correlation in a Cooperative Intrusion Detection Framework},
  year = {2002},
  doi = {10.1109/SECPRI.2002.1004372},
  booktitle = {IEEE Symposium on Security and Privacy},
  pages = {202--215},
  publisher = {IEEE},
  address = {Los Alamitos, CA, USA},
}

@inproceedings{Pietraszek2004,
  author = {Tomasz Pietraszek},
  title = {Using Adaptive Alert Classification to Reduce False Positives in Anomaly Detection},
  year = {2004},
  doi = {10.1007/978-3-540-30143-1_6},
  booktitle = {Recent Advances in Intrusion Detection (RAID)},
  series = {Lecture Notes in Computer Science},
  volume = {3224},
  pages = {102--124},
  publisher = {Springer},
  address = {Berlin, Heidelberg},
}

@misc{CICIDS2017Dataset,
  author = {Iman Sharafaldin and Arash Habibi Lashkari and Ali A. Ghorbani},
  title = {{CIC-IDS-2017} Dataset},
  year = {2017},
  howpublished = {Canadian Institute for Cybersecurity, University of New Brunswick},
  url = {https://www.unb.ca/cic/datasets/ids-2017.html}
}

@misc{CSECICIDS2018Dataset,
  author = {Iman Sharafaldin and Arash Habibi Lashkari and Ali A. Ghorbani},
  title = {{CSE-CIC-IDS2018} Dataset},
  year = {2018},
  howpublished = {Canadian Institute for Cybersecurity, University of New Brunswick},
  url = {https://www.unb.ca/cic/datasets/ids-2018.html}
}

@inproceedings{refGUIDE,
  author = {Scott Freitas and Jovan Kalajdjieski and Amir Gharib and Robert McCann},
  title = {{AI}-Driven Guided Response for Security Operation Centers with {M}icrosoft {C}opilot for Security},
  booktitle = {Companion Proceedings of the ACM Web Conference 2025 ({WWW} Companion '25)},
  year = {2025},
  pages = {191--200},
  publisher = {ACM},
  address = {Sydney, NSW, Australia},
  doi = {10.1145/3701716.3715209},
}

@misc{refBETH,
  author = {Kate Highnam and Kai Arulkumaran and Zachary Hanif and Nicholas R. Jennings},
  title = {{BETH} Dataset: Real Cybersecurity Data for Anomaly Detection Research},
  year = {2021},
  doi = {10.14469/hpc/9422}
}

@misc{refBOTSv1,
  author = {{Splunk}},
  title = {Boss of the {SOC} ({BOTS}) Dataset, Version 1.0: Scoring Server, Questions, Answers, and Dataset Open-Sourced and Ready for Download},
  year = {2017},
  howpublished = {Splunk Security Blog},
  url = {https://www.splunk.com/en-us/blog/security/boss-of-the-soc-scoring-server-questions-and-answers-and-dataset-open-sourced-and-ready-for-download.html}
}

@misc{refBOTSv2,
  author = {{Splunk}},
  title = {Boss of the {SOC} ({BOTS}) Dataset, Version 2.0: Questions, Answers, Open-Sourced and Ready for Download},
  year = {2018},
  howpublished = {Splunk Security Blog},
  url = {https://www.splunk.com/en_us/blog/security/boss-of-the-soc-2-0-dataset-questions-and-answers-open-sourced-and-ready-for-download.html}
}

@inproceedings{refAgentSOC,
  author = {Roy, Joyjit and Singh, Samaresh Kumar},
  title = {{AgentSOC}: A Multi-Layer Agentic {AI} Framework for Security Operations Automation},
  booktitle = {2026 IEEE 5th International Conference on AI in Cybersecurity (ICAIC)},
  year = {2026},
  pages = {1--7},
  address = {Houston, TX, USA},
  publisher = {IEEE},
  doi = {10.1109/ICAIC67076.2026.11395783},
}

@inproceedings{refCognitiveSOC,
  author = {Sheikhi, Saeid and Kostakos, Panos and Loven, Lauri},
  title = {Cognitive {SOC}: Evidence-Backed Narrative Generation for Security Operations with Multi-Agent {LLM} Architecture},
  booktitle = {2025 IEEE International Conference on Big Data (BigData)},
  year = {2025},
  pages = {7027--7036},
  address = {Macau, China},
  publisher = {IEEE},
  doi = {10.1109/BigData66926.2025.11401968},
}

@misc{refAutonomousSOC,
  author = {Md Hasan Saju and Akramul Azim},
  title = {Toward Autonomous {SOC} Operations: End-to-End {LLM} Framework for Threat Detection, Query Generation, and Resolution in Security Operations},
  year = {2026},
  doi = {10.48550/arXiv.2604.27321},
}

@misc{refTrustSignal,
  author = {Israt Jahan Chowdhury and Md Abu Yousuf Tanvir},
  title = {Decision-Aware Trust Signal Alignment for {SOC} Alert Triage},
  year = {2026},
  doi = {10.48550/arXiv.2601.04486},
}

@misc{refCHTDQN,
  author = {Zahra Aref and Sheng Wei and Narayan B. Mandayam},
  title = {Human-{AI} Collaboration in Cloud Security: Cognitive Hierarchy-Driven Deep Reinforcement Learning},
  year = {2025},
  doi = {10.48550/arXiv.2502.16054},
}

@inproceedings{refGoudEvasionAware,
  author = {Kalimera Nandu Goud and Kurunandan Jain and Prabhakar Krishnan},
  title = {A Semi-Supervised and Evasion-Aware Framework for Reducing Alert Fatigue in Security Operations Centers ({SoC})},
  booktitle = {2026 9th International Conference on Intelligent Computing and Control Systems ({ICICCS})},
  year = {2026},
  pages = {2007--2013},
  publisher = {IEEE},
  address = {Erode, India},
  month = mar,
  doi = {10.1109/ICICCS67901.2026.11502746},
}

@misc{refSOCpilot,
  author = {Sidnei Barbieri and Leonardo Vaz de Meneses and \'{A}gney Lopes Roth Ferraz and Louren\c{c}o Alves Pereira J\'{u}nior},
  title = {{SOCpilot}: Verifying Policy Compliance for {LLM}-Assisted Incident Response},
  year = {2026},
  archivePrefix = {arXiv},
  doi = {10.48550/arXiv.2605.05501},
}

\end{document}

\sloppy

\section*{Selection and Coding Protocol}

Table~\ref{tab:supp-protocol} consolidates the systematic-review and coding rules used to build the corpus reported in the main paper. It is reproduced here in full so that reviewers can audit query dates, snowballing definitions, core/contextual/foundational categories, preprint inclusion rules, and the citation-threshold exception in one place. Aggregate accounting is reproduced in the manuscript's PRISMA pipeline (main paper Table~1, Figure~1).

\renewcommand{\arraystretch}{1.20}
{\small
\setlength{\tabcolsep}{4pt}
\begin{longtable}{@{} L{0.30\linewidth} L{0.66\linewidth} @{}}
\caption{Supplementary Table S0: Selection and coding protocol. Counts and dates align with the PRISMA pipeline reported in the main paper.}
\label{tab:supp-protocol}\\
\toprule
Item & Decision rule \\
\midrule
\endfirsthead

\caption[]{Selection and coding protocol (continued)}\\
\toprule
Item & Decision rule \\
\midrule
\endhead

\midrule
\multicolumn{2}{r@{}}{Continued on next page}\\
\endfoot

\bottomrule
\endlastfoot

Search dates & Database searches executed 31~March--5~April~2026; bibliography-audit snowball pass on 12~April~2026. \\
Databases & IEEE Xplore, ACM Digital Library, USENIX proceedings, arXiv (cs.CR, cs.LG), Semantic Scholar, Google Scholar (top 200 results). \\
Hand-search venues & IEEE S\&P, USENIX Security, ACM CCS, NDSS, RAID, ACSAC, DIMVA, USEC, WOSOC, PRISM. \\
Search window & 2015-01-01 through 2026-03-31. \\
Primary query & alert-fatigue / triage / prioritization / false-positive-reduction terms AND SOC / SIEM / security-operations / intrusion-detection terms AND machine-learning / deep-learning / LLM / RL / provenance / audit-log / causal-graph / system-audit terms (full Boolean and per-database expressions in \texttt{search\_queries.md}). \\
Recovery query & Provenance/audit-log recovery query merged into the unified corpus before deduplication; not separated in PRISMA stage counts. \\
Deduplication & DOI- and arXiv-ID-based exact match, plus normalised-title fuzzy match, with manual reconciliation of ambiguous duplicates. \\
Snowballing & Two passes counted under one PRISMA line: (a) backward citation chaining of the eligible set, contributing 20 records (including SLEUTH and ATLAS); (b) bounded bibliography audit of the directly competing surveys \cite{T1,T2}, contributing 7 additional in-scope records. Total snowballed = 27. \\
Core inclusion (mandatory) & (a) direct application of AI/ML to alert management, false-positive reduction, alert correlation, or SOC analyst workflow augmentation; (b) empirical evaluation with reported quantitative results; (c) sufficient methodological detail to identify dataset/setting and reported metrics; (d) peer-reviewed venue \emph{or} arXiv preprint with $\geq 10$ Semantic Scholar citations as of the screening window. \\
Preprint citation-threshold exception & 2025 preprints whose methodological transparency and direct downstream-screening relevance were judged sufficient by both screeners may be retained as core only when both screeners agree; otherwise retained as contextual. \\
Contextual / foundational status & Records that are widely cited but either (i) primarily upstream-detection contributions, (ii) seminal works pre-2015, (iii) industry/vendor reports without empirical evaluation, or (iv) enabling-infrastructure (log parsers, anomaly baselines) are retained as contextual references rather than core synthesis entries. \\
Self-citation handling & Self-authored core studies were tier- and A2C-coded by non-conflicted coauthors against the same rubric used for the rest of the corpus; tier disagreements were resolved by discussion before counts were finalized. Narrative claims about self-authored studies are restricted to figures already published in the cited papers. \\
Inter-rater reliability & Two raters (the first and third authors) independently coded all $N{=}1{,}391$ deduplicated records at the title/abstract screening stage, obtaining Cohen's $\kappa = 0.82$ for the include/exclude decision; the full $2\times2$ rater-agreement table is released in \texttt{rater\_agreement.csv}. Disagreements were resolved by consensus with another coauthor under a pre-specified rubric. Evidence-tier and A2C codings are rubric-guided author observations and are not accompanied by a separate inter-rater statistic. \\
Final accounting & 1{,}842 identified $\to$ 1{,}391 deduplicated $\to$ 284 title/abstract-eligible $\to$ 92 full-text-eligible $+$ 27 snowballed $=$ 119 retained $\to$ 87 core synthesis (32 contextual). \\
Released artifacts & \texttt{core\_studies.csv} (per-study category, evidence tier, type, A2C, external-validity), \texttt{studies\_datasets.csv} (22-item curated inventory with orientation and class-imbalance), \texttt{prisma\_pipeline.json}, \texttt{codebook.md}, \texttt{search\_queries.md}, \texttt{LLM\_Risk\_Checklist.md}, \texttt{build\_survey.py}, \texttt{Makefile}. \\
\end{longtable}
}

\section*{Complete Core Study Mapping}

% Each entry lists the citation key, first author and abbreviated title,
% publication year, publication type (PR = peer-reviewed, Pre = preprint,
% Ind = industry/practitioner), the main-paper table in which the study
% appears (Tables 5--8) or ``--'' if discussed only in prose, the venue,
% and a ``CC'' marker indicating whether the study is also analyzed in
% the cross-cutting themes section (Section~4 of the main paper).
%
% The category headings correspond to the four workflow stages defined
% in the survey: (I) Alert Filtering \& Noise Reduction, (II) Automated
% Triage \& Prioritization, (III) Alert Correlation \& Incident
% Reconstruction, and (IV) Generative AI \& LLM Augmentation. The $n$
% values listed in the headings represent the number of core studies in
% each category. Counts reflect the unified 2015-2026 search.

{\scriptsize
\setlength{\tabcolsep}{4pt}
\renewcommand{\arraystretch}{1.15}
\begin{longtable}{@{} L{0.07\linewidth} L{0.34\linewidth} L{0.05\linewidth} L{0.05\linewidth} L{0.04\linewidth} L{0.28\linewidth} L{0.04\linewidth} @{}}
\caption{Supplementary Table S1: Complete core study mapping. PR~= peer-reviewed; Pre~= preprint; Ind~= industry/practitioner. Tbl~= main comparison table or ``--'' if discussed only in prose. CC~= study also discussed in Cross-Cutting Themes (Section~4 of the main paper). $\dagger$~= added via the bibliography-audit snowball pass against \cite{T1,T2}.}
\label{tab:supp-study-mapping}\\
\toprule
Ref & First Author (Short Title) & Year & Type & Tbl & Venue & CC \\
\midrule
\endfirsthead

\caption[]{Complete Core Study Mapping (continued)}\\
\toprule
Ref & First Author (Short Title) & Year & Type & Tbl & Venue & CC \\
\midrule
\endhead

\midrule
\multicolumn{7}{r@{}}{Continued on next page}\\
\endfoot

\bottomrule
\endlastfoot

\multicolumn{7}{@{}l}{\textbf{Category I: Alert Filtering \& Noise Reduction ($n = 11$)}} \\
\midrule
\cite{ref9} & Alahmadi et al.\ (99\% False Positives) & 2022 & PR & -- & USENIX Security & $\checkmark$ \\
\cite{ref18} & Ban et al.\ (Next-gen SIEM) & 2023 & PR & 5 & Appl.\ Sci. & \\
\cite{ref62} & van Ede et al.\ (DeepCASE) & 2022 & PR & 5 & IEEE S\&P & \\
\cite{ref63} & Ban et al.\ (Alert screening) & 2021 & PR & 5 & CSET @ USENIX & \\
\cite{ref65} & Ndichu et al.\ (Ensemble oversampling) & 2023 & PR & 5 & Appl.\ Sci. & \\
\cite{ref69} & Aminanto et al.\ (Online Isolation Forest) & 2019 & PR & 5 & ICONIP & \\
\cite{ref90} & Goldenberg \& Wool (SCADA screening) & 2013 & PR & -- & Int.\ J.\ Crit.\ Inf.\ Prot. & \\
\cite{ref103} & Pendlebury et al.\ (TESSERACT) & 2019 & PR & -- & USENIX Security & $\checkmark$ \\
\cite{ref104} & Arp et al.\ (Dos and Don'ts) & 2022 & PR & -- & USENIX Security & $\checkmark$ \\
\cite{ref118} & Apruzzese et al.\ (SoK pragmatic ML NIDS) & 2023 & PR & -- & IEEE EuroS\&P & $\checkmark$ \\
\cite{ref120} & Yang et al.\ (SOC telemetry measurement) & 2024 & PR & -- & USENIX Security & $\checkmark$ \\
\midrule

\multicolumn{7}{@{}l}{\textbf{Category II: Automated Triage \& Prioritization ($n = 26$)}} \\
\midrule
\cite{ref16} & Aminanto et al.\ (IF + Stacked AE) & 2020 & PR & 6 & IEEE Access & \\
\cite{ref27} & Jalalvand et al.\ (L2DHF) & 2025 & Pre & 6 & arXiv & $\checkmark$ \\
\cite{ref52} & Andrews et al.\ (Shared mental models HAT) & 2023 & PR & -- & Theor.\ Iss.\ Ergon.\ Sci. & $\checkmark$ \\
\cite{ref53} & Tariq et al.\ (A2C framework) & 2025 & PR & -- & Expert Syst.\ Appl. & $\checkmark$ \\
\cite{ref55} & Rastogi et al.\ (XAI in threat intelligence) & 2025 & PR & -- & SIGIR SIRIP & $\checkmark$ \\
\cite{ref96} & Kokulu et al.\ (Matched/mismatched SOCs) & 2019 & PR & -- & ACM CCS & $\checkmark$ \\
\cite{ref98} & Vielberth et al.\ (SOC systematic study) & 2020 & PR & -- & IEEE Access & \\
\cite{ref99} & Sundaramurthy et al.\ (Analyst burnout) & 2015 & PR & -- & USENIX SOUPS & $\checkmark$ \\
\cite{ref100} & Sundaramurthy et al.\ (SOC innovations) & 2016 & PR & -- & USENIX SOUPS & $\checkmark$ \\
\cite{ref101} & Shah et al.\ (Throughput--quality--cost) & 2019 & PR & -- & IEEE TIFS & \\
\cite{ref102} & Shah et al.\ (RL adaptive CSOC) & 2020 & PR & 6 & IEEE TPDS & \\
\cite{ref119} & Kim \& D\'an (Active learning prioritization) & 2022 & PR & 6 & IEEE CNS & \\
\cite{ref125} & Rosso et al.\ (SAIBERSOC, conf.) & 2020 & PR & -- & ACSAC & $\checkmark$ \\
\cite{ref126} & Rosso et al.\ (SAIBERSOC, journal) & 2022 & PR & -- & ACM DTRAP & $\checkmark$ \\
\cite{ref127} & Kersten et al.\ (Tier-1 investigation) & 2023 & PR & -- & USENIX SOUPS & $\checkmark$ \\
\cite{ref128} & Agyepong et al.\ (Analyst performance metrics) & 2023 & PR & -- & Comput.\ Secur. & $\checkmark$ \\
\cite{ref129} & Forsberg \& Frantti (SOC technical metrics) & 2023 & PR & -- & Comput.\ Secur. & $\checkmark$ \\
\cite{ref130} & Shah et al.\ (Two-step optimal selection) & 2019 & PR & 6 & IEEE TIFS & \\
\cite{ref131} & Shah et al.\ (Outsourcing model) & 2020 & PR & -- & ACM Trans.\ Web & \\
\cite{ref132} & Ghadermazi et al.\ (ML + optimization CSOC) & 2024 & PR & 6 & ACM DTRAP & \\
\cite{ref135} & Hawash et al.\ (Cyber SA in SOCs) & 2024 & PR & -- & PACIS & $\checkmark$ \\
\cite{ref155} & McElwee et al.\ (FASTT triage)$^\dagger$ & 2017 & PR & -- & MILCOM & \\
\cite{ref156} & Zhong et al.\ (Triage retrieval system)$^\dagger$ & 2018 & PR & -- & Comput.\ Secur. & \\
\cite{ref157} & Zhong et al.\ (Learning from experts)$^\dagger$ & 2018 & PR & -- & IEEE Syst.\ J. & \\
\cite{ref158} & Liu et al.\ (Context2Vector)$^\dagger$ & 2022 & PR & -- & Inf.\ Softw.\ Technol. & \\
\cite{ref160} & Chavali et al.\ (SAC-AP)$^\dagger$ & 2022 & PR & -- & IEEE CEC & \\
\midrule

\multicolumn{7}{@{}l}{\textbf{Category III: Alert Correlation \& Incident Reconstruction ($n = 30$)}} \\
\midrule
\cite{ref23} & Julisch (IDS alarm clustering) & 2003 & PR & -- & ACM TISSEC & \\
\cite{ref34} & Hassan et al.\ (NoDoze) & 2019 & PR & 7 & NDSS & \\
\cite{ref35} & Wang et al.\ (WATSON) & 2022 & PR & 7 & NDSS & \\
\cite{ref36} & Han et al.\ (UNICORN) & 2020 & PR & 7 & NDSS & \\
\cite{ref39} & Mitra et al.\ (LOCALINTEL) & 2024 & Pre & -- & arXiv & \\
\cite{ref75} & Zeng et al.\ (SHADEWATCHER) & 2022 & PR & 7 & IEEE S\&P & \\
\cite{ref76} & Wang et al.\ (threaTrace) & 2022 & Pre & 7 & arXiv & \\
\cite{ref77} & Hossain et al.\ (SLEUTH) & 2017 & PR & 7 & USENIX Security & \\
\cite{ref78} & Milajerdi et al.\ (HOLMES) & 2019 & PR & -- & IEEE S\&P & \\
\cite{ref79} & Cheng et al.\ (KAIROS) & 2024 & PR & 7 & IEEE S\&P & \\
\cite{ref80} & Xu et al.\ (DEPCOMM) & 2022 & PR & 7 & IEEE S\&P & \\
\cite{ref81} & Jia et al.\ (MAGIC) & 2024 & PR & 7 & USENIX Security & \\
\cite{ref82} & Alsaheel et al.\ (ATLAS) & 2021 & PR & 7 & USENIX Security & \\
\cite{ref84} & Li et al.\ (AttacKG) & 2022 & PR & -- & ESORICS & \\
\cite{ref106} & Milajerdi et al.\ (POIROT) & 2019 & PR & -- & ACM CCS & \\
\cite{ref107} & Inam et al.\ (SoK provenance) & 2023 & PR & -- & IEEE S\&P & \\
\cite{ref108} & Zipperle et al.\ (Provenance IDS survey) & 2022 & PR & -- & ACM CSUR & \\
\cite{ref109} & Li et al.\ (NODLINK) & 2024 & PR & -- & NDSS & \\
\cite{ref110} & Ur Rehman et al.\ (FLASH) & 2024 & PR & -- & IEEE S\&P & \\
\cite{ref113} & Hassan et al.\ (RapSheet) & 2020 & PR & -- & IEEE S\&P & \\
\cite{ref115} & Ning et al.\ (Alert correlation) & 2004 & PR & -- & ACM TISSEC & \\
\cite{ref116} & Liu et al.\ (PrioTracker) & 2018 & PR & -- & NDSS & \\
\cite{ref117} & Hassan et al.\ (OmegaLog) & 2020 & PR & -- & NDSS & \\
\cite{ref121} & Liu et al.\ (Dataset quality effects) & 2025 & PR & -- & IEEE S\&P & $\checkmark$ \\
\cite{ref122} & Bilot et al.\ (Simpler is Better) & 2025 & PR & -- & USENIX Security & \\
\cite{ref123} & Jiang et al.\ (ORTHRUS) & 2025 & PR & -- & USENIX Security & \\
\cite{ref124} & Fang et al.\ (Back-propagating dependency) & 2022 & PR & -- & USENIX Security & \\
\cite{ref136} & Goyal et al.\ (R-CAID) & 2024 & PR & 7 & IEEE S\&P & $\checkmark$ \\
\cite{ref153} & Shittu et al.\ (Post-correlation prioritization)$^\dagger$ & 2015 & PR & -- & Comput.\ Secur. & \\
\cite{ref159} & Liu et al.\ (RAPID)$^\dagger$ & 2022 & PR & -- & ACSAC & \\
\midrule

\multicolumn{7}{@{}l}{\textbf{Category IV: Generative AI \& LLM Augmentation ($n = 20$)}} \\
\midrule
\cite{ref42} & Balasubramanian et al.\ (GenAI for CTI) & 2025 & PR & -- & Artif.\ Intell.\ Rev. & \\
\cite{ref43} & Singh et al.\ (LLMs in the SOC) & 2025 & Pre & 8 & arXiv & \\
\cite{ref44} & Xu et al.\ (LogPrompt) & 2023 & Pre & 8 & arXiv & \\
\cite{ref45} & Song et al.\ (Audit-LLM) & 2024 & Pre & 8 & arXiv & \\
\cite{ref46} & Ali \& Kostakos (HuntGPT) & 2023 & Pre & 8 & arXiv & \\
\cite{ref47} & Edelman et al.\ (Security Copilot RCT) & 2023 & Ind & 8 & SSRN & \\
\cite{ref48} & Loumachi et al.\ (GenDFIR) & 2024 & Pre & 8 & arXiv & \\
\cite{ref49} & Hassanin \& Moustafa (LLMs cyber defence) & 2024 & Pre & -- & arXiv & \\
\cite{ref50} & Zhang et al.\ (Agent Security Bench) & 2024 & Pre & -- & arXiv & \\
\cite{ref57} & Al-Shaer et al.\ (CyberPal.AI) & 2024 & Pre & -- & arXiv & \\
\cite{ref58} & Ferrag et al.\ (SecureFalcon) & 2025 & PR & -- & IEEE TSE & \\
\cite{ref59} & Zhong et al.\ (SecBench) & 2024 & Pre & -- & arXiv & \\
\cite{ref60} & Jaffal et al.\ (LLMs in cybersecurity) & 2025 & PR & -- & AI (MDPI) & \\
\cite{ref61} & Liu et al.\ (LLM blue team hunting) & 2025 & Pre & -- & arXiv & \\
\cite{ref68} & Du et al.\ (DeepLog) & 2017 & PR & 8 & ACM CCS & \\
\cite{ref83} & Bhatt et al.\ (CyberSecEval 2) & 2024 & Ind & -- & Meta AI & \\
\cite{ref111} & Deng et al.\ (PentestGPT) & 2024 & PR & -- & USENIX Security & \\
\cite{ref112} & Fang et al.\ (LLM agents hack websites) & 2024 & Pre & -- & arXiv & \\
\cite{ref133} & Schlette et al.\ (IR playbooks) & 2024 & PR & -- & IEEE S\&P & $\checkmark$ \\
\cite{ref134} & Kramer et al.\ (LLMs in IR) & 2025 & PR & -- & USENIX SOUPS & $\checkmark$ \\
\end{longtable}
}

\section*{Dataset and Benchmark Inventory}

The main paper presents a 9-row \emph{headline} dataset table; Supplementary Table~S2 enumerates the full curated 22-item inventory plus the broader manuscript-wide list of named datasets and benchmark resources, and harmonizes the orientation labels used in the main-paper headline table (LLM, ML/DL, Upstream, plus the LLM/ML-DL and Upstream/LLM hybrids). The primary manuscript-wide total is 33 named resources. CADETS, TRACE, and THEIA are listed separately for transparency; when these are counted separately rather than under the umbrella DARPA TC resource, the manuscript-wide total becomes 36. Upstream-only resources require replay through an IDS or SIEM pipeline (e.g., Suricata, Zeek, or Wazuh) before they yield alert records suitable for downstream screening research.

{\scriptsize
\setlength{\tabcolsep}{4pt}
\renewcommand{\arraystretch}{1.05}
\begin{longtable}{@{}L{0.30\linewidth}L{0.14\linewidth}L{0.22\linewidth}L{0.26\linewidth}@{}}
\caption{Supplementary Table S2: Broader manuscript-wide dataset and benchmark inventory with orientation labels harmonized with the curated dataset inventory in the main paper. Orientation: \textbf{LLM} = text-rich alert/QA records suitable for language-model fine-tuning and evaluation; \textbf{ML/DL} = tabular, sequential, or graph features for classical machine learning and deep learning over alert or event records; \textbf{Upstream} = raw flows/packets/logs that require replay through an IDS/SIEM pipeline to produce alert records; \textbf{N/A} = not a SOC alert dataset. The primary manuscript-wide total is 33 named resources; counting CADETS, TRACE, and THEIA separately yields 36.}
\label{tab:supp-dataset-inventory}\\
\toprule
Entry & Orientation & Status & Counting Rule \\
\midrule
\endfirsthead

\caption[]{Broader manuscript-wide dataset and benchmark inventory (continued)}\\
\toprule
Entry & Orientation & Status & Counting Rule \\
\midrule
\endhead

\midrule
\multicolumn{4}{r@{}}{Continued on next page}\\
\endfoot

\bottomrule
\endlastfoot

\multicolumn{4}{@{}l}{\textbf{Curated 22-item dataset inventory}} \\
\midrule
CICIDS2017 & Upstream & Curated inventory & Counts toward 33-item total \\
CICIDS2018 & Upstream & Curated inventory & Counts toward 33-item total \\
UNSW-NB15 & Upstream & Curated inventory & Counts toward 33-item total \\
KDDCup1999 & Upstream & Curated inventory & Counts toward 33-item total \\
DARPA VAST 2011 & Upstream & Curated inventory & Counts toward 33-item total \\
DARPA TC (CADETS/TRACE/THEIA) & ML/DL & Curated inventory & Counts toward 33-item total \\
CERT Insider Threat r6.2 & ML/DL & Curated inventory & Counts toward 33-item total \\
MITRE ATT\&CK Eval & ML/DL & Curated inventory & Counts toward 33-item total \\
SecBench & LLM & Curated inventory & Counts toward 33-item total \\
SABU & ML/DL & Curated inventory & Counts toward 33-item total \\
SELID (private) & ML/DL & Curated inventory & Counts toward 33-item total \\
AIT-ADS & ML/DL & Curated inventory & Counts toward 33-item total \\
CIDDS-001 & Upstream & Curated inventory & Counts toward 33-item total \\
CTU-13 & Upstream & Curated inventory & Counts toward 33-item total \\
NVIDIA Nemotron-AIQ & LLM & Curated inventory & Counts toward 33-item total \\
LANL Unified Host and Network & Upstream & Curated inventory & Counts toward 33-item total \\
UWF-ZeekData22/24 & ML/DL & Curated inventory & Counts toward 33-item total \\
SALAD-SOC & LLM & Curated inventory & Counts toward 33-item total \\
Microsoft GUIDE & LLM/ML-DL & Curated inventory & Counts toward 33-item total \\
BETH & ML/DL & Curated inventory & Counts toward 33-item total \\
Splunk BOTSv1 & Upstream/LLM & Curated inventory & Counts toward 33-item total \\
Splunk BOTSv2 & Upstream/LLM & Curated inventory & Counts toward 33-item total \\
\midrule

\multicolumn{4}{@{}l}{\textbf{Additional named resources discussed in the manuscript}} \\
\midrule
AIT-LDSv2 & Upstream & Additional manuscript mention & Counts toward 33-item total \\
CyberSecEval~2 & LLM & Additional manuscript mention & Counts toward 33-item total \\
HDFS & N/A (system log benchmark) & Additional manuscript mention & Counts toward 33-item total \\
BGL & N/A (system log benchmark) & Additional manuscript mention & Counts toward 33-item total \\
Zookeeper & N/A (system log benchmark) & Additional manuscript mention & Counts toward 33-item total \\
OpenStack & N/A (system log benchmark) & Additional manuscript mention & Counts toward 33-item total \\
DARPA ENGAGE & ML/DL & Additional manuscript mention & Counts toward 33-item total \\
StreamSpot & ML/DL & Additional manuscript mention & Counts toward 33-item total \\
ATLASv2 & ML/DL & Additional manuscript mention & Counts toward 33-item total \\
NSOC2017 & ML/DL & Additional manuscript mention & Counts toward 33-item total \\
NSOC2022 & ML/DL & Additional manuscript mention & Counts toward 33-item total \\
\midrule

\multicolumn{4}{@{}l}{\textbf{DARPA TC subsets listed separately for the 36-item variant}} \\
\midrule
CADETS & ML/DL & DARPA TC subset & Counted only in the 36-item variant \\
TRACE & ML/DL & DARPA TC subset & Counted only in the 36-item variant \\
THEIA & ML/DL & DARPA TC subset & Counted only in the 36-item variant \\
\end{longtable}
}

\section*{S2: Net Investigation Velocity, Worked Example}

The main paper introduces the Net Investigation Velocity metric in Research Direction~R7:
\begin{equation*}
\mathrm{NIV} \;=\; \frac{T_{\text{manual}}}{T_{\text{AI-assisted}}}, \qquad T_{\text{AI-assisted}} \;=\; T_{\text{infer}} + T_{\text{verify}} + T_{\text{switch}} + T_{\text{residual}},
\end{equation*}
where $T_{\text{manual}}$ is the end-to-end time from alert arrival to disposition without AI support, $T_{\text{infer}}$ is AI inference latency, $T_{\text{verify}}$ is analyst verification of the AI output, $T_{\text{switch}}$ is context-switching overhead between the AI tool and the analyst's normal workspace, and $T_{\text{residual}}$ is residual hands-on investigation time. NIV $> 1$ indicates genuine acceleration; NIV $\leq 1$ indicates that the AI layer imposes more overhead than it saves.

\noindent
\emph{Fast-verification scenario.} Suppose a Tier-1 analyst's manual triage time is $T_{\text{manual}} = 8$~min; an LLM summarizer adds $T_{\text{infer}} = 6$~s and $T_{\text{switch}} = 30$~s, the analyst spends $T_{\text{verify}} = 90$~s checking the summary, and residual hands-on investigation drops to $T_{\text{residual}} = 4$~min. Then $T_{\text{AI-assisted}} = 6 + 90 + 30 + 240 = 366$~s ($\approx 6.1$~min) and $\mathrm{NIV} = 480/366 \approx 1.31$, a 31\% net speedup.

\noindent
\emph{Slow-verification scenario.} If verification time rises to 5~min because the analyst no longer trusts the summary at first glance, $T_{\text{AI-assisted}} = 6 + 300 + 30 + 240 = 576$~s and $\mathrm{NIV} = 480/576 \approx 0.83$. The AI layer is operationally net-slower despite identical accuracy. The example illustrates why latency, verification cost, and tool-switch overhead must be reported alongside accuracy in SOC AI evaluations.

\section*{S3: Research-Agenda Traceability (R1--R9)}

Table~\ref{tab:supp-agenda-traceability} reproduces the full traceability of the nine-direction research agenda. Each row links a research direction (Section~8 of the main paper) to the open challenges it primarily targets (C1--C6, Section~7), the evidence gap from the cross-category synthesis (Section~6) it would close, and an operational measurement plan.

{\small
\setlength{\tabcolsep}{4pt}
\renewcommand{\arraystretch}{1.20}
\begin{longtable}{@{}L{0.05\linewidth} L{0.20\linewidth} L{0.10\linewidth} L{0.26\linewidth} L{0.31\linewidth}@{}}
\caption{Supplementary Table S3: Traceability of the nine-direction research agenda (R1--R9).}
\label{tab:supp-agenda-traceability}\\
\toprule
ID & Direction & Challenges & Evidence gap closed & Operational measurement plan \\
\midrule
\endfirsthead
\caption[]{Research-agenda traceability (continued)}\\
\toprule
ID & Direction & Challenges & Evidence gap closed & Operational measurement plan \\
\midrule
\endhead
\midrule \multicolumn{5}{r@{}}{Continued on next page}\\
\endfoot
\bottomrule
\endlastfoot
R1 & Federated SOC learning networks & C1, C3 & External-validity gap (1/87 cross-environment) & Per-org QRR-at-FNR delta vs.\ centralized model; non-IID handling; participation-incentive design \\
R2 & Security foundation models & C2, C3 & LLM Tier-1 scarcity (2/20 in Cat~IV) & Triage accuracy on real alerts; hallucination rate; nDCG@$k$; comparison to RAG over the same retrieval base \\
R3 & Production-representative benchmark infrastructure & C3, C6 & Tier-1 scarcity overall (9/87); dataset-orientation gaps & Public, anonymized SOC-alert benchmark; high-fidelity simulator; community leaderboard with shared metric set \\
R4 & Adversarially aware SOC AI & C2 & Robustness gap (rare adversarial evaluation in Cat~II/IV) & ADR under specified evasion; poisoning resilience; game-theoretic adaptive-attacker simulation~\cite{ref61} \\
R5 & Agentic SOC architecture and governance & C5 & Collaboration mode gap (7/87 A2C-Collaboration) & Action-space audit; safe-exploration metric; human-approval triggers; immutable audit log compliance \\
R6 & Multimodal SOC reasoning & C3, C4 & Single-modality bias in Cat~III/IV & Joint log/trace/binary/screenshot reasoning accuracy; cross-modal evaluation harness~\cite{ref60} \\
R7 & Operational latency / NIV & C4 & Sparse scalability evidence in Cat~III/IV & Net Investigation Velocity (NIV); end-to-end latency budget under enterprise loads (worked example in Supplementary Section S2) \\
R8 & Adversarial GenAI / prompt-injection robustness & C2, C5 & LLM-specific threat-model gap & Direct/indirect prompt-injection benchmarks~\cite{ref50}; tool-misuse and context-poisoning audit \\
R9 & Closed-loop feedback mandate & C5, C6 & Closed-loop operationalization gap & Auditable upstream-tuning interface; suppression-drift monitor; attacker-controlled blind-spot test \\
\end{longtable}
}

% Use a natbib-compatible numeric style; supplementary readability does not
% require ACM-formatted entries in the bibliography list.
\bibliographystyle{plainnat}
\bibliography{SOC_Survey}